\newcommand\ie{\emph{i.e.}}
\newcommand\eg{\emph{e.g.}}

\documentclass{aa}  
\usepackage{graphicx}
\usepackage{booktabs}
\usepackage{txfonts}
\usepackage{lipsum}
\usepackage{subcaption}         

\usepackage{lscape}             

\usepackage{placeins}

\begin{document}

	\title{Multimessenger spectroscopy}

	\author{Kristen Lackeos\inst{1,}\inst{2}\corrauth{kristen.lackeos@dzastro.de}       
	}
	
	\institute{Deutsches Zentrum für Astrophysik (DZA), Postpl. 1, 02826, Görlitz, Deutschland
		\and Max-Planck-Institut f{\"u}r Radioastronomie (MPIfR), Auf dem H{\"u}gel 69, 53121, Bonn, Deutschland\\}

	\abstract
	{}
	{We establish a new multimessenger method for scheduling optical spectroscopic measurements of ultracompact double white dwarf binaries observed with the Laser Interferometer Space Antenna (LISA). The central idea is to use the orbital parameters inferred from the gravitational wave (GW) signal to predict the argument of latitude $u(t)$, which sets the phase dependence of the spectroscopic radial velocity. With $u(t)$ known in advance, radial velocities can be measured at selected orbital phases. The resulting multimessenger ephemeris gives complementary access to binary parameters that can be difficult to determine from either messenger alone, in particular the individual component masses and the source distance.}
	{The link between the GW and the spectroscopic descriptions is $u(t)=\Phi_{\rm GW}(t)/2+\delta_{e,\omega}(t)+\mathcal{O}(e^2)$, where $\delta_{e,\omega}$ is the first-order equation of the centre correction for an eccentric orbit with an apsidal advance. It vanishes for a circular orbit, where the argument of latitude is half the GW phase. The LISA parameter posteriors determine $\Phi_{\rm GW}(t)$, and for an eccentric orbit, the correction $\delta_{e,\omega}(t)$ is required to recover the true orbital position. Knowledge of the GW phase allows the optical observer to schedule targeted observations for a particular orbital phase, for example at quadrature or conjunction. We propose a linear regression of the radial velocities against $\cos u$. For double-lined spectroscopic binaries, the two slopes give the radial velocity semi-amplitudes $K_1$ and $K_2$. Combining the semi-amplitudes with Kepler's third law, the orbital inclination and period measured by LISA allows one to derive both component masses and a source distance. We derived small-eccentricity gravitational waveforms for ultracompact binaries, which are linear in the Laplace--Lagrange parameters. We performed a targeted search for binaries in circular and eccentric orbits in simulated LISA data. For each system we forecast the time window over which the predicted $u(t)$ remains accurate enough to schedule phase-resolved spectroscopic observations. Within this window, the velocity error incurred from the uncertainty in the predicted argument of latitude remains below the measurement uncertainty of the optical radial velocities.}
	{Because the masses are derived from the velocity amplitudes rather than from the frequency drift $\dot f_{\rm GW}$, they avoid the biases that affect masses inferred from $\dot f_{\rm GW}$ alone, whether from tides, mass transfer, or acceleration along the line of sight. For a representative four-year LISA observation, the ephemeris remains sufficiently accurate to support phase-resolved spectroscopy as part of a multimessenger observing programme for years after the GW observation ends. When propagated four years beyond the midpoint of that observation, the timing uncertainty associated with the predicted orbital phase is $0.3~\mathrm{s}$ for the eccentric system and $0.9~\mathrm{s}$ for the circular system, values that are competitive with that of optical timing ephemerides.}
	{}

	
	\maketitle
	\nolinenumbers

	\section{Introduction}
	\label{sec:intro}
	Over the past several decades, the science of close binary systems \citep{2014LRR....17....3P, 2022ARep...66S...5C} has advanced significantly through the use of multi-wavelength observations. In this paper we focus on a subclass of stellar binaries called ultracompact binaries (UCBs), which have orbital periods ranging from minutes to greater than one hour. In the X-ray band, analogous systems are classified as ultracompact X-ray binaries, a designation sometimes extended to systems with periods exceeding one hour \citep{ 1987ApJ...312L..17S, 1999A&A...349L...1I, 2002MNRAS.332L...7R, 2010ApJ...724..417G, 2010PASP..122.1133S, 2018arXiv180701060N, 2018PhRvL.121m1105T, 2020ApJ...900L...8C, 2021MNRAS.503.3540C,
		2023A&A...677A.186A,
		2023ApJ...944...83Q,
		2024ApJ...961..110Q,
		2025A&A...700A.107G, 2025arXiv251202360M}. Historically, X-ray surveys have led to the discovery of the shortest period systems because accreting compact objects (neutron stars, black holes) are identifiable from a single observation. On the other hand, when observed photometrically, short-period white dwarf (WD) binaries are unremarkable in individual exposures. Identifying them as systems of short orbital period requires time-domain surveys capable of resolving periodic variability on minute-to-hour timescales.
	Today, the discovery and characterisation of the UCB population is driven by
	optical photometric and spectroscopic observations
	\citep{2010ApJ...717L.108K,2010ApJ...711L.138R,Burdge2019a,
		2020ApJ...905...32B,2024A&A...683L..10K,2026ApJ..1000..237C, 2026ApJ..1010..121L};
	targeted spectroscopic surveys, including the Supernova Ia Progenitor survey
	(SPY), the Extremely Low Mass survey, and the double-lined double
	white dwarf survey (DBL)
	\citep{2020A&A...638A.131N,2020ApJ...889...49B,2024MNRAS.532.2534M};
	and astrometric surveys
	\citep{2021A&A...649A...1G,2024ApJ...963..100K}.
	The Vera C. Rubin Observatory's Large-aperture Synoptic Survey Telescope Deep Drilling Field programme promises to expand this population by reaching faint UCBs inaccessible to current surveys \citep{2023PASP..135j5002H}.
	
	In the near future, UCBs will also be discovered and monitored with space-based gravitational wave (GW) detectors such as TianQin \citep{2016CQGra..33c5010L}, LISA \citep{amaroseoane2017laserinterferometerspaceantenna,2023LRR....26....2A},  the Deci-hertz Interferometer Gravitational wave Observatory (DECIGO) \citep{2021PTEP.2021eA105K}, the Lunar Gravitational Wave
	Antenna \citep{2025JCAP...01..108A}, and a combination of detectors \citep{2026arXiv260301330Y}. Moreover, UCBs will be observed at various stages of evolution \citep{2026arXiv260724951W}. Radio observations of UCBs remain rare, and the vast majority of known systems were optically or X-ray detected. Coherent radio emission has been detected in a handful of accreting systems, such as HM~Cancri \citep[e.g.][]{ 2002MNRAS.332L...7R, 2002A&A...386L..13I, Seto:2023rfh}, where the origin of the radio emission remains debated.\footnote{\cite{Zhou2024_J0737_LISA} discusses the detectability of binary pulsars in the LISA frequency band.} Population synthesis models predict tens to hundreds of millions of compact double WD binaries (dWD) in the Galaxy \citep{2001A&A...375..890N, 2017A&A...602A..16T,2017MNRAS.470.1894K,2022MNRAS.511.5936K}. More than 300 dWDs have been spectroscopically confirmed to date, with more than 20 predicted to be detectable with LISA based on their measured orbital parameters \citep{2011ApJ...727....3K,2011MNRAS.418L.157K,2011MNRAS.413L.101K,2012ApJ...757L..21H,2014MNRAS.444L...1K,2016ApJ...824...46B,2020ApJ...905...32B,2020ApJ...892L..35B,2020ApJ...905L...7B,2021ApJ...921..160C,2021ApJ...918L..14K,2022ApJ...933...94B,2023LRR....26....2A,2023MNRAS.525.1814M,2023MNRAS.518.5123M,2024ApJ...977..262C,2024NatAs...8..491L,2025ApJ...991...65B}. See \cite{2024MNRAS.532.2534M,2025MNRAS.541.3494M} for the discoveries and results of the DBL, which subsequently also produced a maintained database of compact dWD binaries.\footnote{https://github.com/JamesMunday98/ClosedWDbinaries} Recent surveys, including from \cite{2025arXiv250515580V} with the Zwicky Transient Facility, continue to expand this sample of optically bright, spectroscopically accessible systems, which make up a small part of the larger WD population that contribute foreground noise in the LISA frequency band.
	
	This paper addresses two topics: (a) the additional information that GW ephemerides provide for spectroscopic observations of dWDs in circular and eccentric orbits, and (b) new approaches for including small eccentricities in GW data analysis of UCBs. The majority of LISA's detectable UCBs are expected to be circular,\footnote{Astrophysical models predict the overwhelming majority of LISA UCBs will be circular. Even so, machine learning has been proposed as a means to detect binary systems with eccentricity in global fits that use circular waveforms \cite{2026A&A...714A..76T}.} but \cite{2025arXiv250515580V} find that residual eccentricities at the $10^{-3}-10^{-2}$ level are realised in nature, making unified treatment of both circular and eccentric orbits essential for comprehensive LISA GW data analysis.
	
	We explore the use of phase-coherent orbital ephemerides that LISA constructs for UCBs as a fiducial reference for scheduling spectroscopy. We use the term `observing baseline' as the time interval between spectroscopic observing epochs, which may be separated by months or years. Here, `phase-coherent' means that a single phase model tracks the orbital motion continuously across the observing baseline, preserving an unambiguous cycle count and enabling prediction of the orbital phase at later epochs. This in turn reduces the observational burden that has historically made radial velocity (RV) orbit determination for these systems costly. To frame this question more precisely, it is instructive to review how phase-connected ephemerides are established and used across different observational wavelengths.

	The construction of phase-coherent ephemerides from electromagnetic (EM) observations is well established in binary star astronomy. Perhaps the most familiar example is through timing a radio pulsar. One obtains a train of regular pulses that serve as fiducial markers for the neutron star's spin. By measuring pulse times of arrival and fitting a timing model (parametrized by rotational, positional, and, possibly, binary parameters), one obtains a continuous spin-phase prediction with an unambiguous cycle count (pulse number) over decades. This is what radio astronomers call a `phase-connected' solution or pulsar ephemeris. Fermi Large Area Telescope pulsar observations use radio timing ephemerides to phase-fold sparse gamma-ray photons over multi-year baselines \citep{2008A&A...492..923S}. For redback pulsars, where binary interaction disrupts radio timing, the direction is reversed: The uninterrupted gamma-ray photon stream supplies the phase-coherent reference that radio observations cannot maintain \citep{2015ApJ...807...18P}. In each case, the external reference isolates a signal that would otherwise be inaccessible. Spectral measurements at a targeted orbital phase, namely the `targeted-$u$' method proposed here, facilitate an analogous strategy, with LISA's coherent GW phase model as the reference for RV follow-up.
	
	For LISA, matched filtering against a coherent waveform model tracks the GW phase $\Phi_{\rm GW}(t)$ referenced to the barycentric coordinate time (TCB), the timescale adopted for LISA data \citep{2026arXiv260322377B}.  The waveform model is determined by eight parameters: the amplitude ($\mathcal A$), frequency ($f_{\rm GW, 0}$), frequency derivative ($\dot f_{\rm GW}$), initial phase ($\phi_0$) and polarisation angle ($\psi$) of the GW; the sky location, the ecliptic longitude and latitude $(\phi,\theta)$; and the cosine of orbital inclination ($\cos\iota$). The subscript $0$ marks the
	epoch ($t_0$) at the start of the LISA observation, from which the phase model is
	measured. Because the signal phase evolves predictably, matched filtering coherently
	combines $N_{\rm cycles}$ signal cycles, yielding a signal-to-noise ratio
	(${\rm S/N}$) that scales as $\sqrt{N_{\rm cycles}}$
	\citep{2019CQGra..36j5011R}. Over multi-year baselines, the accumulated GW phase extends
	coherently between epochs. Although the accumulated phase advance,
	$\Delta\Phi_{\rm GW}=\Phi_{\rm GW}(t_\beta)-\Phi_{\rm GW}(t_\alpha)$,
	may encompass hundreds of thousands of GW cycles, its uncertainty remains
	much smaller than $2\pi$, so both the integer cycle count and the residual
	phase within a cycle are known. The mechanism differs from radio timing (here, phase tracking of orbital motion
	rather than pulse-counting of spin), but in both cases the phase remains connected across the full baseline.
	
	Spectroscopic binary orbits are parametrized by the binary period; projected semi-major axes $a_1\sin\iota$, with $\iota$ as the orbital inclination (and $a_2\sin\iota$ in the case where spectra from the primary and secondary star are observable); eccentricity, initial position of periastron, $\omega_1$ of the primary star;\footnote{Also called the `argument of periastron' or `argument of pericentre'.} and a temporal phase reference, such as the time of periastron passage ($T_0$). For eclipsing binaries, the binary period, mid-eclipse time of the primary (and possibly the secondary), and inclination are measurable. When two eclipses are timed per orbit, it is also possible to produce a lower limit on the eccentricity. By fitting photometric and RV data as functions of observation time, binary modelling recovers orbital ephemerides \citep{2006ASPC..349...71W}. Fitting timing models (including period derivatives) maintains phase-connected solutions over multi-year baselines \citep{2024ApJ...977..262C}. For WD UCBs that are not eclipsing,\footnote{\cite{2025arXiv250400548S} shows how eclipses depend on stellar radii, orbital inclination, and separation.} LISA's GW monitoring provides a phase-coherent ephemeris that eclipse timing would otherwise supply (Appendix~Sect.~\ref{app:appendixA}).
	
	In the LISA approach to defining barycentric orbital ephemerides for scheduling spectroscopic observations, the GW phase $\Phi_{\rm GW}(t)$ maps to the orbital phase via 
	\begin{equation}
		u(t) = \Phi_{\rm GW}(t)/2 + \mathcal{O}(e),
		\label{eq:phase_map}
	\end{equation}
	where $u$ is the orbital phase measured from the ascending node. The origin of $\Phi_{\rm GW}$ is set by the same node convention \citep{2003PhRvD..67b4015C}. Generally, for eccentric orbits the orbital phase is called the argument of latitude.\footnote{The term `argument of latitude' is historical; $u(t)$ is the argument (input angle) from which the body’s celestial latitude above or below the reference plane is determined. At the nodes ($u = 0, \pi$), the body crosses the reference plane, while intermediate values of $u$ correspond to excursions above or below it. Different observational disciplines adopt different phase conventions. The radial velocity is expressed in terms of the true anomaly, $\nu(t)$, and the argument of periastron, $\omega$, with phase zero often defined at periastron passage or arbitrary phase offset. In light curve analyses, the orbital phase is sometimes defined from conjunction rather than periastron, leading to shifted definitions of phase angles (e.g. $\phi(t) = \nu(t) + \omega - \tfrac{\pi}{2}$ for the longitude measured from conjunction). In the latter two cases, the terminology shifts to `longitude'.} Here, $\Phi_{\rm GW}/2$ is the mean orbital phase, while $u$ is the true orbital phase. The $\mathcal{O}(e)$ term is the leading difference between them and vanishes for a circular orbit. The $\mathcal{O}(e)$ term, or equation of the centre to $\mathcal{O}(e)$, is constrained by LISA GW data analysis through separate `eccentricity parameters' and is not accounted for with $\Phi_{\rm GW}$. This mapping enables phase-resolved spectroscopic scheduling (with targeted-$u$ observation) over multi-year baselines without repeated observations distributed across orbital phase (hereafter, dense RV sampling). 
	
	Even with dense RV sampling, it is possible for the S/N to be limited. \citet{2020ApJ...905L...7B} obtained $312$ Keck/LRIS exposures of the $8.8$ min eclipsing dWD ZTF~J2243+5242 but were unable to measure RV semi-amplitudes, citing the faintness of the source, the short exposures required to preserve temporal resolution, and a large readout duty cycle. \citet{2020ApJ...905L...7B} argue that for the broader UCB population the Vera C. Rubin Observatory and LISA will discover, phase-resolved spectroscopy of more than a handful will be impossible without substantial time on a very large telescope such as the Extremely Large Telescope (ELT). The question here becomes whether GW phase information, when available, will help alleviate this burden.
	
	\citet{2020ApJ...892L..35B} show that for the non-eclipsing single-lined system J2322+0509, with an orbital period of 1201 s, the phase of the RV orbit supplies the timing observable normally provided by eclipses. Using a predicted GW orbital decay, $\dot P_b=-2.2\times10^{-12}~{\rm s~s^{-1}}$, \citet{2020ApJ...892L..35B} estimate that with $\pm 10$ s epoch errors and RV epoch measurements obtained every other year, a $5\sigma$ measurement of $\dot P_b$ over a 14-year baseline is projected. At such low $f_{\rm GW}$, LISA still provides the coherent orbital phase needed for ephemeris prediction, but the additional phase evolution due to $\dot f_{\rm GW}$ is comparatively small over a 4-year mission. A long optical RV timing baseline provides an independent constraint on the quadratic term in the phase, whose uncertainty limits how far the ephemeris can be predicted forward. The source has $f_{\rm GW}=2/P_b=1.66~{\rm mHz}$, which is near the low-frequency end of the known verification binary sample \citep{2025ApJ...994..152L}, where the scaling for a detached binary driven by gravitational radiation, $\dot f_{\rm GW}\propto M_c^{5/3} f_{\rm GW}^{11/3}$ with chirp mass $M_c$, makes the chirp difficult to resolve over the mission.  \citet{2020ApJ...892L..35B} recommend an optical timing baseline be established well before LISA starts observing these systems, which is crucial for the eventual joint EM-GW analysis.
	
	For systems where the GW phase forecast is already accurate enough to schedule spectroscopy, the targeted-$u$ method developed here reverses the role of the timing reference. Rather than using dense RV coverage to construct the orbital clock, one uses the parameters inferred from the GW signal to place sparse spectra at predicted orbital phases. Knowledge of the observing phase of each spectrum allows the RV semi-amplitudes to be measured efficiently, providing a route to physical parameters that LISA alone does not generally determine, including the individual component masses and, when it is not otherwise constrained by the GW measurement, the source distance. The resulting masses test the physical interpretation of the observed orbital evolution, while the EM constraints provide priors for subsequent LISA analyses.

	Ultracompact binary systems are the most numerous GW sources in the LISA band \citep{2019MNRAS.490.5888L}, and the list of compelling science cases is steadily growing \citep{2023A&A...678A.123L}. Combined optical and GW observations of compact binaries provide complementary constraints on binary evolution, compact stellar remnants, WD structure, pathways to fast optical transients, and component mass and source distance determination.\footnote{\url{https://ag2025.astronomische-gesellschaft.de/posters/Kupfer_Thomas_91.pdf}} Maximising the scientific return from these systems demands continued development of state-of-the-art methods, not only in instrumentation and analysis techniques but also in advantageous observing and scheduling
	strategies. \cite{2025arXiv251212075M} emphasise that phase-resolved spectroscopy is essential to constraining and obtaining accurate masses.
	
	\citet{2025arXiv251214800P} note that resolving the narrow H$\alpha$ core in dWDs requires optical spectroscopic resolutions of $R > 20000$, restricting spectroscopic characterisation to targets with magnitudes $G \lesssim 18$--$20$, even with ANDES on the ELT, and they project at most $\sim 250$ dWDs will be characterised this way by the early 2040s. Sensitivity is not the only constraint. Even when a full orbital campaign is feasible, the RV curve is not uniformly usable. Spectral lines from the two stellar components blend near conjunction and are maximally separated at quadrature \citep{2022ARep...66S...5C}. A GW-derived orbital ephemeris predicts when these favourable phases occur, allowing spectroscopic observations to be targeted rather than distributed across the orbit. We pursue a resource-efficient orbital phase predictor built from joint GW and optical observations of UCBs. Namely, the GW phase $\Phi_{\rm GW}$ provides a route to phase-resolved spectroscopy without requiring eclipses or densely sampled RV curves. 
	
	Recent observations have shown that compact post-common-envelope dWDs need not be strictly circular. \citet{2025arXiv250515580V} report the first robust non-zero eccentricity measurements in short-period dWDs, finding $e \sim 2\times 10^{-3}/\cos\omega$ in two systems out of seven. Although longer orbital periods further complicate individual LISA detection due to overlapping confusion noise, these binaries indicate that residual eccentricities at the $10^{-3}$-$10^{-2}$ level are realised in nature and that apsidal motion in compact WD binaries is an observational, not just theoretical, possibility \citep{2007ApJ...665L..59W}. Evidence from other compact isolated binaries reinforces this. An eccentricity of $\sim 0.145$ has been inferred in the neutron star-black hole merger event GW200105 \citep{2026ApJ..1000L...2M}. The authors also discuss several mechanisms that possibly give rise to non-negligible orbital
	eccentricity. Together these results strengthen the case for analysis strategies that treat even mild eccentricities in the LISA dWD population. Motivated by this, we have developed methods to measure eccentricity with LISA and to use the resulting orbital phase information to guide multimessenger spectroscopy for both circular and eccentric ultracompact orbits. We describe the eccentricity through the Laplace--Lagrange (LL) parameters $\eta=e\sin\omega$ and $\kappa=e\cos\omega$, which appear in both the GW phase mapping and the optical RV model.

	In Sect.~\ref{sec:argument_of_lat} we introduce the Keplerian RV curve, ${\rm v_r}(t)$; the argument of latitude, $u$ (Table~\ref{tab:table1}); and the concepts of scheduling optical observations at a particular $u$ and the RV difference observable, $\Delta {\rm v}$. In Sect.~\ref{subsec:expected_magnitudes} we present order-of-magnitude estimates of quantities important for assessing the feasibility of targeted- and fixed-$u$ methods proposed for UCBs. We also bound the leading systematics: the gradual shift in semi-amplitude from inspiral, the blurring of the line profile caused by orbital motion during the interval over which spectra are combined, and barycentric drift from galactic acceleration. In Sect.~\ref{sec:applicationsU} we develop applications for the targeted-$u$ methodology for UCBs in circular and eccentric orbits and show how the component masses and the source distance are derived. In Sect.~\ref{sec:horizon_forecast} we discuss the mapping from the GW parameters to the argument of latitude $u$ for eccentric and circular orbits. We discuss the GW parameterisation for UCBs in eccentric orbits and the additional degrees of freedom this incurs in the waveform model. We then perform a targeted search for binaries in eccentric and circular orbits in simulated LISA data. Using the resulting parameter covariance matrix to compute $\sigma_u(t)$, we define the observational time window over which the orbital phase $u(t)$ extrapolated from the recovered GW parameters remains accurate to within the tolerance set by the RV measurement uncertainty. Finally, in Sect.~\ref{sec:mass_diagram_illustration} we place four loci in the $(M_1,M_2)$ plane. Namely, $K_1/K_2$ fixes the mass ratio, while $K_1+K_2$, together with the LISA inclination, period, and eccentricity, fixes the total mass. Their intersection determines the component masses without assuming what drives the orbital evolution. We compare this intersection with the remaining two loci, which are derived from the secular orbital evolution measured by LISA. The chirp mass follows from $f_{\rm GW}$, $\dot f_{\rm GW}$, and $e$ if the frequency evolution is driven solely by gravitational radiation reaction. A second estimate of the total mass follows from $f_{\rm GW}$, $\dot\omega$, and $e$ if the apsidal advance is purely relativistic.

	\section{Argument of latitude}
	\label{sec:argument_of_lat}
	
	The optically observable Keplerian RV curve is the sum of two terms:
	\begin{equation}
		{\rm v}_r(t) = \gamma + {\rm v}_{\rm b}(t).
		\label{eq:vr_radial_v}
	\end{equation}
	The systemic radial velocity is $\gamma$, which includes the line-of-sight (l.o.s.) velocity of the binary centre of mass and constant spectroscopic offsets, such as gravitational redshift.
	In the absence of a measurable change in the l.o.s. velocity of the binary centre of mass, $\gamma$ is constant. In a double-lined (SB2) system, the intrinsic offsets differ between the two stars, so the measured systemic offsets are written as $\gamma_1$ and $\gamma_2$. A change in the centre-of-mass contribution between epochs appears as a common shift in $\gamma_1$ and $\gamma_2$. We denote this change by $\Delta {\rm v}_{\rm CM}$.
	Following standard spectroscopic binary notation \citep{2025arXiv250400548S}, we express the radial velocities in terms of the primary star's barycentric orbital elements. For a binary with components labelled 1 (primary) and 2 (secondary), the l.o.s. orbital contributions, ${\rm v}_b$ are expressed in terms of the two absolute (barycentric) orbits, because RV curves trace each component's motion about the system centre of mass. These orbits are co-planar, have the same eccentricity, and share a common line of apsides \citep{2006ASPC..349...71W},
	\begin{equation}
		\begin{aligned}
			{\rm v}_{\rm b,1}(t) &= K_1[\cos(\omega_1+\nu_1(t)) + e\cos\omega_1] \\
			&= K_1[\cos u_1(t) + e\cos\omega_1], \\
			{\rm v}_{\rm b,2}(t) &= -K_2[\cos(\omega_1+\nu_1(t)) + e\cos\omega_1] \\
			&= -K_2[\cos u_1(t) + e\cos\omega_1],
		\end{aligned}
	\end{equation}
	where $\nu_1(t)$ is the true anomaly of star 1 in its barycentric orbit, $\omega_1$ is star 1's argument of periastron, and $u_1 = \omega_1 + \nu_1$ 
	is star 1's argument of latitude measured from the ascending node. For $e=0$, where periastron is undefined, we retain $u_1$ as the orbital angle measured from the ascending node. Both radial velocities are expressed in terms of the same $(u_1, \omega_1, e)$ from star 1's barycentric orbit, with the negative sign for star 2 arising from the relation $\omega_2 = \omega_1 + \pi$ between the two stars' arguments of periastron. Throughout, ${\rm v}_r>0$ corresponds to recession (redshift) and ${\rm v}_r<0$ to approach (blueshift).
	
	The semi-amplitudes $K_1$ and $K_2$ are defined as
	\begin{equation}
		K_j = \frac{2\pi a_j \sin \iota}{P_b\sqrt{1-e^2}}, \quad j \in \{1,2\},
		\label{eq:semiamplitude_eq}
	\end{equation}
	where $a_j$ is the semi-major axis of star $j$ in its barycentric orbit, $P_b$ the orbital period, $\iota$ the inclination, and $e$ the eccentricity 
	(identical for both stars). The mass ratio $q = M_2/M_1$ relates the semi-amplitudes as $K_1 = q \cdot K_2$ and the semi-major axes as $a_2 = a_1/q$.
	
	For systems where just one star's spectral component is observed, we use the notation $K = K_1$ and $u = u_1$, dropping subscripts for clarity.
	The true anomaly $\nu(t)=\nu_1=\nu_2$ completes a full cycle each orbit, while $\omega$ evolves secularly due to relativistic periastron advance. On the orbital timescale, $\omega$ is effectively constant and $u(t) = \omega + \nu(t)$ traces the star's position in its orbit. 
	
	Using the GW-derived orbital ephemeris, spectroscopy is scheduled at a chosen argument of latitude $u_\star$. Radial velocity extrema (maximum recession and approach) occur at $u=0$ and $u=\pi$ (respectively at the ascending and descending nodes), exactly for all $e$, 
	while RV zero crossings satisfy $\cos u = -e\cos\omega$. For $e \ll 1$,
	\begin{equation}
		u \simeq \frac{\pi}{2} + e\cos\omega
		\quad\text{or}\quad
		u \simeq -\frac{\pi}{2} - e\cos\omega.
		\label{eq:RVzero_cross}
	\end{equation}
	At these zero crossings, the l.o.s. orbital component does not contribute to the sum, isolating the systemic motion.
	
	Radial velocities inferred from optical spectra correspond to the sum in Eq.~\eqref{eq:vr_radial_v}, measured relative to laboratory rest wavelengths. We use `quadrature' for the RV extrema at $u = 0, \pi$ and `conjunction'  for the RV zero crossings at $u = \pm\pi/2$ (exact for circular orbits, with $\mathcal{O}(e)$ corrections for eccentric orbits).
	See Table~1 in \citet{2025arXiv250400548S} for an overview of the system parameters accessible through individual techniques or joint observations of eclipsing, spectroscopic, and astrometric binaries.
	\begin{table*}[!tp]
	\caption{Notation used in this work.}
	\label{tab:table1}
	\centering
	\setlength{\tabcolsep}{4pt}
		\begin{tabular}[t]{@{}p{1.3cm}p{\dimexpr0.48\textwidth-1.3cm-2\tabcolsep\relax}@{}}
			\hline\hline
			Symbol & Definition \\
			\hline
			$P_{\rm b}$ & Periastron to periastron orbital period \\
			$a$ & Semi-major axis \\
			$\iota$ & Orbital inclination \\
			$e$ & Orbital eccentricity \\
			\noalign{\smallskip}
			$\omega$ & Argument of periastron \\
			$\dot\omega$, $\dot\omega_{\rm tot}$, $\dot\omega_{\rm LISA}$ & Apsidal advance rate; $\dot\omega_{\rm tot}$ is the total, comprising relativistic, tidal, and rotational terms, and $\dot\omega_{\rm LISA}$ is its value recovered from the GW data \\
			$\nu$ & True anomaly \\
			$\mathcal{M}$ & Mean anomaly \\
			$n_{\rm b}$ & Anomalistic mean motion, $2\pi/P_{\rm b}=\dot{\mathcal M}$ \\
			$\bar n_{\rm b}$ & Mean orbital phase rate, $\dot\Phi_{\rm asc}=\pi f_{\rm GW}=n_{\rm b}+\dot\omega$ \\
			$u$ & Argument of latitude, $\omega+\nu$, measured from the ascending node \\
			$u_\star$ & Target argument of latitude at which a spectroscopic observation is scheduled \\ 
			$t_\star$, $t_{\star,k}$ & Epoch at which the predicted argument of latitude reaches $u_\star$, indexed by $k$ over successive passages \\
			$T_0$ & Time of periastron passage \\
			$T_{\rm asc}$ & Mean orbital phase reference epoch, $\Phi_{\rm asc}(T_{\rm asc})=0$ \\
			$t_0$, $t_{\rm mid}$ & Reference epochs: the start of the LISA observation and its midpoint, $t_{\rm mid}=t_0+T_{\rm obs}/2$ \\
			$(\eta,\kappa)$ & Laplace--Lagrange parameter pair $(e\sin\omega(t),~e\cos\omega(t))$; subscripts `LISA' and $s$ denote LISA-inferred and session values, respectively \\
			$\Phi_{\rm asc}$ & Mean orbital phase measured from the ascending node,
			$\mathcal{M}+\omega(t)$; abbreviated $\Phi$ in Appendix~\ref{app:appendixD}.
			Reduces to $\bar n_{\rm b}(t-T_{\rm asc})$ for constant $\bar n_{\rm b}$ \\
			\noalign{\smallskip}
			$\delta_{e,\omega}$ & Equation of the centre to $\mathcal{O}(e)$, or first order eccentricity correction to the argument of latitude, $2\kappa\sin\Phi_{\rm asc}-2\eta\cos\Phi_{\rm asc}$ \\
			$M_j$ & WD mass of component $j$ \\
			$R_j$ & Stellar radius of component $j$ \\
			$d$ & Distance to the source \\
			$q$ & Mass ratio, $M_2/M_1=K_1/K_2$ \\
			$M_{\rm tot}$ & Total binary mass, $M_1+M_2$ \\
			$M_{\rm c}$ & Chirp mass, $(M_1 M_2)^{3/5}/(M_1+M_2)^{1/5}$ \\
			\hline
		\end{tabular}
		\hfill
		\begin{tabular}[t]{@{}p{1.3cm}p{\dimexpr0.48\textwidth-1.3cm-2\tabcolsep\relax}@{}}
			\hline\hline
			Symbol & Definition \\
			\hline
			$K$ & Half-amplitude of the RV curve \\
			${\rm v}_{\rm CM}$ & Centre-of-mass velocity of the binary \\
			$z$ & Gravitational redshift, constant per star \\
			$\gamma$ & Systemic velocity, ${\rm v}_{\rm CM}+z$ \\
			${\rm v}_r$ & Line-of-sight radial velocity of a binary component \\
			${\rm v}_b$ & Orbital velocity \\
			\noalign{\smallskip}
			$\sigma_{\rm spec}$ & RV uncertainty of a single spectrum \\
			$\sigma_{\rm v}$ & Single-epoch RV measurement uncertainty, $\sigma_{\rm spec}/\sqrt{N_{\rm RV}}$ for $N_{\rm RV}$ spectra \\
			$\sigma_\Delta$ & Uncertainty in the two-epoch RV difference, $\sqrt{2}~\sigma_{\rm v}$ \\
			$\sigma_u$ & Uncertainty in the predicted argument of latitude from the LISA posterior \\
			$\delta u_{\rm tol}$ & Phase tolerance on the predicted argument of latitude, set by the RV measurement uncertainty \\
			$\alpha$ & Factor multiplying the phase tolerance \\
			$T_{\rm valid}$ & Phase-coherence horizon for targeted spectroscopy \\
			$\tau_{\rm chirp}$ & GW frequency evolution timescale, $f_{\rm GW}/\dot f_{\rm GW}$ \\
			$\delta t$ & Timing uncertainty of a scheduled observation \\
			$\Delta T$ & Baseline between two spectroscopic epochs \\
			\noalign{\smallskip}
			$\Phi_{\rm GW}$ & GW phase of the dominant $n=2$ mode \\
			$\phi$ & Value of $\Phi_{\rm GW}$ at the reference epoch \\$f_{\rm GW}$ & GW frequency, dominant mode \\
			$\dot f_{\rm GW}$ & Time derivative of the GW frequency \\ 
			$n_{\rm GW}$ & GW angular frequency, $2\pi f_{\rm GW}=2\bar n_{\rm b}$ \\ 
			$\dot n_{\rm GW}$ & Time derivative of $n_{\rm GW}$, $2\pi\dot f_{\rm GW}$ \\
			$\mathcal{A}$ & GW strain amplitude \\
			$\psi$ & GW polarisation angle \\
			$z_\omega$ & Apsidal cycles accumulated over the LISA observation, $(\dot\omega/2\pi)T_{\rm obs}$ \\
			\hline
		\end{tabular}
		\tablefoot{A subscript $0$ denotes a quantity at the
		analysis epoch $t_0$, as sampled by the LISA Markov Chain Monte Carlo (MCMC), while a subscript
		${\rm mid}$ denotes the corresponding quantity at the midpoint reference
		epoch $t_{\rm mid}$, obtained in post-processing. Reported parameter values
		refer to $t_{\rm mid}$.}
		\end{table*}

	The GW phase model allows radial velocities obtained at widely separated
	epochs to be compared at the same argument of latitude. These fixed-$u$
	differences provide access to secular changes in the orbital terms and to
	changes in the barycentric velocity. We consider
	two epochs, at times $t_\alpha$ and $t_\beta$, satisfying
	\begin{equation}
		u(t_\alpha)=u(t_\beta).
	\end{equation}
	For circular orbits, this corresponds to the same orbital phase at both
	epochs. For eccentric orbits, $u=\omega+\nu$, so the fixed-$u$ condition
	implies
	\begin{equation}
		\nu(t_\beta)-\nu(t_\alpha)
		=
		-\left[\omega(t_\beta)-\omega(t_\alpha)\right].
	\end{equation}
	
	Evaluating Eq.~\eqref{eq:vr_radial_v} at the same $u_\star$ in both epochs and subtracting the earlier from the latter, the differenced $\cos(\omega+\nu)$ terms cancel, leaving the two-epoch RV difference,
	\begin{equation}
		\begin{aligned}
			\Delta {\rm v}_r(u_\star; t_\alpha, t_\beta)
			&{}= \Delta {\rm v}_{\rm CM} + \Delta K\cos u_\star \\
			&{}+ K(t_\beta)~e(t_\beta)\cos\omega(t_\beta) - K(t_\alpha)~e(t_\alpha)\cos\omega(t_\alpha).
		\end{aligned}
		\label{eq:fixedu_exact}
	\end{equation}
	The apsidal drift contribution depends on the difference between $\kappa(t) = e\cos\omega(t)$ at each epoch. The barycentric RV drift, $\Delta {\rm v}_{\rm CM} = {\rm v}_{\rm CM}(t_\beta) - {\rm v}_{\rm CM}(t_\alpha)$, has contributions from physical acceleration of the binary centre of mass, for example due to a wide tertiary companion or the Galactic potential, and from kinematic effects such as perspective acceleration. For systems with constant 3D velocity, a changing l.o.s. geometry yields a non-zero $\Delta {\rm v}_{\rm CM}$ \citep{1970SvA....13..562S}. Systematics in the data also contribute to $\gamma$. 
	
	For detached UCBs the secular change in the RV semi-amplitude between the two epochs, $\Delta K = K(t_\beta) - K(t_\alpha)$, is set mainly by GW-driven orbital inspiral. In Sect.~\ref{subsec:expected_magnitudes} we show that $|\Delta K \cos u_\star|$ is bounded below the spectroscopic precision floor and the apsidal drift amplitude for the UCBs of interest. We subsequently dropped the $\Delta K$ contribution in what follows and worked with the reduced form of the two-epoch RV difference,
	\begin{equation}
		\Delta {\rm v}_r(u; t_\alpha, t_\beta) 
		= \Delta {\rm v}_{\rm CM} + K[e(t_\beta)\cos\omega(t_\beta) - e(t_\alpha)\cos\omega(t_\alpha)].
		\label{eq:fixedu_no_K_evolution}
	\end{equation}
	
	Precision Doppler treatments of close binaries account for post-Keplerian effects that modify the Keplerian RV curve. These include the relativistic advance of periastron, the quadratic Doppler effect from orbital motion, and the gravitational redshift associated with the companion potential \citep{1999ApJ...523..771K,Zucker_2006,2022ARep...66S.412C}. For an eccentric orbit, apsidal advance contributes to the $\mathcal{O}(e)$ correction in the mapping between the GW phase and the argument of latitude in Eq.~\eqref{eq:phase_map}. Its secular evolution can be measured with spectroscopy or with GWs, and can also be traced through long-baseline eclipse timing.
	The constant parts of the quadratic Doppler shift and the gravitational shifts due to the companion's potential are retained in the component offsets $z_j$ discussed below. For eccentric orbits, the varying orbital speed and binary separation also produce small phase-dependent contributions to these shifts, causing distortions of the Doppler curve that require a precision of $\sim 1~{\rm ms^{-1}}$ to measure \citep{2022ARep...66S.412C}, much better than what is possible for the optical spectroscopic measurements of UCB components considered here (Sect.~\ref{subsec:expected_magnitudes}).
	
	\section{Expected signal magnitudes}
	\label{subsec:expected_magnitudes}
	
	In this section we present order-of-magnitude estimates meant to motivate the prospect of measuring targeted-$u$ observables of Sect.~\ref{sec:applicationsU}. We introduce the quantities relevant to multimessenger spectroscopy. The GW phase uncertainty $\sigma_{\phi_{\rm GW}} \sim \text{\ss}/(S/N)$, with $\text{\ss}\gtrsim1$ accounting for parameter correlations \citep{2003PhRvD..67b4015C}, maps to an uncertainty $\delta t$ in the predicted time at which the binary reaches a specified argument of latitude,
	\begin{equation}
		\delta t = \frac{\text{\ss}~ P_b}{4\pi~(S/N)}
		\approx 0.25
		\left(\frac{\text{\text{\ss}}}{1}\right)
		\left(\frac{P_b}{5~\mathrm{min}}\right)
		\left(\frac{100}{S/N}\right)~\mathrm{s}.
		\label{eq:timing_precision_Cutler}
	\end{equation}
	Eclipse arrival times are measured to $\simeq 0.5~{\rm s}$ precision with high cadence cameras on large telescopes \citep{2021MNRAS.507..350D, 2025arXiv250515580V}. Unlike the estimates for optical phase tracking below, this expression is not an extrapolation over a future baseline $\Delta T$. The observing duration is accounted for through the accumulated coherent LISA S/N, set by the number of GW cycles tracked during the mission. This should be regarded as a simple estimate of the timing uncertainty associated with the LISA GW reference phase. The full forecast depends on the covariance among phase, frequency, and frequency derivative, as well as the eccentricity parameters, when present. These correlations propagate into the future phase uncertainty $\sigma_u(t)$. A more complete scheduling horizon analysis in Sect.~\ref{sec:horizon_forecast} propagates the full covariance to identify the admissible observing windows and the epochs of best timing precision.
	
	To compare the GW timing error $\delta t$ with its optical counterpart $\delta t_{\rm optical}$, we estimate the error accumulated when an optically derived orbital ephemeris is propagated beyond its reference epoch.
	Optical timing uncertainty depends on contributions from phase offset, frequency, and frequency derivative uncertainties:
	\begin{equation}
		\delta t_{\rm optical}(\Delta T)\simeq
		\left[
		\left(\frac{\sigma_{\phi_0}}{f}\right)^2
		+
		\left(\frac{\sigma_f}{f}\Delta T\right)^2
		+
		\left(\frac{\sigma_{\dot f}}{2f}\Delta T^2\right)^2
		\right]^{1/2},
		\label{eq:optical_timing_prec}
	\end{equation}
	where $\sigma_{\phi_0}$, $\sigma_f$, and $\sigma_{\dot f}$ are the
	1$\sigma$ uncertainties in those quantities, respectively, with $\phi_0$ in units of
	cycles. This expression neglects covariance terms between the fitted
	ephemeris parameters. For HM~Cnc, two decades of optical timing by
	\cite{2023MNRAS.518.5123M} give
	$\sigma_f/f\simeq9.6\times10^{-9}$,
	$\sigma_{\dot f}/f\simeq3.2\times10^{-17}~{\rm s}^{-1}$,
	and a reference phase uncertainty of $\sigma_{\phi_0}=0.004$ cycles.
	The frequency evolution terms give $\delta t\sim1.2$~s after
	$\Delta T=4$~yr, while including the reference phase uncertainty
	and treating the parameter uncertainties as independent gives
	$\delta t\sim1.8$~s. For comparison, an optically adjusted X-ray ephemeris for HM Cnc of \citet{2010ApJ...711L.138R} gives $\sigma_f/f\simeq3.2\times10^{-8}$ and
	$\sigma_{\dot f}/f\simeq6.4\times10^{-16}~{\rm s}^{-1}$.
	The underlying X-ray timing solution had
	$\sigma_{\phi_0}=0.0014$ cycles, which produces
	$\delta t\sim6.5$~s after $\Delta T=4$~years under the same
	independent parameter assumption. 
	For J0651+2844, the optical eclipse timing of
	\citet{2012ApJ...757L..21H} gives
	$\sigma_f/f\simeq7.2\times10^{-8}$,
	$\sigma_{\dot f}/f\simeq3.7\times10^{-15}~{\rm s}^{-1}$,
	and a reference phase uncertainty of
	$\sigma_{\phi_0}\simeq9.5\times10^{-4}$ cycles.
	The frequency evolution terms alone yield $\delta t\sim30.9$~s after
	$\Delta T=4$~yr.
	
	The role of LISA in the present work is not to replace eclipse or stable photometric timing in systems where those clocks exist. Rather, for optically accessible LISA binaries, LISA supplies an independent, phase-coherent GW phase model that determines the argument of latitude at the spectroscopic epochs. In this role, the GW phase prediction partly substitutes for the reference otherwise built from phase-resolved spectroscopy, reducing the optical coverage required to construct the local RV solution. When an optical timing baseline exists, it supplies an additional clock. When such a baseline is absent or incomplete, the GW-derived phase model provides the phase reference for targeted-$u$ spectroscopy.
	
	\subsection{Numerical estimates}
	For small values of eccentricity ($e\lesssim 0.1$), a UCB system with $P_b=5~\mathrm{min}$ and $M_1=M_2=0.6M_{\odot}$ has equal RV semi-amplitudes,
	\begin{equation}
		\begin{aligned}
			K_{1,2}/\sin \iota = {}& 747\left(\frac{M_2}{0.6M_{\odot}}\right)\left(\frac{1.2M_{\odot}}{M_1+M_2}\right)^{2/3} \\
			&\times \left(\frac{5~{\rm min}}{P_b}\right)^{1/3}\left(\frac{1}{1-e^2}\right)^{1/2} {\rm km~s^{-1}},
		\end{aligned}
		\label{eq:semiamp_est}
	\end{equation}
	with $M_{\rm tot}=M_1+M_2$.
	The general relativistic apsidal motion at the leading post-Newtonian (PN) order is
	\begin{equation}
		\dot{\omega_{\rm GR}} = \frac{3}{(1-e^2)}\left(\frac{2\pi}{P_b}\right)^{5/3}\left(\frac{G M_{\rm tot}}{c^3}\right)^{2/3}.
		\label{eq:omega_dot_GR}
	\end{equation}
	For the $P_b=5~\mathrm{min}$, $M_{\rm tot}=1.2M_\odot$ example with $0<e\ll1$, over a baseline $\Delta T$ the accumulated advance is
	$$\Delta\omega_{\rm GR} = 4.1\left(\frac{M_{\rm tot}}{1.2M_{\odot}}\right)^{2/3}\left(\frac{5{\rm min}}{P_b}\right)^{5/3}\left(\frac{1}{1-e^2}\right)\left(\frac{\Delta T}{1{\rm mo}}\right){\rm rad}.$$
	The apsidal motion has, in addition to the relativistic component $\dot\omega_{\rm GR}$, Newtonian components due to stellar rotation and tides \citep{1985PAZh...11..536S,2008PhRvL.100d1102W}. Expressions for $\dot\omega_{\rm GR}$, $\dot\omega_{\rm tide}$, and $\dot\omega_{\rm rot}$ for dWDs are found in \cite{2012ApJ...745..137V}.
	For our equal mass binary, the contribution to $\dot\omega$ from tides is comparable in magnitude to the GR contribution. Estimating the tidal quadrupole contribution by using the fitted WD tidal deformability factor $k_iR_i^5$ from \cite{2012ApJ...745..137V} $\mathcal K(M_i)= -0.632+0.370M_i^{-1.709}$ -- where $M_i$ is in units of solar mass ($M_{\odot}$) and $\mathcal K$ is in units of $10^{-10}R_\odot^5$, where $R_\odot$ is the solar radius -- gives, for equal mass binaries,\footnote{\cite{2022PhRvD.106b3012P} show that `the relativistic correction to the observable tidal deformability is negligible for low mass WDs but becomes increasingly important for more massive white dwarfs.'}
	\begin{equation}
		\begin{aligned}
			\Delta\omega_{\rm tide} \simeq {}& 3.7
			\left[
			\frac{\left(\frac{q}{1}\right)\mathcal K(M_1) + \left(\frac{q}{1}\right)^{-1}\mathcal K(M_2)}{0.508}
			\right] \\
			&\times \left(\frac{M_{\rm tot}}{1.2M_\odot}\right)^{-5/3}
			\left(\frac{5~{\rm min}}{P_b}\right)^{13/3}
			f_{\rm tide}(e)
			\left(\frac{\Delta T}{1~{\rm mo}}\right)
			{\rm rad},
		\end{aligned}
		\label{eq:del_om_tide}
	\end{equation}
	where $q=M_2/M_1$,
	$$
	f_{\rm tide}(e)=
	\frac{1+\frac{3}{2}e^2+\frac{1}{8}e^4}{(1-e^2)^5}.
	$$
	The $\dot\omega$ of an unequal mass binary, for example, with $M_1=0.347 M_{\odot}$ and $M_2=0.285 M_{\odot}$ is dominated by the tidal contribution, with $\Delta \omega_{\rm tide}\simeq100$ rad over a month's time. We do not estimate $\Delta\omega_{\rm rotation}$ here; we note, however, that for the dWDs we consider, it is smaller than $\Delta\omega_{\rm GR}$ by more than an order of magnitude. A LISA measurement of apsidal motion must be interpreted as a measurement of the total apsidal advance, including the GR contribution, the tidal contribution, and the rotational contribution. The latter depends on stellar rotation and is typically smaller \citep{2008PhRvL.100d1102W}.
	
	In the fixed-$u$ velocity difference of Eq.~\eqref{eq:fixedu_no_K_evolution}, the contribution from apsidal evolution is linear in $e$: 
	\begin{equation}
		\begin{aligned}
			\Delta {\rm v}_{\rm aps}
			{} & = K_1e[\cos\omega(t_\beta)-\cos\omega(t_\alpha)] \\
			& = -2 K_1 e\sin\!\left(\omega_1+\frac{\Delta\omega}{2}\right)\sin\!\left(\frac{\Delta\omega}{2}\right),
		\end{aligned}
		\label{apsidal_drift_est0}
	\end{equation}
	where $|\Delta {\rm v}_{\rm aps}| \le 2 K_1 e$. Combining this upper limit with the precision of the velocity from a single spectrum, $\sigma_{\rm spec}\gtrsim 15~{\rm km~s^{-1}}$, gives the lower bound on eccentricity amenable to the targeted-$u$ method from a single spectrum pair, $e\gtrsim0.01$. Co-adding $N$ spectra at each epoch lowers the uncertainty as $N^{-1/2}$ and lowers the bound by the same factor, at the cost of a correction for the finite exposure window discussed below. Using the same binary as was used for Eq. \eqref{eq:semiamp_est}, the characteristic upper bound to apsidal drift for any initial $\omega_0$ is
	\begin{equation}
		|\Delta {\rm v}_{\rm aps}| \le 15 \left(\frac{e}{0.01}\right)\left(\frac{\sin \iota}{1.0}\right)~\mathrm{km\ s^{-1}}.
		\label{eq:apsidal_drift_est1}
	\end{equation}
	The analogous signal for the companion has an amplitude of $2K_2e$.
	
	In Eq.~\eqref{eq:fixedu_exact} a secular term due to the evolution of semi-amplitude $K$ appears. Because $K \propto n_{\rm b}^{1/3}$, inspiral driven by gravitational radiation evolves the semi-amplitude at a rate of $\dot K/K = \dot f_{\rm GW}/(3f_{\rm GW}) = (3\tau_{\rm chirp})^{-1}$, with a characteristic chirp timescale, i.e.
	\begin{equation}
		\tau_{\rm chirp} \;=\; \frac{f_{\rm GW}}{\dot f_{\rm GW}}
		\;\simeq\; 10^{5}~\mathrm{yr}~
		\left(\frac{M_c}{0.52~M_\odot}\right)^{-5/3}
		\left(\frac{P_b}{5~\mathrm{min}}\right)^{8/3}.
		\label{eq:tau_chirp}
	\end{equation}
	Over a baseline $T$, this contributes  $|\Delta K \cos u_\star| = K~T/(3\tau_{\rm chirp})$. For the fiducial $0.6+0.6~M_\odot$ system (using the estimate for the semi-amplitude from Eq.~\eqref{eq:semiamp_est} above), this evaluates to
	\begin{equation}
		\begin{aligned}
			|\Delta K \cos u_\star|
			{} & \simeq \\
			& \hspace{-1.5em} 2.5\times 10^{-3}
			\left(\frac{K}{747~\mathrm{km~s^{-1}}}\right)
			\left(\frac{T}{1~\mathrm{yr}}\right)
			\left(\frac{10^{5}~\mathrm{yr}}{\tau_{\rm chirp}}\right)
			\left(\frac{\sin \iota}{1.0}\right)
			~\mathrm{km~s^{-1}},
			\label{eq:DeltaK_bound}
		\end{aligned}
	\end{equation}
	which is four orders of magnitude below the spectroscopic floor $\sigma_{\rm spec}\sim15~{\rm km~s^{-1}}$. Comparing the contribution from the secular change in $K$ with the maximum apsidal contribution in Eq.~\eqref{eq:apsidal_drift_est1} gives
	\begin{equation}
		\frac{|\Delta K \cos u_\star|}{|\Delta {\rm v}_{\rm aps}|_{\rm max}}
		\;\le\; \frac{T}{6~e~\tau_{\rm chirp}}
		\;\simeq\; 1.7\times 10^{-4}
		\left(\frac{T}{1~\mathrm{yr}}\right)
		\left(\frac{10^{5}~\mathrm{yr}}{\tau_{\rm chirp}}\right)
		\left(\frac{10^{-2}}{e}\right),
		\label{eq:DeltaK_vs_aps}
	\end{equation}
	so the $\Delta K$ term remains negligible compared to the characteristic apsidal velocity scale $\Delta {\rm v}_{\rm aps}$ over the eccentricity range relevant in this case, $e\gtrsim 10^{-2}$. This estimate assumes orbital decay dominated by gravitational radiation. In tidally locked systems, an additional tidal contribution modifies the frequency derivative by a factor $(1+r_{\rm tide})$, where $r_{\rm tide}= \dot P_{\rm tide}/\dot P_{\rm GW}$ is the ratio of the tidal and GW contributions to the orbital period derivative \citep{2025PhRvD.112d3013L}. Since both contributions are negative for detached inspiralling systems, $r_{\rm tide}>0$, with the chirp timescale shortened by the same factor. The requirement that neither WD fills its Roche lobe bounds $r_{\rm tide}\lesssim 0.27$ for the fiducial component masses, independent of period over $P_b=4$-$10$~min, so $\tau_{\rm chirp}$ is reduced by at most $\sim20\%$. The semi-amplitude drift remains negligible at the spectroscopic floor, and we are justified in dropping the $\Delta K\cos u_\star$ term from Eq.~\eqref{eq:fixedu_exact} and proceeding with Eq.~\eqref{eq:fixedu_no_K_evolution}.

	A finite exposure of duration $T_{\rm exp}$ averages ${\rm v}_r$ over a curved orbital arc. The bias is largest at quadrature, where the velocity reaches $K_1$,
	\begin{equation}
		\begin{aligned}
			\Delta {\rm v}_{\rm smear}
			&= K_1\left[
			1-\frac{\sin(n_b T_{\rm exp}/2)}
			{n_b T_{\rm exp}/2}
			\right]
			\\
			&\simeq \frac{K_1(n_b T_{\rm exp})^2}{24}
			\qquad (n_b T_{\rm exp}\ll 1).
		\end{aligned}
		\label{eq:Deltav_smear}
	\end{equation}
	For the fiducial system, this evaluates to
	\begin{equation}
		\Delta {\rm v}_{\rm smear}\simeq 1.4
		\left(\frac{K_1}{750~\mathrm{km~s^{-1}}}\right)
		\left(\frac{T_{\rm exp}}{10~\mathrm{s}}\right)^{\!2}
		\left(\frac{5~\mathrm{min}}{P_b}\right)^{\!2}~\mathrm{km~s^{-1}}.
		\label{eq:Deltav_smear_est}
	\end{equation}
	Since the period and integration time are known, the factor is a fixed number and divides out of the measured velocities, leaving the recovered semi-amplitude unbiased. The quantity applied to the observed semi-amplitudes $K_{\rm obs}$ is the inverse of the attenuation factor, $K = K_{\rm obs}/\mathrm{sinc}(n_b T_{\rm exp}/2)$. For a $P_b = 5$ min binary observed in $10$ s exposures, each integration spans $1/30$ of the orbit and the correction is $1/\mathrm{sinc}(\pi/30)\simeq1.002$, or $0.2\%$. Reaching the S/N needed to measure a velocity likely requires co-adding spectra over a wider phase range. For a co-addition spanning a third of the orbit, the window is $\Delta u = 2\pi/3$ and the correction grows to $1/\mathrm{sinc}(\pi/3)\simeq1.21$, a $21\%$ upwards correction applied equally to $K_1$ and $K_2$.

	For an isolated binary, galactic and perspective accelerations are on the order of $a_{\rm CM}\sim 10^{-10}~\mathrm{m~s^{-2}}$, so
	\begin{equation}
		\Delta {\rm v}_{\rm CM} \sim a_{\rm CM}\Delta T \;\lesssim\;
		3\times10^{-3}\left(\frac{a_{\rm CM}}{10^{-10}{\rm m~s^{-2}}}\right)\left(\frac{\Delta T}{1{\rm yr}}\right)~\mathrm{m~s^{-1}},
		\label{eq:DeltaVcm_est}
	\end{equation}
	which is negligible compared to the $\rm km\ s^{-1}$ apsidal drift term Eq.~\eqref{eq:apsidal_drift_est1}. It is possible for compact tertiaries to produce much larger drift contributions.
	A third body of mass $M_3\sim 1M_\odot$ at $a_{\rm out}\sim 5~\mathrm{AU}$ produces $a_{\rm CM}\sim 2.3\times10^{-4}~\mathrm{m~s^{-2}}$. For the case when $\Delta T\ll P_{\rm out}$,
	$$
	\Delta {\rm v}_{\rm CM} \sim 0.6\left(\frac{M_3}{1M_{\odot}}\right)\left(\frac{5 {\rm AU}}{a_{\rm out}}\right)^2\left(\frac{\Delta T}{1{\rm mo}}\right)~\mathrm{km\ s^{-1}}.
	$$
	In such cases $\Delta {\rm v}_{\rm CM}$ can become comparable to the intrinsic apsidal signal; however, \cite{2025A&A...704A.156R} predict that UCBs observed with LISA and hosting tertiary companions will typically reside in hierarchical triples with wide outer orbits, with the semimajor axis of the third body on the order of $1\times10^3$ AU \citep{2026AJ....172..266K}. This dramatically reduces the significance of any tertiary imprint on the observed Doppler shifted lines. That is to say, we expect any conclusive detection of a Doppler shift from a bound third star to be an exceptionally rare scenario. A large $\Delta {\rm v}_{\rm CM}$, exceeding that expected from Galactic acceleration, is more likely to reflect either instrumental systematics or a long-duration perturbation associated with a stellar fly-by.
	
	\subsection{Apsidal signal: Dependence on baseline and argument of periastron}
	\label{subsec:ecc_binary_aps_motion}
\begin{figure*}[t]
	\sidecaption
	\includegraphics[width=12cm]{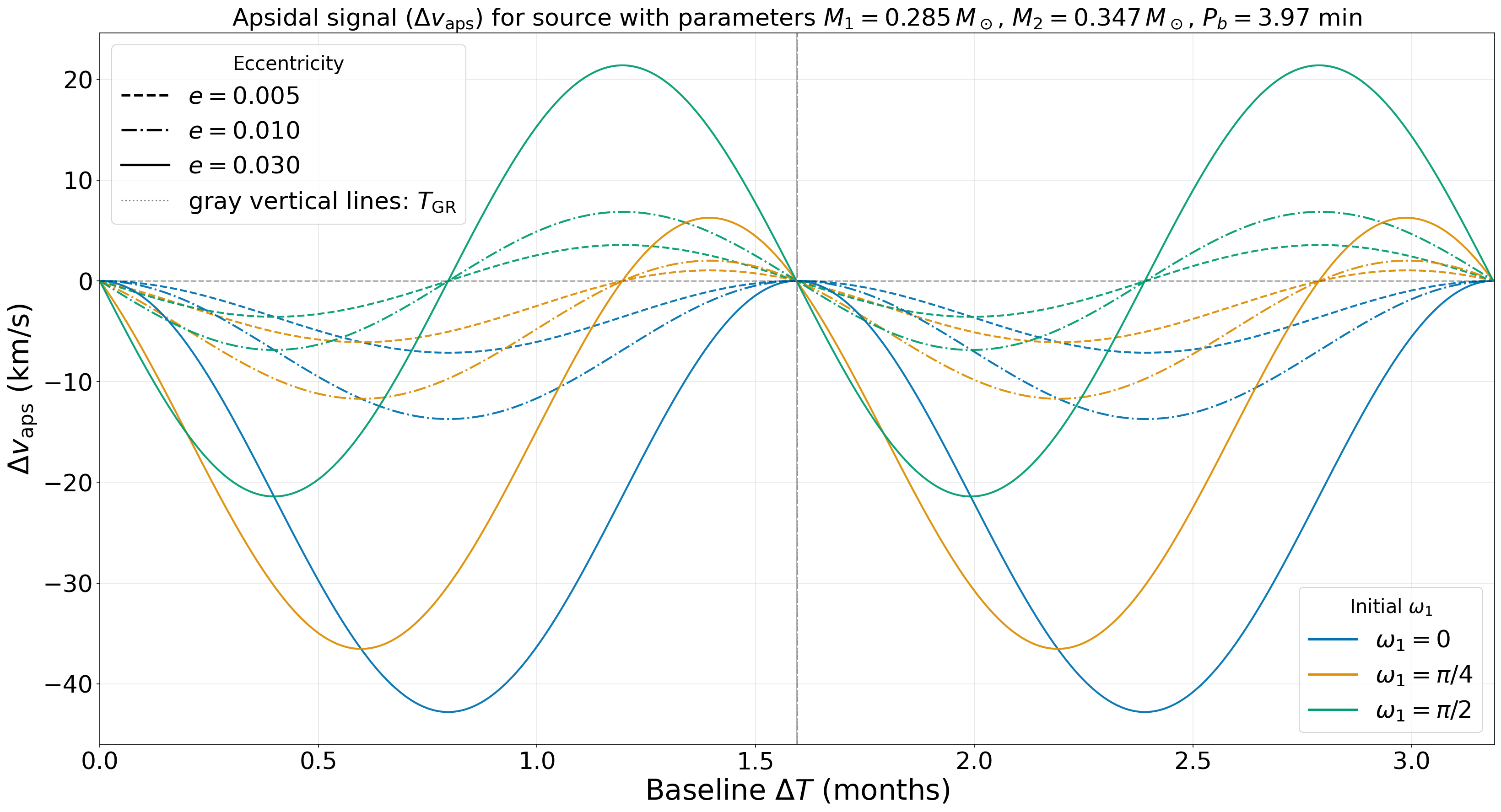}
		\caption{
			Apsidal velocity difference $\Delta {\rm v}_{\rm aps}$ versus observation baseline for the target WD binary with
			$M_1 = 0.347~M_\odot$, $M_2 = 0.285~M_\odot$, $P_b = 3.97$~min, and $\iota = 84.7^\circ$. Different colours show different initial periastron arguments $\omega_1$, while line styles indicate the eccentricities $e = 0.005,  0.01$, and $0.03$. The vertical dashed line marks the apsidal advance timescale. The signal oscillates on the apsidal timescale $T_{GR}$, and its dependence on
			$\omega$ determines which baselines yield large signals and which fall near nulls. When LISA resolves $(\eta,\kappa)$, the system's $\omega$ is known,
			allowing a favourable baseline to be selected a priori. For $(\eta,\kappa) \simeq (0.0068, 0.0068)$, $\omega_0 \simeq \pi/4$.
			At $e \sim 10^{-2}$, the peak velocity shift due to apsidal motion, $2 K_1 e \approx 12~{\rm km~s^{-1}}$, is comparable to typical spectroscopic precision, $\sigma_{\rm spec} \sim 15~{\rm km~s^{-1}}$. Single-epoch pairs are then comparable to the noise level. For the smallest eccentricities, detectability is controlled by the number and scheduling of fixed-$u$ pairs, not only by the maximum per-pair amplitude.}
		\label{fig:apsidal_signal_omega1_dependence}
	\end{figure*}

	As a running example throughout this paper, we adopt a binary system selected from an observationally driven population of Galactic binaries for LISA\footnote{https://gitlab.in2p3.fr/korol/observationally-driven-population-of-galactic-binaries} \citep{2022MNRAS.511.5936K}. The authors use orbital and mass distributions anchored to the local WD sample observed in the Sloan Digital Sky Survey and the SPY survey \citep{2012ApJ...751..143M,2018MNRAS.476.2584M} rather than to distributions anchored by binary population synthesis. The parameters, $M_1 = 0.347~M_\odot$, $M_2 = 0.285~M_\odot$, $P_b = 3.97$~min, and $\iota = 84.7^\circ$, are reused in subsequent numerical estimates. The source was modelled as a detached dWD with eccentricity $e=0.0096$. Its inclination
	is approximately edge-on with RV semi-amplitudes ($K_1 \simeq 585$ and $K_2 \simeq 713~\mathrm{km~s^{-1}}$) sitting well above the precision of a single spectrum, $\sigma_{\rm spec}\sim15~{\rm km~s^{-1}}$. For this source, with $\dot\omega=\dot\omega_{\rm GR}$, the apsidal timescale  is $T_{\rm GR} \approx 1.6$~months, setting the baseline on which the $\Delta {\rm v}_{\rm aps}$ signal oscillates,
	Fig.~\ref{fig:apsidal_signal_omega1_dependence}. Including the contribution from tides, $\dot\omega=\dot\omega_{\rm GR}+\dot\omega_{\rm tide}$ shortens the apsidal timescale considerably to $< 1$ week, for the same fiducial system.
	
	Figure~\ref{fig:apsidal_signal_omega1_dependence} shows that the magnitude of the apsidal signal $\Delta {\rm v}_{\rm aps}(t)$ depends on both the observing baseline $\Delta T=t_\beta-t_\alpha$ and the initial argument of periastron $\omega_0=\operatorname{arctan_2}(\eta_0,\kappa_0)$.  The signal oscillates on the apsidal timescale, so a baseline chosen without regard to $\omega_0$ might produce $\Delta{\rm v_{aps}}$ measurements near a null. This is particularly relevant to Sect.~\ref{sec:delta_vcm_sb2_ecc}, where monitoring apsidal evolution is used to constrain $\Delta{\rm v}_{\rm CM}$.

	\subsection{Summary of expected signal magnitudes}
	\label{subsec:magnitude_summary}
	
	Before moving to applications for targeted-$u$ observations, we summarise items from this section of particular importance to the remaining text.
	In this paper we present a multimessenger method for scheduling optical spectroscopic measurement of UCBs using GW-derived ephemerides. In order to assess whether GW-derived ephemerides are competitive with their optically derived counterparts, we provide an estimate for optical timing uncertainties in Eq.~\eqref{eq:optical_timing_prec}. With three examples, we find a representative range for $\delta t_{\rm optical}$ between $1-30$~s. In Sect.~\ref{sec:horizon_forecast}, we find that the timing uncertainties derived from the GW parameter covariance are comparable over the relevant observing baselines following the LISA observation.

	Equations~\eqref{eq:omega_dot_GR} and \eqref{eq:del_om_tide} show that apsidal motion in short-period dWDs is rapid, with $\Delta\omega$ reaching radians over month or shorter baselines and the tidal contribution exceeding $\Delta\omega_{\rm GR}$. The accumulated $\Delta\omega$ sets the fixed-$u$ apsidal velocity difference through Eq.~\eqref{apsidal_drift_est0}, which is bounded by $2Ke$ and passes through maxima and nulls as the baseline changes. LISA measures the total apsidal rate, which includes relativistic, tidal, and rotational contributions, so we treat $\dot\omega$ as an independent parameter in the subsequent GW data analysis of the binary in an eccentric orbit.
	
	Finite exposure times attenuate the measured velocity amplitudes through finite exposure averaging. For the fiducial $5$ min binary, a $10$~s exposure produces only a $0.2\%$ correction, but co-adding spectra over one third of an orbit increases this to $21\%$. Because the attenuation factor is fixed by the known exposure window, it is corrected when recovering the semi-amplitudes. If neglected, measured semi-amplitudes are underestimated.

	\section{Applications of targeted argument of latitude observables}
	\label{sec:applicationsU}
	
	A LISA-determined binary phase ephemeris provides another route to obtaining phase-resolved spectra for UCBs in both circular and eccentric orbits. We distinguish SB2 systems, in which spectral lines tracing both stellar components are detected, from single-lined (SB1) systems, in which only one component's spectral lines are visible. First we consider component mass measurements for SB2 systems in circular (Sect.~\ref{sec:sb2_circular}) and eccentric (Sect.~\ref{sec:theory_ecc}) orbits. We then turn to applications involving Eq.~\eqref{eq:fixedu_no_K_evolution} to measure $\Delta\kappa$ and $\Delta {\rm v}_{CM}$ from optical observations, Sect.~\ref{sec:delta_kappa_consistency} and Sect.~\ref{sec:DeltaVCM}, respectively.
	
	For an SB2, phase-targeted spectroscopy combined with the orbital inclination and period measured by LISA provides the information needed to determine the component masses through Kepler's third law. The method is developed below.
	
	For eccentric orbits we make use of the LL parameters \citep{10.1046/j.1365-8711.2001.04606.x}
	\begin{equation}
		(\eta,\kappa)=(e\sin\omega,e\cos\omega),
		\label{eq:eta_kappa_main}
	\end{equation}
	which remain well defined as $e\to0$, whereas $\omega$ itself does not. At a fixed argument of latitude $u$, the leading $K\cos u$ term is the same at both epochs and cancels in the difference, Eq.~\eqref{eq:fixedu_no_K_evolution}. What remains in each component is a common barycentric drift and a change in $\kappa$ with opposite signs for the two stars. For circular orbits, the GW phase determines the argument of latitude, $u(t)=\Phi_{\rm GW}(t)/2$, while for eccentric orbits the $\mathcal{O}(e)$ correction $\delta_{e,\omega}$ is included in the phase mapping to recover $u(t)$. In both cases, this allows velocity measurements at known orbital phases to be converted into semi-amplitudes.

	\subsection{Double-lined systems}
	\label{sec:caseSB2}
	
	In this type of system, one is able to measure spectral lines from both stars, so the optical velocities determine the mass ratio, while LISA supplies the orbital phase needed to place each spectroscopic measurement on the RV curve and recover the individual semi-amplitudes without reconstructing the full curve.
	Here, `phase resolved' refers to spectra assigned to a known orbital phase interval. In the ideal targeted case, this interval is short, so each spectrum is treated as a velocity measurement at the LISA predicted value $u_\star$. The exposure time is nevertheless set by the S/N required to reach the desired RV precision. For faint systems, longer integrations produce spectra averaged over a known phase exposure window centred on $u_\star$. We assume $T_{\rm exp}\ll P_b$ and one RV measurement per component per spectrum. When phase-resolved spectra are binned across some phase range, say a third of the orbit, the GW-derived phase model $u(t)$ still supplies the central value of $u$ and the width of the window. Corrections are applied to remove the effect of velocity smearing, Eq. \eqref{eq:Deltav_smear}, so the measured parameters are not biased. In the first half of Appendix~Sect.~\ref{app:appendixB}, we discuss corrections for finite
	phase integration, which attenuates measured velocity amplitudes when spectra are integrated or co-added over a non-negligible fraction of the orbit.
	
	\subsubsection{Mass ratios from paired radial velocities}
	\label{subsec:wilson1941}
	
	\citet{1941ApJ....93...29W} observed that momentum conservation in the barycentric frame places every simultaneous velocity pair $({\rm v}_1,{\rm v}_2)$ onto the line
	\begin{equation}
		{\rm v}_1=\gamma(1+q)-q~{\rm v}_2 ,
		\label{eq:wilson_line}
	\end{equation}
	regardless of orbital phase, where $q=M_2/M_1$ and $\gamma$ is the systemic velocity. Two spectra at distinct phases determine the mass ratio from the
	slope, $-q$, and the systemic velocity from the intercept, $\gamma(1+q)$.\footnote{Obtaining the mass ratio in this way is mentioned in \citet{2025arXiv250400548S}, also.} Here we have assumed $\gamma$ is common to both components ($\gamma_1=\gamma_2=\gamma$). A differential offset between the two stars
	(e.g. from gravitational redshift) shifts the intercept to $\gamma_1+q\gamma_2$ but leaves the slope $-q$ invariant,
	\begin{equation}
		{\rm v}_1 = \gamma_1 + q~\gamma_2 - q~{\rm v}_2 .
		\label{eq:wilson_line_general}
	\end{equation}
	
	The \citet{1941ApJ....93...29W} method (WIL41) fixes $q$ without knowing the orbital phase, but it does not determine the
	individual semi-amplitudes $K_1$ and $K_2$. That information is lost when the argument of latitude is eliminated. The GW-derived phase model supplies $u(t)$, so the same velocity measurements are placed on the RV curves and used to determine the individual semi-amplitudes $K_1$ and $K_2$, not only their ratio. 
	
	\subsubsection{Component masses for circular orbits}
	\label{sec:sb2_circular}
	
	The GW phase model extends the WIL41 velocity--velocity relation, Eq.~\eqref{eq:wilson_line}, by retaining the argument of latitude that WIL41 eliminates. For circular orbits, the radial velocities of the two components are
	\begin{equation}
		{\rm v}_{r,1}(u)=\gamma_1+K_1\cos u ,
		\qquad
		{\rm v}_{r,2}(u)=\gamma_2-K_2\cos u .
		\label{eq:v1v2_circular_rv}
	\end{equation}
	The constant offsets $\gamma_j={\rm v}_{\rm CM}+z_j$ collect the systemic velocity and the epoch-independent contributions to the apparent radial velocity. For close dWDs in general, $z_j$ is expected to be dominated by the WD surface gravitational redshift, $GM_j/(R_jc)$, expressed as an equivalent velocity shift. For UCBs additional constant contributions from the companion potential and the second-order Doppler shift from the orbital motion become important and are discussed below. The only time-dependent orbital factor is $\cos u(t)$. Defining $x=\cos u$, each component velocity is a straight line in $x$, with intercept $\gamma_j$ and slope $\pm K_j$, where $x$ is an independent variable whose value at each epoch is predicted by the GW phase model. The intercepts are not equal in general. Their difference $\gamma_1-\gamma_2$ records the differential constant spectroscopic offset $z_1-z_2$. WIL41 cannot recover this difference because eliminating $x$ collapses both intercepts into the single combination $\gamma_1+q\gamma_2$ (Eq.~\eqref{eq:wilson_line_general}).
	The GW-derived phase model breaks the intercept degeneracy by predicting $x=\cos u$ at each epoch, so that $\gamma_1$ and $\gamma_2$ are measured separately rather than only through the combination $\gamma_1+q\gamma_2$. The degeneracy in the decomposition
	$\gamma_j={\rm v}_{\rm CM}+z_j$ remains. The transformation $({\rm v}_{\rm CM}, z_j)\to({\rm v}_{\rm CM}+\delta,z_j-\delta)$ leaves both
	intercepts unchanged. An external estimate of $(z_1+z_2)/2$, for example from a WD mass-radius relation, is needed to infer ${\rm v}_{\rm CM}$ from $(\gamma_1+\gamma_2)/2$.
	
	LISA supplies $u(t)$ in advance (Appendix~Sect.~\ref{app:appendixA2}), so $x(t)$ is known at
	every epoch.  Spectra at several epochs with distinct $x$ determine $\gamma_1,\gamma_2,K_1,K_2$ from two independent linear regressions. 
	For measurements with comparable RV precision, the semi-amplitudes are most precisely determined by sampling both quadratures, near $u=0$ and $u=\pi$, where $x\simeq +1$ and $x\simeq -1$ provide the largest velocity separation and the smallest first-order sensitivity to phase errors (Appendix~Sect.~\ref{app:appendixC1}).
	
	Eliminating $x$ between the two component velocities gives Eq.~\eqref{eq:wilson_line_general}, namely the velocity--velocity relation of the WIL41 method. The
	slope is the optical mass ratio and the intercept is $\gamma_1+q\gamma_2$. The distinction is that WIL41 uses this velocity–velocity relation alone, whereas the GW-derived phase model also supplies the value of $x=\cos u$ for each spectrum.
	The same data is viewable as two complementary projections. The velocity–velocity projection gives $q$ and the combination $\gamma_1+q\gamma_2$. The velocity–phase projection gives the two semi-amplitudes $K_1,K_2$ together with the individual intercepts $\gamma_1,\gamma_2$.
	
	For a circular SB2, the preferred phases are the two RV extrema, $u=0$ and $u=\pi$. Two spectra at the extrema are the smallest dataset that determines the two slopes and the two intercepts. At these phases $|\cos u|=1$ maximises the l.o.s. excursion $\pm K_j$ around each intercept. One spectrum at each quadrature phase gives the semi-amplitudes,
	\begin{equation}
		K_1=\tfrac{1}{2}\left[{\rm v}_{r,1}(0)-{\rm v}_{r,1}(\pi)\right],
		\qquad
		K_2=\tfrac{1}{2}\left[{\rm v}_{r,2}(\pi)-{\rm v}_{r,2}(0)\right],
		\label{eq:sb2_half_diff}
	\end{equation}
	with 
	\begin{equation}
		\gamma_1=\tfrac12[{\rm v}_{r,1}(0)+{\rm v}_{r,1}(\pi)],
		\qquad
		\gamma_2=\tfrac12[{\rm v}_{r,2}(0)+{\rm v}_{r,2}(\pi)].
		\label{eq:sb2_half_sum}
	\end{equation}
	
	With $K_1$ and $K_2$ in hand, the orbital inclination and period measured by LISA fix the absolute scale. For a circular orbit, the orbital phase advances at $\pi f_{\rm GW}$, corresponding to an orbital period $P_b=2/f_{\rm GW}$. The total semi-major axis follows from Eq.~\eqref{eq:semiamplitude_eq},
	\begin{equation}
		a=\frac{(K_1+K_2)\sqrt{1-e^2} P_b}{2\pi\sin\iota}, 
		\label{eq:semimajor_axis_j}
	\end{equation}
	where $e=0$, and Kepler's third law gives the total mass,
	\begin{equation}
		M_{\rm tot}=\frac{4\pi^2 a^3}{G P_b^2}.
		\label{eq:Kepler_mtot}
	\end{equation}
	The mass ratio then splits this into the individual component masses,
	\begin{equation}
		M_1=\frac{M_{\rm tot}}{1+q},
		\qquad
		M_2=qM_1.
		\label{eq:M1M2_cir_and_ecc}
	\end{equation}
	Here $q$ comes from $K_1/K_2$, while $M_{\rm tot}$ is obtained by combining the optical semi-amplitudes with the inclination and orbital period $P_b$ provided from the GW side. With both component masses in hand, the chirp mass ($M_c$) follows, and the GW frequency and amplitude then constrain the source distance.
	
	The difference $\gamma_1-\gamma_2=z_1-z_2$ removes the common barycentric
	velocity and isolates the differential constant spectroscopic offset
	between the two components. Of the terms that survive, only the surface
	gravitational redshift depends on the stellar radii. The other terms depend on the component masses and their separation. The companion's gravitational potential at the orbital separation contributes $G(M_2-M_1)/(ac)$ to $\gamma_1-\gamma_2$ and the constant second-order Doppler shift from the orbital motion contributes $G(M_2-M_1)/(2ac)$. Together they give $3G(M_2-M_1)/(2ac)$ (from the relativistic RV terms of \citealt{2007ApJ...654L..83Z}). \citet{2017MNRAS.466.1575R} use the measured $\gamma_1-\gamma_2$ as a consistency check on the gravitational redshift predicted from their inferred masses, separation, and a WD mass-radius relation. They neglect the second-order Doppler term, which is justified at their orbital periods of hours. At the periods considered here, the combined contribution independent of the radii reaches ${\sim}20\%$ of $G(M_1/R_1-M_2/R_2)/c$ for $P_b\simeq5$~min, of which the second-order Doppler shift supplies one third. In the targeted-$u$ approach, the component masses and separation are instead determined from the measured orbital velocities together with the LISA GW parameters. With $M_1$, $M_2$, and $a$ known, subtracting $3G(M_2-M_1)/(2ac)$ from the measured $\gamma_1-\gamma_2$ isolates $G(M_1/R_1-M_2/R_2)/c$, providing an independent check on the radii predicted by an adopted WD mass-radius relation.

	\citet{1941ApJ....93...29W} notes that, although two spectra suffice in principle to determine $q$ and $\gamma$, with Eq.~\eqref{eq:wilson_line_general}, limited RV precision makes this minimum solution unsatisfactory. Rather, an overdetermined velocity--velocity regression provides parameter uncertainties from the residuals and exposes discordant measurements. For the same practical reasons, Eqs.~\eqref{eq:sb2_half_diff} and \eqref{eq:sb2_half_sum} are regarded as the two spectrum limit of the ${\rm v}$--$\cos u$ regression, while $K_j$ and $\gamma_j$ should in practice be inferred from all velocities with a known argument of latitude. \citet{1941ApJ....93...29W} also notes that $q$ and $\gamma$ are determined independently of uncertainties in the remaining orbital elements, a consequence of eliminating the phase. In our methodology we are able to retain the orbital phase in the regression, since it is provided by an external clock $\Phi_{\rm GW}$, and this allowed us to determine $K_1$ and $K_2$ individually rather than only their ratio. As a consequence, the semi-amplitudes inherit uncertainty from the LISA ephemeris through the prediction of $u(t)$, while the remaining orbital elements do not affect the determination of the slopes. This uncertainty is propagated from the LISA posterior through $\sigma_u$ (Appendix~Sect.~\ref{app:appendixC1}).
	
	Because the velocity curve is stationary at $u=0,\pi$, a targeting
	error $\delta u_{\rm t}$ affects the velocity at second order only. It
	attenuates both semi-amplitudes by a common factor, which cancels
	exactly in $q$ but biases each $K_j$ low by
	$K_j\sigma_u^2/2$. The intercepts $\gamma_j$ scatter by the same amount
	with no preferred sign. Eq.~\eqref{eq:tol_quad} bounds this residual at
	quadrature. Finite exposures act on the semi-amplitudes similarly,
	Eq.~\eqref{eq:Deltav_smear}, with the phase spread across the integration
	in place of a single offset. The two attenuations multiply as
	$\cos\delta u_{\rm t}\times\mathrm{sinc}(n_bT_{\rm exp}/2)$. If the
	finite-exposure correction is not applied, $q$ remains unchanged, but
	both semi-amplitudes are underestimated by the same factor $\Lambda_K$,
	so the inferred component masses are each scaled by $\Lambda_K^3$.

	Without a sufficiently accurate GW prediction of the orbital phase, the value of $\cos u(t)$ associated with each spectroscopic observation must instead be determined electromagnetically. In a non-eclipsing system, this mapping is obtainable from the phase of the RV orbit itself \citep{2020ApJ...892L..35B}, but doing so requires time series spectroscopy to build and maintain the orbital ephemeris before targeted spectra are assigned to selected orbital phases. This RV route remains indispensable for longer period or lower S/N systems where LISA detects the source but does not
	provide a useful constraint on $\dot f_{\rm GW}$. For LISA-resolved UCBs with a sufficiently predictive phase model, however, the phase prediction removes or shortens this preparatory stage. The first targeted spectra are then scheduled at chosen values of $u$ and yield $K_j$ and $\gamma_j$. This matters most for the non-eclipsing UCBs that make up the bulk of the LISA-resolved dWD population \citep{2023A&A...678A.123L}, where no eclipse ephemeris is available to play the same role.
	
	\subsubsection{Component masses for eccentric orbits}
	\label{sec:theory_ecc}
	
	The spectroscopic fit equations for circular orbits Eq.~\eqref{eq:v1v2_circular_rv} carry over to eccentric orbits under the substitution $\cos u\rightarrow F$. The Keplerian radial velocities are written
	\begin{equation}
		{\rm v}_{r,1}(t)=\gamma_1+K_1 F(t) ,
		\qquad
		{\rm v}_{r,2}(t)=\gamma_2-K_2 F(t) ,
		\label{eq:v1v2_ecc_rv}
	\end{equation}
	with
	\begin{equation}
		F(t)=\cos u(t)+\kappa(t) ,
		\qquad
		\kappa(t)=e\cos\omega(t) ,
		\label{eq:F_definition}
	\end{equation}
	where $\kappa$ is the LL parameter of
	Eq.~\eqref{eq:eta_kappa_main}, evaluated at each observing epoch.
	
	For a circular binary, the GW phase alone fixes the observing phase. An
	eccentric binary requires three further quantities from the LISA analysis. The
	LL pair $(\eta_0,\kappa_0)$ of Eq.~\eqref{eq:eta_kappa_main},
	evaluated at the reference epoch $t_0$, and the apsidal advance rate
	$\dot\omega$. Eccentricity appears in the waveform at first order through
	$(\eta,\kappa)$, while apsidal advance rotates that pair, so all three are
	independent parameters of the eccentric GW model
	(Sect.~\ref{subsec:magnitude_summary}; the parameterisation is set out in
	Sect.~\ref{sec:horizon_forecast_ecc}). Here we assume that their posteriors have
	been obtained for the binary.
	
	These parameters serve two purposes in what follows.
	First, they determine the $\mathcal{O}(e)$ correction $\delta_{e,\omega}$ of
	Table~\ref{tab:table1}, so that the GW phase model,
	Eq.~\eqref{eq:phase_map}, gives the argument of latitude $u(t)$ at each
	observing epoch. Second, they provide the value of $\kappa$ required to
	construct $F$ at each observing epoch. When $e\neq0$,
	$(\eta_0,\kappa_0)$ determine $e$ and $\omega_0$, while the
	evolution of $\omega(t)$ follows from $\dot\omega$.
	
	Eq.~\eqref{eq:F_definition} defines $F$ in terms of $u$ and $\kappa$. The value of $\kappa$ at each spectroscopic epoch is predicted from the LISA parameters rather than determined from the spectrum itself.
	It is propagated from the LISA reference epoch by rotating $(\eta_0,\kappa_0)$, assuming constant $\dot\omega$ and fixed $e$. We write
	$\kappa_{\rm LISA}(t)$ for this predicted value, as distinct from the true
	$\kappa(t)$, and evaluate $F$ with $\kappa_{\rm LISA}(t)$ throughout.
	
	With $u(t)$ and $\kappa_{\rm LISA}(t)$ both predicted,
	Eq.~\eqref{eq:v1v2_ecc_rv} has the same linear structure as
	Eq.~\eqref{eq:v1v2_circular_rv} for circular orbits. Each component velocity is
	a straight line in the single variable $F$, with intercept $\gamma_j$ and slope
	$\pm K_j$.

	Spectra at several phases with distinct $F$ values within an observing session determine $\gamma_1,\gamma_2,K_1,K_2$ from two independent linear regressions. The same corrections to the semi-amplitudes discussed for circular orbits apply here, so the recovered $K_j$ are corrected in exactly the same way. On the orbital timescale, the LISA prediction is effectively constant, $\kappa_{\rm LISA}(t)\simeq\kappa_{\rm LISA}(t_s)=\kappa_s$, while $u$ varies over $2\pi$, so $F\in[-1+\kappa_s,~1+\kappa_s]$. The constant $\kappa_s$ shifts $F$ without changing its variation, so it is degenerate with the fitted intercept and cannot affect the slope. An error $\delta\kappa_s$ in the LISA-predicted value of $\kappa_s$ shifts the fitted intercepts by $\delta\gamma_1=-K_1\delta\kappa_s$ and $\delta\gamma_2=+K_2\delta\kappa_s$, without changing the recovered semi-amplitudes.
	The two phases $0$ and $\pi$ provide the maximum peak-to-peak velocity separation.
	
	Spectra obtained at quadrature phases give the semi-amplitudes from Eq.~\eqref{eq:sb2_half_diff}, as in the circular case. Once the semi-amplitudes are derived, the systemic velocities of both stellar components are also obtained,
	\begin{equation}
		\begin{aligned}
			\gamma_1 &=
			\tfrac{1}{2}\left[{\rm v}_{r,1}(0)+{\rm v}_{r,1}(\pi)-2\kappa_s K_1\right],\\
			\gamma_2 &=
			\tfrac{1}{2}\left[{\rm v}_{r,2}(\pi)+{\rm v}_{r,2}(0)+2\kappa_s K_2\right].
		\end{aligned}
		\label{eq:sb2_half_sum_ecc}
	\end{equation}
	Although two spectra are sufficient in principle to determine the
	semi-amplitudes and systemic velocities, the limited RV
	precision makes this solution unsatisfactory for the same reasons
	discussed in Sect.~\ref{sec:sb2_circular} for circular orbits.
	
	For co-added spectra, recovery of the systemic velocities from Eq.~\eqref{eq:sb2_half_sum_ecc} additionally requires replacing $\kappa_s K_j$ by $\Lambda_\kappa\kappa_s K_j$, where both $\kappa_s$ and $K_j$ are the intrinsic values. Omitting this finite-bin
	correction does not affect the semi-amplitudes or component masses, but
	biases the measured $\gamma_1$ and $\gamma_2$. This modification to
	Eq.~\eqref{eq:sb2_half_sum_ecc} applies only when the spectra are integrated
	over a finite phase interval.
	
	From this point, the component mass determination proceeds as in the circular case. The semi-major axis is obtained from Eq.~\eqref{eq:semimajor_axis_j} using the measured value of $e$ from LISA, the total mass by Eq.~\eqref{eq:Kepler_mtot}, and the component masses by Eq.~\eqref{eq:M1M2_cir_and_ecc}. The resulting total mass is determined from the measured orbital velocities and LISA GW parameters, rather than from $\dot f_{\rm GW}$ or $\dot\omega$. For the eccentric orbit, the anomalistic binary period $P_b$, defined as the interval from periastron to periastron \citep{2001PhRvL..87y1101S}, follows from the LISA measurements of $f_{\rm GW}$ and $\dot\omega_{\rm LISA}$ through $n_b=\bar n_b-\dot\omega_{\rm LISA}$, with $\bar n_b=\pi f_{\rm GW}$. The GR contribution to the apsidal advance is then fixed by this total mass, $e$, and $P_b$. Subtracting it from the total apsidal advance measured by LISA,
	$\dot\omega_{\rm LISA}-\dot\omega_{\rm GR}$, isolates the remaining astrophysical contribution, including tidal and rotational distortions associated with the finite radii of the WDs.

	\subsubsection{A multimessenger consistency test of eccentricity evolution}
	\label{sec:delta_kappa_consistency}
	
	Under the fixed-$e$ and constant-$\dot\omega$ apsidal model used in the UCB waveform, the LL parameter evolves as
	\begin{equation}
		\kappa_{\rm LISA}(t)
		=
		\kappa_{\rm ref}\cos[\dot\omega(t-t_{\rm ref})]
		-
		\eta_{\rm ref}\sin[\dot\omega(t-t_{\rm ref})] .
		\label{eq:kappa_lisa_prediction}
	\end{equation}
	The LISA posterior constrains $(\eta, \kappa)$ at $t_{\rm ref}$ alone.
	Values at other times follow from the assumed rotation at constant $\dot\omega$
	and fixed $e$. We write $(\eta_{\rm LISA},\kappa_{\rm LISA})$ in place of $(\eta,\kappa)$ to mark this
	extrapolation.
	
	The predicted change between two observing epochs is
	\begin{equation}
		\Delta\kappa_{\rm LISA}
		=
		\kappa_{\rm LISA}(t_\beta)
		-
		\kappa_{\rm LISA}(t_\alpha) .
		\label{eq:delta_kappa_LISA}
	\end{equation}
	Referring the rotation to the first epoch, with
	$\Theta=\dot\omega~\Delta T$ and $\Delta T=t_\beta-t_\alpha$,
	Eq.~\eqref{eq:kappa_lisa_prediction} gives
	\begin{equation}
		\Delta\kappa_{\rm LISA}
		=
		-2\sin\frac{\Theta}{2}
		\left[
		\kappa_{\rm LISA}(t_\alpha)\sin\frac{\Theta}{2}
		+
		\eta_{\rm LISA}(t_\alpha)\cos\frac{\Theta}{2}
		\right] .
		\label{eq:delta_kappa_rewrite}
	\end{equation}
	The bracket equals $e\sin(\omega_\alpha+\Theta/2)$, so
	$|\Delta\kappa_{\rm LISA}|\leq2e$ and the predicted change oscillates on the
	apsidal timescale.
	
	Within a single observing session, the fits of Eq.~\eqref{eq:v1v2_ecc_rv} determine $K_1$ and $K_2$. We refer to these as intra-session measurements. Differences between radial-velocity measurements made at the same $u$ in observing sessions separated by a long baseline, which we call inter-session differences, contain $\Delta\kappa$ together with any change in ${\rm v}_{\rm CM}$. Neglecting the small secular evolution of the semi-amplitudes quantified in Sect.~\ref{subsec:expected_magnitudes},
	\begin{equation}
		\begin{aligned}
			\Delta {\rm v}_1
			&=
			\Delta {\rm v}_{\rm CM}
			+
			K_1\Delta\kappa,\\
			\Delta {\rm v}_2
			&=
			\Delta {\rm v}_{\rm CM}
			-
			K_2\Delta\kappa.
		\end{aligned}
		\label{eq:dv_ecc}
	\end{equation}
	The inter-epoch differences remove constant velocity offsets specific to each component, including the gravitational redshifts, while subtracting the differences between the two components additionally removes any velocity shift common to both components, namely $\Delta {\rm v}_{\rm CM}$, and we find
	\begin{equation}
		\Delta\kappa_{\rm spec}
		=
		\frac{\Delta {\rm v}_1-\Delta {\rm v}_2}{K_1+K_2} .
		\label{eq:delta_kappa_spec}
	\end{equation}
	Here $K_1$ and $K_2$ are the semi-amplitudes corrected for the finite exposure averaging discussed in Appendix~Sect.~\ref{app:appendixB}.
	
	For equal uncertainties $\sigma_{\rm v}$ on the individual velocity measurements, standard error propagation gives an uncertainty
	$2\sigma_{\rm v}/(K_1+K_2)$ for $\Delta\kappa_{\rm spec}$ (Appendix~Sect.~\ref{app:appendixC}). 
	The LISA prediction is sharper by a factor of five to fifty over the baselines of Table~\ref{tab:table4}, so the comparison is limited by the spectroscopy. Since
	$|\Delta\kappa|\leq2e$, and for the fiducial source ($e\simeq0.0096$, $K_1+K_2\simeq1298~{\rm km~s^{-1}}$) $2e<\sigma(\Delta\kappa_{\rm spec})$, a single pair cannot distinguish the predicted apsidal change from zero, and of order tens of pairs are needed at the fiducial velocity precision $\sigma_{\rm v}=\sigma_{\rm spec}\sim 15~{\rm km~s^{-1}}$, with one spectrum per epoch.
	
	If the two epochs are not sampled at exactly the same $u$, the $\cos u$ term in Eq.~\eqref{eq:v1v2_ecc_rv} does not cancel in $\Delta{\rm v_1}-\Delta{\rm v_2}$ completely. At the RV
	extrema, however, the leading phase-matching term vanishes and the
	remaining contribution is second order in the phase errors~Appendix~Sect.~\ref{app:appendixC11}. For
	independent errors with characteristic uncertainty $\sigma_u$, the root mean square (rms)
	contribution to $\Delta\kappa_{\rm spec}$ is of order $\sigma_u^2$ and carries no dependence on the velocity semi-amplitudes. At the epochs considered
	here, this contribution is well below both the spectroscopic uncertainty
	and the uncertainty in $\Delta\kappa_{\rm LISA}$. At quadrature, the comparison between $\Delta\kappa_{\rm spec}$ and $\Delta\kappa_{\rm LISA}$ provides a consistency test of the GW-predicted evolution of $\kappa(t)$.
	
	\subsection{Barycentric velocity changes}
	\label{sec:DeltaVCM}
	
	Population synthesis predicts that LISA-detected UCBs with bound tertiaries will reside primarily in wide outer orbits with third-body separations of $\sim10^3$~AU \citep{2025A&A...704A.156R,2026AJ....172..266K}, producing l.o.s. accelerations far below the spectroscopic floor (Eq.~\eqref{eq:DeltaVcm_est}). Detection of a tertiary system through fixed-$u$ spectroscopy is expected to be rare rather than routine. Constraining $a_{\rm los}$ is nevertheless worth pursuing, because an unrecognised acceleration can be mistaken for intrinsic frequency evolution \citep{2018PhRvD..98f4012R} and thereby bias the chirp mass inferred from $\dot f_{\rm GW}$ \citep{2021MNRAS.502.4199X}. Using a population of simulated dWDs with tertiaries, \citet{2021MNRAS.502.4199X} estimate that for approximately $9\%$ of LISA dWDs with measurable chirp masses, the acceleration induced by the tertiary biases the inferred chirp mass by 5--30\%, mostly at $f_{\rm GW}$ between 2 and 4~mHz.

	To the first order in ${\rm v}_{\rm los}/c$, with ${\rm v}_{\rm los}$ as the l.o.s. velocity of the binary barycentre, the frequency derivative measured by LISA can be written as
	\begin{equation}
		\dot f_{\rm GW}
		=
		\dot f_{\rm GW~only}
		+\dot f_{\rm tide}
		+\dot f_{\rm MT}
		+\cdots
		+\dot f_{\rm acc},
		\label{eq:fdot_decomposition}
	\end{equation}
	where $\dot f_{\rm MT}$ denotes the contribution from mass transfer. The l.o.s. acceleration contribution is
	\begin{equation}
		\dot f_{\rm acc}
		=
		\frac{f_{\rm GW}}{c}a_{\rm los},
		\label{eq:fdot_acc}
	\end{equation}
	where $a_{\rm los}>0$ corresponds to acceleration towards the observer.
	The term $\dot f_{\rm GW~only}$ is the contribution from gravitational radiation alone. Every term in Eq.~\eqref{eq:fdot_decomposition} except $\dot f_{\rm acc}$ arises from the orbital evolution. For detached circular binaries with approximately constant $a_{\rm los}$, separating $\dot f_{\rm GW~only}$ and $\dot f_{\rm acc}$ from the GW signal alone requires a measurement of $\ddot f_{\rm GW}$, which LISA measures for few dWDs \citep{2025PhRvD.111d4023E}. In a detached eccentric binary, a measurement of the apsidal advance gives the total mass, since $\dot\omega_{\rm GR}$ depends on $M_{\rm tot}$, $f_{\rm GW}$, and $e$ rather than on $\dot f_{\rm GW}$, and so remains available where $\dot f_{\rm GW}$ is too small to measure \citep{2001PhRvL..87y1101S}. \citet{2023PhRvD.107d3009X} use the total mass
	obtained this way to separate $\dot f_{\rm acc}$ from $\dot f_{\rm GW~only}$. 
	For dWDs the apsidal advance is not $\dot\omega_{\rm GR}$ alone (Sect.~\ref{subsec:expected_magnitudes}),
	\begin{equation}
		\dot\omega_{\rm tot}
		=
		\dot\omega_{\rm GR}
		+
		\dot\omega_{\rm tide}
		+
		\dot\omega_{\rm rot}
		+\cdots ,
		\label{eq:omega_dot_tot}
	\end{equation}
	and the tidal and rotational terms prevent the inversion used by \citet{2023PhRvD.107d3009X}. Here $\dot\omega_{\rm tot}$ is instead an observable, and the component masses supply both its interpretation and the chirp mass (Sect.~\ref{sec:horizon_forecast_ecc}).
	
	An independent chirp mass constrains $\dot f_{\rm GW~only}$ and improves the recovered acceleration by orders of magnitude, provided the orbital evolution is assumed to be driven by gravitational radiation alone \citep{2018PhRvD..98f4012R,2025PhRvD.111d4023E}. If tides or mass transfer contribute, their terms in Eq.~\eqref{eq:fdot_decomposition} are instead absorbed into the inferred $a_{\rm los}$ when gravitational radiation alone is assumed. Fixed-$u$ spectroscopy provides
	\begin{equation}
		a_{\rm los}
		\simeq
		-\frac{\Delta {\rm v}_{\rm CM}}{\Delta T},
		\qquad
		\dot f_{\rm acc}
		\simeq
		-\frac{f_{\rm GW}}{c}
		\frac{\Delta {\rm v}_{\rm CM}}{\Delta T}
		\label{eq:fdot_acc_spectroscopy}
	\end{equation}
	independently of $\dot f_{\rm GW}$, given that $a_{\rm los}$ is approximately constant across the GW and optical baselines. When the same observations also determine the component masses, both $\dot f_{\rm GW~only}$ and $\dot f_{\rm acc}$ are constrained independently, leaving
	\begin{equation}
		\dot f_{\rm GW}
		-\dot f_{\rm GW~only}
		-\frac{f_{\rm GW}}{c}
		\frac{\Delta {\rm v}_{\rm CM}}{\Delta T}
		=
		\dot f_{\rm tide}
		+\dot f_{\rm MT}
		+\cdots .
		\label{eq:fdot_remaining}
	\end{equation}
	The method developed here requires spectral lines that trace the orbital motion of at
	least one component, a sufficiently accurate ephemeris to compare
	common values of $u$ between epochs, and spectroscopic offsets that remain
	stable over the baseline. Interacting systems generally fail the first
	requirement. For example, when the optical spectrum is dominated by emission from the
	accretion region, the measured radial velocities follow the emitting region rather
	than a stellar photosphere. The direct-impact accretors among the known
	ultracompact systems, including HM~Cnc \citep{2010ApJ...711L.138R}, fall
	into this category. Short
	orbital periods compound the difficulty, since exposures long enough for
	adequate signal cover a significant fraction of the orbit
	(Eq.~\eqref{eq:Deltav_smear}). Even where photospheric lines are present,
	changes in the line-forming regions between epochs shift the measured
	centroids and produce an apparent non-zero $\Delta {\rm v}_{\rm CM}$
	\citep{2010MNRAS.401.1857V,2017MNRAS.470.4473P}.
	The constructions below apply wherever the measured velocities trace the
	stellar orbits, which is the usual case for non-interacting systems.

	\subsubsection{Circular SB1 and SB2 systems: Fixed-$u$ differencing}
	\label{sec:delta_vcm_sb1_sb2_circular}
	
	The regression of Eq.~\eqref{eq:v1v2_circular_rv} determines the intercept $\gamma_j={\rm v}_{\rm CM}+z_j$. A fixed-$u$
	difference across two epochs removes the constant offset $z_j$. For
	component $j$,
	\begin{equation}
		\Delta {\rm v}_j
		=
		\Delta \gamma_j
		=
		\Delta {\rm v}_{\rm CM},
		\label{eq:cir_deltaVcm}
	\end{equation}
	provided that the same argument of latitude is sampled at both epochs
	and secular changes in $K_j$ are negligible, Eq.~\eqref{eq:DeltaK_bound}. The temporal baseline must
	be long enough for the resulting common l.o.s. drift due to external
	acceleration, $\Delta {\rm v}_{\rm CM}$, to be measurable.
	For the expected UCB population, this term is usually negligible. The cancellation provided by component subtraction isolates $\Delta {\rm v}_{\rm CM}$ regardless of its magnitude.

	\subsubsection{SB2 systems in eccentric orbits: Separating apsidal and barycentric drift}
	\label{sec:delta_vcm_sb2_ecc}
	
	For eccentric orbits, fixed-$u$ differencing across two epochs retains an
	additional contribution from the change in $\kappa=e\cos\omega$. Two
	epochs scheduled at the same argument of latitude
	(Sect.~\ref{sec:argument_of_lat}) yield, for
	component $j\in\{1,2\}$,
	\begin{equation}
		\Delta {\rm v}_j
		=
		\Delta\gamma_j
		+
		(-1)^{j-1}K_j
		\Big([e\cos\omega]_{t_\beta}-[e\cos\omega]_{t_\alpha}\Big),
		\label{eq:SB2_fixedu_diff}
	\end{equation}
	where, as in the circular case, $\Delta\gamma_1=\Delta\gamma_2=\Delta{\rm v}_{\rm CM}$ for constant offsets $z_j$.

	For the multi-epoch analysis, it is useful to eliminate the apsidal contribution $\Delta\kappa=[e\cos\omega]{t_\beta}-[e\cos\omega]{t_\alpha}$ by working with the measured velocity differences. Combining the two component equations in Eq.~\eqref{eq:SB2_fixedu_diff} to remove $\Delta\kappa$ gives an equivalent WIL41 relation for the velocity differences,
	\begin{equation}
		\Delta {\rm v}_1=\Delta {\rm v}_{\rm CM}(1+q)-q\Delta {\rm v}_2 .
		\label{eq:SB2_fixedu_diff_WIL41}
	\end{equation}
	When $\Delta {\rm v}_1$ is plotted against $\Delta {\rm v}_2$, the fixed-$u$ pairs lie on a line with slope $-q$ and intercept $(1+q)\Delta {\rm v}_{\rm CM}$; equivalently, the inverse plot has slope $-1/q$, as in the standard SB2 form \citep{1941ApJ....93...29W}.  Any component-dependent offset that is constant in time cancels in $\Delta {\rm v}_j$, so the intercept $(1+q)\Delta {\rm v}_{\rm CM}$ depends only on the shared barycentric drift over the baseline. If $\Delta {\rm v}_{\rm CM}$ is negligible, the line passes through the origin, and a single fixed-$u$ pair gives $q=-\Delta {\rm v}_1/\Delta {\rm v}_2$. More generally, several distinct two-epoch pairs with a common baseline duration $\Delta T$ can be used to fit the slope and intercept together. Pairs obtained at separate epochs sample distinct apsidal phases and hence generally different values of $\Delta\kappa$, while, over a span short compared with the outer orbital period, the common baseline gives approximately the same $\Delta {\rm v}_{\rm CM}$. In WIL41 the points spread along the line because simultaneous pairs are taken at different orbital phases, so $\cos u$ varies from pair to pair. Here every pair is taken at the same $u$, and the spread instead comes from the secular change in $\kappa=e\cos\omega$ between epochs. In both cases the mass ratio is robust to effects that scale the common orbital factor, $\cos u$ there and $\Delta\kappa$ here, identically for the two components. Finite exposure averaging, for example, modifies the phase-dependent term in the same way for both stars, Eq.~\eqref{eq:Deltav_smear}, while $q$ remains unchanged.
	
	The barycentric drift is recovered from the intercept as a second observable. Once the line is determined, the intercept equals $\Delta {\rm v}_{\rm CM}(1+q)$, so
	\begin{equation}
		\Delta {\rm v}_{\rm CM}=\text{intercept}/(1+q)
		\label{eq:intercept_Deltav_cm}
	\end{equation}
	after $q$ is obtained from the slope. This is the analogue of the WIL41 recovery of the systemic velocity from the intercept, except that $\Delta {\rm v}_{\rm CM}$ measures
	the change in systemic velocity rather than its static value, providing a probe of Galactic acceleration or unseen tertiary companions (Sect.~\ref{sec:DeltaVCM}).
	
	Multi-epoch RV differences have also been used in WD population studies, for example by \citet{2012ApJ...751..143M}, who employed the maximum difference from sparsely sampled epochs to extract population statistics. In our case, the RV difference is instead scheduled at a known orbital phase determined by the GW phase and used to constrain WD binary parameters.

	\subsubsection{SB1 systems in eccentric orbits: Separating apsidal and barycentric drift}
	\label{sec:delta_vcm_sb1_ecc}
	
	Here we treat the only case not covered so far, an SB1 system in an eccentric orbit. Because only one component velocity is available, there is no opposite-sign apsidal contribution from a second component with which to eliminate $\Delta\kappa$, as in Eq.~\eqref{eq:SB2_fixedu_diff_WIL41}. An alternative is to use measurements at the RV zero crossings. These phases are difficult to use in SB2 systems because the component lines approach one another and blend, whereas in an SB1 only one set of spectral lines is present.
	
	For an SB1 system in an eccentric orbit, the semi-amplitude $K$ is first determined from Eq.~\eqref{eq:sb2_half_diff}. If the radial velocity is then measured at the same $u$ at two epochs, the difference includes the barycentric drift $\Delta{\rm v}_{\rm CM}$ together with the change in the eccentric contribution $K\Delta\kappa$. Subtracting the latter using the LISA prediction for $\Delta\kappa$ recovers $\Delta{\rm v}_{\rm CM}$. An alternative is to use the LISA prediction for $\kappa(t)$ to schedule observations at phases where the orbital contribution to the radial velocity vanishes,
	\begin{equation}
		\cos u_z=-\kappa .
	\end{equation}
	The intra-session regression still provides the semi-amplitude and intercept of the visible component, while the zero crossings isolate the barycentric contribution.
	To the first order in eccentricity, Eq.~\eqref{eq:RVzero_cross} gives
	\begin{equation}
		u_{z,+}=\frac{\pi}{2}+\kappa,
		\qquad
		u_{z,-}=\frac{3\pi}{2}-\kappa .
	\end{equation}
	For a common phase-scheduling offset $\delta u$, averaging the two
	zero-crossing velocities gives
	\begin{equation}
		\frac{
			{\rm v}_r(u_{z,+}+\delta u)
			+
			{\rm v}_r(u_{z,-}+\delta u)
		}{2}
		=
		\gamma+\mathcal O(eK\delta u^2).
		\label{eq:rvzero_average}
	\end{equation}
	The linear velocity error cancels, so the average recovers $\gamma$ at
	each epoch with an error of order $eK\delta u^2$. Both crossings are
	accessible within one observing session for orbital periods of
	approximately $5$--$60$~min. Differencing their averages between
	epochs removes the static offset $z_j$ and gives
	$\Delta{\rm v}_{\rm CM}$ independently of the apsidal contribution.

	\subsection{Obtaining and constraining component masses}
	\label{sec:sb2_chirp_compare}
	
	A multimessenger approach to component masses proposes the use of the optical SB2 mass ratio together with the LISA chirp mass \citep{2019arXiv190401601K,2023MNRAS.525.4121J}. The mass ratio itself $q=M_2/M_1$ can, for example, be obtained from SB2 radial velocities \citep{1941ApJ....93...29W,2025arXiv250400548S,2002MNRAS.332..745M}. For purely GW orbital angular momentum loss, the chirp mass combined
	with $q$ supplies component masses. Namely, $f_{\rm GW}$ and
	$\dot f_{\rm GW}$ determine the chirp mass $M_c$ and
	\begin{equation}
		M_1=M_c\frac{(1+q)^{1/5}}{q^{3/5}},
		\qquad
		M_2=qM_1.
		\label{eq:M_c_q}
	\end{equation}
	When mass transfer counteracts the inspiral driven by GW emission, the
	chirp mass inferred from a positive $\dot f_{\rm GW}$, assuming GW
	losses alone, sets a lower bound on the true chirp mass, as discussed
	for HM~Cnc by \citet{2023MNRAS.518.5123M}. In contrast,
	\citet{2025PhRvD.112d3013L} show that, for tidally locked detached
	binaries, the frequency derivative receives a tidal contribution
	$\dot f_{\rm GW}=\dot f_{\rm GW~only}(1+r_{\rm tide})$, so the
	inferred chirp mass is the effective tidal value
	$\mathcal M_{\rm tide}=M_c(1+r_{\rm tide})^{3/5}$, which sets an
	upper bound on the true $M_c$. For most detached dWDs, \citet{2025PhRvD.112d3013L} point out the infeasibility of breaking the tide-chirp mass degeneracy by measuring the second frequency derivative $\ddot f_{\rm GW}$ from photometry over a long baseline, and instead break it by using the GW strain amplitude, which is independent of $r_{\rm tide}$, together with a photometric distance. The strain amplitude encodes the chirp mass through
	\begin{equation}
		\mathcal{A} = \frac{2(M_cG)^{5/3}(\pi f_{\rm GW})^{2/3}}{c^4~d},
		\label{eq:GW_amplitude}
	\end{equation}
	where distance $d$ and $f_{\rm GW}$ are determined electromagnetically, isolating $M_c$ without reference to $\dot{f}_{\rm GW}$ (see also \cite{2025arXiv251025653T}). Comparing this 
	amplitude-derived $M_c$ with the $\dot{f}_{\rm GW}$-derived value, $\mathcal M_{\rm tide}$, then gives $r_{\rm tide}$.
	
	Another GW route uses the circular 0.5~PN modes studied by \citet{2025PhRvL.135f1402S}. With an external parallax distance $d$, the
	amplitude $\mathcal A$ of the strong quadrupole mode in Eq.~\eqref{eq:GW_amplitude} gives the chirp mass, since $\mathcal A\propto \mathcal M_c^{5/3}/d$. The amplitudes of the 0.5~PN modes then determine the scaling parameter $s=\beta\Delta\propto (M_1-M_2)M_{\rm tot}^{-2/3}$. In principle, the pair $(\mathcal M_c,s)$ separates the two component masses for nearby bright systems in which the 0.5~PN modes are detectable.
	
	Masses inferred from the orbital velocity amplitudes measured with phase-resolved spectroscopy are not tied to a particular interpretation of $\dot f_{\rm GW}$. For SB2 systems in circular and eccentric orbits, targeted phases allow one to measure phase-resolved velocities to determine $K_1$ and $K_2$.  Applications extend beyond the regime of a chirp from gravitational radiation alone, including semi-detached UCBs detected with LISA and systems where the secular drift in $K$ is not considered numerically small.
	
	Component masses are also attainable from spectroscopic SB2 systems. An example of dWD component mass determination is given by \citet{2017MNRAS.466.1575R}, who combined the mass ratio; the relative depths of the two Doppler-shifted H$\alpha$ absorption lines, one associated with each WD; and spectral model fitting to constrain the individual masses.

	With a combination of photometry and spectroscopy, one also obtains component masses, for example, see \cite{2012ApJ...757L..21H,2023ApJ...950..141K,2023ApJ...959..114K}. A best scenario for constraining component masses is with an SB2, double-eclipsing system. An example of dWD component mass determination for such a system using multiple constraints is given by \citet{2019Natur.571..528B}, for ZTF J1539+5027. The authors combine three loci in the $(M_1, M_2)$ plane: a mass-radius relation for the hot primary and constraints from light curve modelling, the spectroscopic RV semi-amplitudes, and the chirp mass inferred from the 
	measured orbital decay with allowance for a tidal contribution.
	
	For eclipsing, unequal mass SB1 systems, the light travel delay between primary and secondary eclipses (namely, the offset of the secondary from half a period) gives the mass ratio when combined with a single RV amplitude and the period, allowing the individual masses to be recovered without a second spectroscopic orbit \citep{2010ApJ...717L.108K}. In binary systems with eccentricity, this offset is contaminated by an eccentricity term $\propto e\cos\omega$, so $e$ must be independently constrained, typically from the radial velocities, before timing is used to provide a mass constraint \citep{2010ApJ...717L.108K,2012ApJ...753..101B}. These eclipse timing constraints target the individual component masses of a single system. 
	The majority of UCBs detected with LISA will not be eclipsing binaries. For these systems neither eclipse timing nor eclipse offset eccentricity constraints are available, and dense RV campaigns remain expensive in telescope time even for bright targets. For example, \cite{2025PhRvD.112d3013L} notes that RV monitoring, in principle, achieves similar goals, but only through extended orbital campaigns with future facilities such as the Thirty Meter Telescope, Giant Magellan Telescope, and Extremely Large Telescope, which is an impractical solution for oversubscribed facilities.
	
	Obtaining component masses for SB1, non-eclipsing systems requires a joint analysis of spectroscopic, astrometric, and photometric constraints. For example, \citet{2020ApJ...892L..35B} use composite binary models (two synthetic WD spectra weighted by their radii) to yield both masses, through interpolation of the atmospheric parameters on He-core WD tracks. Hubble Space Telescope ultraviolet spectroscopy further sharpens this by constraining the hot primary, whose parameters are then held fixed in the same joint fit to recover the cool secondary \citep{2025ApJ...991...65B}.

	\subsection{Determining $\eta$ and $\kappa$ in optical and gravitational wave astronomy}
	\label{sec:eta_kappa_sum}
	
	In eclipsing systems, the secondary eclipse offset $\Delta t_2$ from phase 0.5, after Roemer\footnote{The Roemer delay is the shift in eclipse timing caused by the finite light travel time across the orbit. At the two conjunctions the eclipsed star sits at different distances from us, so the secondary eclipse is displaced from phase 0.5 even for a circular orbit.} subtraction, satisfies $ \Delta t_2 \simeq 2P_b\kappa/\pi.$ A single epoch gives the lower limit $e \ge |\kappa|$. If $e$ remains approximately constant, $\Delta t_2$ traces a sinusoid of amplitude $2 P_b e/\pi$ and period $2\pi/\dot\omega$ \citep{2025arXiv250515580V}, and a resolved apsidal cycle would yield $e$ and $\dot\omega$ separately. Over baselines short compared to the apsidal period, only the local slope is accessible, leaving $e$, $\omega$, and $\dot\omega$ degenerate without longer monitoring or an independent measurement of $\dot\omega$. We show this is provided by LISA's GW phase, which delivers a total $\dot\omega_{\rm tot}$ that includes contributions from stellar structure, \eg~from tides. Targeted-$u$ spectroscopy breaks the same degeneracy independently. Subtracting the velocity change of one star from that of the other returns $\Delta\kappa$, Eq.~\eqref{eq:delta_kappa_spec}, without reference to $\dot\omega$ or to the assumed apsidal model, so it measures the evolution of $\kappa$ rather than inferring it (Sect.~\ref{sec:delta_kappa_consistency}). The other parameter, $\eta=e\sin\omega$, is instead inferred from the unequal durations of the primary and secondary eclipses, or indirectly from O$-$C timing residuals\footnote{Observed minus Calculated} as apsidal motion changes $\kappa(t)$ \citep{2023MNRAS.522.1310T,2010exop.book...55W, 2010ApJ...717L.108K,2025arXiv250515580V}.
	
	In eclipsing binaries, the two photometric observables that constrain $\eta$ and $\kappa$ separately are not equally accessible, despite both arising at the same order in $e$. The displacement of the secondary eclipse from $P_b/2$ is $(2 P_b/\pi)~\kappa$, while the duration difference between primary and secondary eclipses is of order $\eta~W_{\rm eclipse}$ \citep{1981AJ.....86..102P}. Although both effects are first order in $e$, their observable amplitudes are set by different timescales: the orbital period for $\kappa$, the eclipse duration for $\eta$. In UCBs the eclipse occupies only a small fraction of the orbit, so at $|\kappa| \simeq |\eta|$ the timing offset exceeds the width difference by a factor $P_b/W_{\rm eclipse}$, roughly an order of magnitude. At $e \sim 10^{-3}$ the width difference falls below one second, smaller than the arrival time precision achievable for an individual eclipse even with fast cameras on large telescopes \citep[$\simeq 0.5$ s;][]{2025arXiv250515580V}. 
	
	Measuring eccentricity parameters $(\eta,\kappa)$ from spectroscopic orbit fitting is even more difficult for UCBs. For close binary systems, \cite{2022ARep...66S...5C} summarises two classical spectroscopic orbit methods for estimating orbital elements from RV curves. For small eccentricities $e<0.5$, the eccentricity information is encoded in detailed structure in the RV curve. One method uses characteristic areas bounded by a densely sampled RV curve, while another avoids area measurements and recovers $(\eta,\kappa)$ from characteristic orbital phases, $\phi=(t-t_0)/P_b$, identified on the RV curve, for an arbitrary initial time $t_0$. The latter is more closely related to targeted-$u$ phase measurements, but the information flow is different. In the targeted-$u$ programme, LISA supplies the argument of latitude for scheduled spectroscopic measurements. In the classical spectroscopic construction, the characteristic orbital phases are inferred from an existing, though incomplete, RV curve. In the two methods described by \cite{2022ARep...66S...5C}, the eccentricity information is obtained from the spectroscopic orbit itself. For short-period UCBs this is generally impractical, because phase-resolved spectroscopy is observationally demanding. Moreover, it is unclear what is the smallest eccentricity amenable to the two classical approaches.
	
	It is natural to ask whether LISA constrains $(e,\omega)$. Sampling $(e,\omega)$ is challenging at small eccentricity because the boundary $e\geq0$ biases noisy measurements towards nonzero eccentricity \citep{1971AJ.....76..544L,2021AJ....161..241F}. It is  preferable to work with the non-singular combinations $(\eta,\kappa)=(e\sin\omega,e\cos\omega)$, which remain well defined as $e\to0$ and avoid the ill-defined argument of periastron in the $(e,\omega)$ parameterisation \citep{2008PhRvD..77d4013P}, although this does not by itself remove the underlying statistical bias \citep{2013PASP..125...83E}. To first order in eccentricity, the LISA waveform is linear in $(\eta,\kappa)$ (Appendix~Sect.~\ref{app:appendixE}). LISA constrains the two components $e\sin\omega$ and $e\cos\omega$, rather than $e$ and $\omega$ separately.\footnote{In radio pulsar timing, $(\eta,\kappa,T_{\rm asc})$ resolve a different  problem. The Roemer delay depends on $T_0$ and $\omega$ only  through the combination $T_0-\omega/n_b$, making them individually  poorly constrained when $e$ is small \citep{10.1046/j.1365-8711.2001.04606.x}.} The motivation here is to supply LISA-derived orbital phase information and eccentricity parameters $(\eta(t),\kappa(t))\rightarrow (e,\omega(t))$ to optical astronomers for the purpose of targeted spectroscopic measurements allowing one to place new constraints on the component masses of the binary system.
	
	\subsection{Summary of applications}
	
	We have introduced a common framework for targeted-$u$
	spectroscopy of SB2 and SB1 binaries in circular and eccentric
	orbits. In both cases, the LISA parameter posteriors define the argument of latitude
	$u(t)$ at which each spectrum is taken, Eq.~\eqref{eq:phase_map}. The GW phase fixes only the mean phase measured from the ascending node,
	$\Phi_{\rm asc}=\Phi_{\rm GW}/2$, which equals $u(t)$ for a circular orbit. For
	an eccentric orbit, the LISA posteriors for $(\eta,\kappa,\dot\omega)$ provide
	the LL pair and its apsidal evolution, allowing
	$(\eta,\kappa)$ to be evaluated at the epoch of each spectrum and the
	$\mathcal{O}(e)$ correction $\delta_{e,\omega}$ of
	Table~\ref{tab:table1} to be applied, as described in
	Appendix~Sect.~\ref{app:appendixA2}. Spectra are then obtained on demand, each carrying an
	orbital phase stamp. The radial velocities measured from them are regressed on
	$\cos u$ for a circular orbit and on $F=\cos u+\kappa$ for an eccentric
	orbit. Velocities measured at several phases within one session determine the
	semi-amplitudes and the velocity intercepts.

	The WIL41 velocity--velocity relation, Eq.~\eqref{eq:wilson_line}, eliminates the orbital phase \citep{1941ApJ....93...29W}. Equations~\eqref{eq:v1v2_circular_rv} and \eqref{eq:v1v2_ecc_rv}, for SB2 systems in circular and eccentric orbits respectively, retain it, because the GW phase solution supplies $u$ for every spectrum. With this restored phase information, the individual semi-amplitudes $K_1$ and $K_2$ are measured separately, rather than only their ratio. Combined with the orbital parameters measured by LISA, these yield the individual component masses $M_1$ and $M_2$.
	
	Two-epoch velocity differences, Eq.~\eqref{eq:fixedu_no_K_evolution}, over sufficiently long baselines are used to probe $\Delta\kappa$ and $\Delta {\rm v}_{CM}$. The velocity differences of the two components provide an independent spectroscopic estimate of $\Delta\kappa_{\rm spec}$ through Eq.~\eqref{eq:delta_kappa_spec}. The systemic velocity offsets and the common barycentric velocity change between epochs both cancel in forming $\Delta\kappa_{\rm spec}$, which is then compared with the change predicted by LISA in $\kappa=e\cos\omega$ between the same epochs,
	$\Delta\kappa_{\rm LISA}=\kappa_{\rm LISA}(t_\beta)-\kappa_{\rm LISA}(t_\alpha)$, defined using Eq.~\eqref{eq:kappa_lisa_prediction}.
	
	For most UCBs, $\Delta{\rm v}_{\rm CM}$ is expected to lie below the
	inter-epoch velocity precision, Eq.~\eqref{app:unc_of_Delta_v}, and a
	null result is itself useful. Such a result constrains the acceleration term in
	Eq.~\eqref{eq:fdot_decomposition} from the spectroscopy alone, so that
	$\dot f_{\rm GW}-\dot f_{\rm GW~only}-\dot f_{\rm acc}$ isolates the sum of the remaining
	orbital contributions. The recovery of $\Delta{\rm v}_{\rm CM}$ for SB1
	and SB2 systems in circular orbits is shown in Sect.~\ref{sec:delta_vcm_sb1_sb2_circular}, with the corresponding treatments
	for SB2 and SB1 systems in eccentric orbits given in
	Sect.~\ref{sec:delta_vcm_sb2_ecc} and Sect.~\ref{sec:delta_vcm_sb1_ecc}, respectively.
	
	Finally, we have compared the LL parameters
	$(\eta,\kappa)$ in photometric, spectroscopic, and LISA measurements~Sect.~\ref{sec:eta_kappa_sum}.
	Eclipse photometry constrains $\kappa$ well through the secondary
	eclipse offset and $\eta$ poorly through the eclipse-width difference,
	and neither is available without eclipses. Spectroscopy provides an
	independent measurement of $\Delta\kappa$ from fixed-$u$ velocity
	differences between epochs, after removing the common barycentric
	drift~Sect.~\ref{sec:delta_kappa_consistency}. LISA supplies both $\eta$ and $\kappa$, together with their evolution, in a form that remains well behaved as $e\to0$.
	
	\section{Phase-coherence horizon forecasting}
	\label{sec:horizon_forecast}
	
	The preceding applications rely on a GW-derived orbital phase model that is sufficiently accurate to assign an argument of latitude to each spectroscopic observation. For a chosen argument of latitude $u_\star$, the phase model predicts the sequence of epochs $t_{\star,k}$ satisfying
	\begin{equation}
		u(t_{\star,k}) = u_\star \pmod{2\pi}.
	\end{equation}
	The index $k$ labels successive passages through the same argument of latitude. Where the choice of passage does not matter, we drop the index and write $t_\star$. These predicted epochs allow spectra to be obtained at selected orbital phases, from which the RV semi-amplitudes can be determined for circular and eccentric orbits, Sect.~\ref{sec:sb2_circular} and Sect.~\ref{sec:theory_ecc}, respectively. They also allow observations to be made at matched phases across separated epochs to measure changes in $\kappa$, Sect.~\ref{sec:delta_kappa_consistency}, and the barycentric velocity, Sect.~\ref{sec:DeltaVCM}.
	The question addressed in this section is how long the uncertainty in the predicted argument of latitude remains small enough for these phase-targeted spectroscopic measurements. The admissible uncertainty is set by the spectroscopic precision. We define the phase tolerance $\delta u_{\rm tol}$ as the largest error in the predicted argument of latitude for which the resulting velocity error stays below the adopted RV uncertainty, either for a single epoch or for a velocity difference between two epochs. This tolerance depends on the ratio $\sigma_{\rm v}/K$, where $\sigma_{\rm v}$ is the single-epoch velocity precision and $K$ is the RV semi-amplitude (Appendix~Sect.~\ref{app:appendixC1}). For a given GW phase solution and anchor $u_\star$, reducing $\sigma_{\rm v}$ tightens the phase tolerance and shortens the time horizon $T_{\rm valid}$ (Eq.~\eqref{eq:T_valid_alpha}) over which the GW ephemeris is accurate enough for scheduling. 
	
	Our focus is the multimessenger exchange of orbital information between GW and optical observations, so we do not review the LISA data analysis of UCBs here; see \cite{2023A&A...678A.123L} and references therein. For circular orbits, the relevant output for the fixed-$u$ method is the GW phase as a function of time,
	\begin{equation}
		\Phi_{\rm GW}(t) = \phi_{\rm ref}
		+ 2\pi\!\left[
		f_{\rm GW,ref}(t-t_{\rm ref})
		+ \tfrac{1}{2}\dot f_{\rm GW}(t-t_{\rm ref})^2
		\right],
		\label{eq:GW_phase_t_ref}
	\end{equation}
	The LISA compact binary gravitational waveform template for circular orbits is a function of eight parameters, first introduced in Sect.~\ref{sec:intro}, ($\mathcal A$, $f_{\rm GW,0}$, $\dot f$, $\phi_{\rm 0}$, $\psi$, $\phi,\theta$, $\cos\iota$). The predicted argument of latitude is $u(t)=\Phi_{\rm GW}/2+\epsilon$, where $\epsilon$ aligns the phase origin of $\Phi_{\rm GW}$ (determined by the LISA inferred GW parameters) with that of $u$. Eq.~\eqref{eq:GW_phase_t_ref} measures the GW phase from a reference epoch, and that epoch is a free choice. Ours is set by the search code, which places it at the start of the observation, $t_0$. This reference time is essential for determining the $\Phi_{\rm GW}\rightarrow u_\star \rightarrow t_\star$ mapping. In what follows, we choose to transform the LISA parameter posteriors for $\phi_0$ and $f_{\rm GW,0}$ to the midpoint of the LISA observation $t_{\rm mid}$ in post-processing.
	
	For eccentric orbits, the LISA parametrisation additionally includes the LL parameters $\eta=e\sin\omega$ and $\kappa=e\cos\omega$, together with the apsidal advance rate $\dot\omega$. These parameters provide the correction $\delta_{e,\omega}(t)$ required to predict the argument of latitude $u(t)$. In what follows, the posteriors for $\eta_0$ and $\kappa_0$ are likewise transformed to $t_{\rm mid}$. The parameterisation adopted for the LISA analysis and the form of $\delta_{e,\omega}$ are given in Sect.~\ref{sec:horizon_forecast_ecc}, while the mapping from the LISA-inferred GW parameters to $u(t)$ is discussed in Appendix~Sect.~\ref{app:appendixA2}.

	\subsection{Estimating a time horizon for eccentric orbits with a simulated LISA dWD analysis}
	\label{sec:horizon_forecast_ecc}
	We forecast a time horizon $T_{\mathrm{valid}}$ over which a phase anchor $u(t)$ remains predictable to within a tolerance set by the RV measurement error, Appendix~Sect.~\ref{app:appendixC1}. The input is the covariance matrix derived from the LISA posteriors resulting from a targeted search for the UCB introduced in Sect.~\ref{subsec:ecc_binary_aps_motion}. The fiducial system has component masses $M_1=0.347~M_\odot$ and $M_2=0.285~M_\odot$, GW frequency $f_{\rm GW,0}\sim8.41$~mHz ($P_b=3.97$~min), inclination $\iota=84.65^\circ$, distance $d=0.577$~kpc, and Galactic coordinates $(l,b)=(-93.56^\circ,-19.19^\circ)$.
	
	We simulated a 4-year GW observation using GBGPU \citep{2007PhRvD..76h3006C, 2018PhRvD..98f4012R, 2022MNRAS.517..697K, Katz_Bayle_lisa_on_gpu_2026} and the LISA Data Challenge `Sangria' analytic noise simulator \citep{LDC_WorkingGroup_software_2022}. We incorporated small eccentricities in the GBGPU simulated waveform templates by adding the
	leading $\mathcal{O}(e)$ sidebands to the dominant circular mode (derived in Appendix~Sect.~\ref{app:appendixD}, defined for small eccentricities in Appendix~Sect.~\ref{app:appendixE}). With $f_{\rm GW,0}$ the dominant $n=2$ frequency, the $n=1$ and $n=3$ sidebands lie near $f_{\rm GW,0}/2$ and $3f_{\rm GW,0}/2$, respectively. These sidebands are linear in the LL parameters $\eta = e\sin\omega$ and $\kappa = e\cos\omega$, sampled at the chain epoch $t_0$ as $(\eta_0,\kappa_0)$ in place of $(e,\omega)$. The apsidal rate $\dot\omega$ was sampled alongside $(\eta_0,\kappa_0)$. 
	
	Waveforms valid at arbitrary eccentricity, with 1~PN orbital dynamics and periastron advance incorporated, are given by  \citet{2007MNRAS.374..721T}. Their prescription is appropriate for LISA sources containing neutron stars or black holes, where eccentricities are potentially substantial and the relativistic periastron advance of two point masses dominates. For the dWDs considered here, with $e \lesssim 10^{-2}$, we used the leading small-eccentricity Newtonian description. At Newtonian quadrupole order, the dominant $f_2=f_{\rm GW,0}$ mode has amplitude $A+\mathcal{O}\Big(e^2A\Big)$, while eccentricity generates sidebands near $f_{\rm GW,0}/2$ and $3f_{\rm GW,0}/2$ with amplitudes $\mathcal{O}(eA)$ \citep{1963PhRv..131..435P,10.1093/mnras/274.1.115,2001MNRAS.325..358P,2016PhRvD..93l4061M}. The $h_+(t)$ and $h_\times(t)$ waveforms (Eqs.~\eqref{app:final_form_h_plus} and~\eqref{app:final_form_h_cross}) retain these leading sidebands to first order in the LL parameters
	$(\eta,\kappa)=(e\sin\omega,e\cos\omega)$, while neglecting terms of relative size $\mathcal{O}(e^2)$. For the injected value $e\simeq 9.6\times10^{-3}$, the neglected relative corrections are at the $\sim 10^{-4}$ level, beyond the accuracy needed for the scheduling forecast developed below. In compact dWDs, apsidal advance is not driven by the dynamics of two point masses alone. Tidal and rotational distortions contribute at a level  comparable to, or for asymmetric systems exceeding, the GR term, (Eq.~\eqref{eq:omega_dot_GR}-\eqref{eq:del_om_tide}).  The periastron advance rate, $\dot{\omega}_{\rm tot}$, was treated as a free waveform parameter and was recovered from the data.
	
	Searches for binaries with asymmetric component masses target a different physical contribution to the waveform \citep{2025PhRvL.135f1402S} than the signal from eccentricity analysed here. Even for a circular orbit, the leading 0.5 PN corrections produce, omitting the subscript that marks the reference epoch here, modes at $f_{\rm GW}/2$ and $3f_{\rm GW}/2$ with amplitudes proportional to the mass asymmetry $\Delta = (M_1-M_2)/(M_1+M_2)$, where $M_1\ge M_2$, and vanish in the equal-mass limit. For Galactic dWDs, \citet{2025PhRvL.135f1402S} applied the leading 0.5~PN corrections to close WD binaries in the mHz band and assessed
	whether the resulting modes could be
	detected in nearby systems. \citet{2026PhRvD.113f3009S} subsequently extended this question to a synthetic Galactic dWD population, showing that these 0.5~PN modes are generally undetectable with LISA except for rare nearby systems, but could be detected for a larger sample by future decihertz observatories such as DECIGO and the Big Bang Observer.
	
	The 0.5~PN modes at $f_{\rm GW}/2$ and $3f_{\rm GW}/2$ have amplitudes of order $\mathcal A\beta\Delta$, where $\beta=(\pi G M_{\rm tot} f_{\rm GW}/c^3)^{1/3}\sim v/c$ is the PN velocity parameter and $\mathcal A$ is the leading-order quadrupole GW amplitude. These modes probe the mass asymmetry $\Delta$, and hence the mass ratio. Although these features occur in overlapping frequency ranges, the small-eccentricity signal considered here has a different origin. It is present already at leading quadrupole order, has amplitudes of order $e\mathcal A$, and is shifted by the apsidal motion. 
	\citet{2025PhRvL.135f1402S} note that apsidal motion shifts the eccentricity-induced Newtonian sidebands away from the 0.5~PN modes near $f_{\rm GW}/2$ and $3f_{\rm GW}/2$, allowing the two contributions to be distinguished in sufficiently long observations.
	\citet{2025PhRvL.135f1402S} also note that apsidal advance shifts the eccentricity-induced Newtonian sidebands away from the circular 0.5~PN modes, producing a small frequency gap that, in UCBs, is likely to include relativistic, tidal, and rotational contributions.
	For $T\gtrsim4~{\rm yr}$, this gap is expected to be resolvable at the matched filter frequency resolution $\sim 1/T$.
	
	We did not include the circular 0.5~PN modes in the present analysis. For the source analysed here, these modes scale as $\mathcal A\beta\Delta$, with $\beta\Delta\simeq4\times10^{-4}$, whereas the eccentricity-induced sidebands scale as $e\mathcal A$ with $e\simeq10^{-2}$. The recovered $n=1,3$ structure is expected to be set primarily by the eccentricity-induced sidebands, with the circular 0.5~PN modes contributing at a much lower amplitude level.
	
	Here we use the frequency shift of the eccentric sidebands caused by apsidal motion for a different purpose. Rather than using it only to separate eccentricity-induced sidebands from circular 0.5~PN modes, we regard the total apsidal advance $\dot\omega_{\rm tot}$ as an observable.
	Once the fixed-$u$ method supplies $M_{\rm tot}$, Eq.~\eqref{eq:Kepler_mtot}, the contribution from the orbital dynamics of two point masses alone, $\dot\omega_{\rm GR}(M_{\rm tot})$, is fixed, and the remainder $\dot\omega_{\rm tot}-\dot\omega_{\rm GR}$ isolates the tidal and rotational terms of Eq.~\eqref{eq:omega_dot_tot}.
	
	This offset affects the waveform as follows. Apsidal advance displaces the sidebands, so that the $n=1$ line splits to $f_{\rm GW}/2\pm\dot\omega/(2\pi)$, and the $n=3$ line shifts to $3f_{\rm GW}/2-\dot\omega/(2\pi)$. Their amplitudes are linear in the LL parameters $(\eta,\kappa)$, with $\dot\omega$ sampled in the targeted MCMC. Using the small-eccentricity \cite{1964PhDT........51P} scaling, $d\ln P_b/d\ln e=18/19$, equivalently $|\dot e|/e=(19/18)\dot f_{\rm GW}^{\rm GR}/f_{\rm GW}$, the fractional change in eccentricity over a 4-year observation is only $|\dot e|T_{\rm obs}/e\sim 2\times10^{-5}$, so the eccentricity $e$ was held fixed in the waveform model.
	
	The analysis here was carried out with the \texttt{Eryn} sampler \citep{Karnesis:2023ras,michael_katz_2023_7705496,2013PASP..125..306F}. For the LISA analysis, to perform a targeted search, we fixed the sky location at its injected value. In practice, the sky location would be supplied by EM observations of the counterpart to arcsecond precision or better, far tighter than is needed for the phase forecast. Scheduling requires only the parameters that fix the argument of latitude $u(t)$. We formed their joint posterior
	\begin{equation}
		\left(
		\phi_{\rm mid},~
		f_{\rm GW,mid},~
		\dot f_{\rm GW},~
		\eta_{\rm mid},~
		\kappa_{\rm mid},~
		\dot\omega
		\right)
		\label{eq:phase_forecast_coords}
	\end{equation}
	from the sampled LISA posterior, referencing the phase model and the LL parameter pair to the midpoint of the LISA observation, $t_{\rm mid}$. The remaining two, $\dot f_{\rm GW}$ and the apsidal advance rate $\dot\omega$, do not depend on the reference epoch. We propagated the covariance of these six parameters through a first-order expansion of $u(t)$ about the posterior mean to obtain the predicted scheduling uncertainty $\sigma_u(t)$.
	
\begin{figure*}[!htbp]
		\centering
		\includegraphics[width=1.0\textwidth]{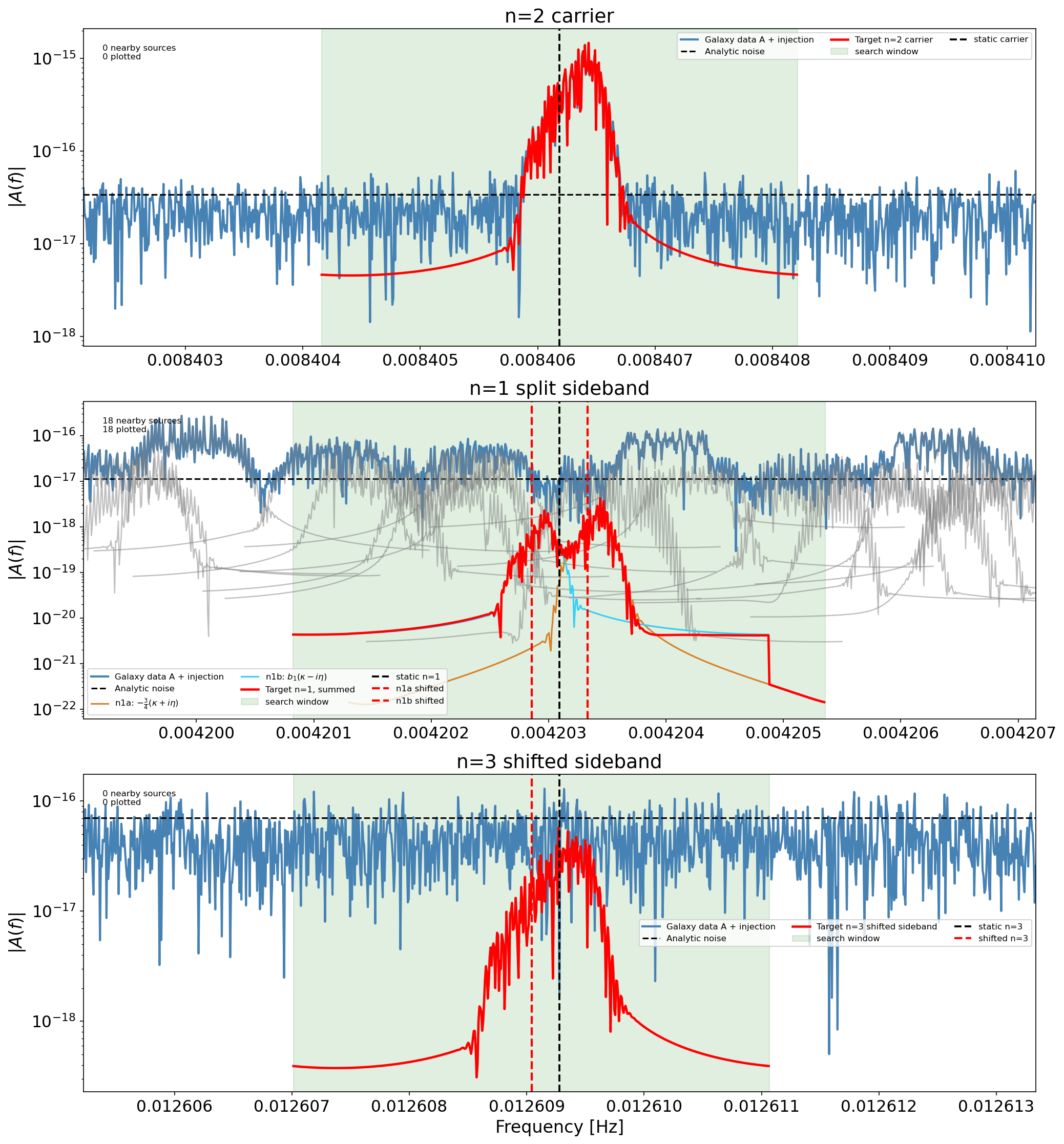}
		\caption{Four-year amplitude spectral densities of the $n=1,2,3$ GW frequency components for a simulated binary in an eccentric orbit. The binary is placed at the sky location of a source with $f_{\rm GW,0}=8.4061864571450453~{\rm mHz}$ and has LL parameters $\kappa=\eta=0.0068$, corresponding to $(e,\omega)\simeq(0.00962,\pi/4)$. Each panel covers the 512 bin analysis window (shaded regions) used in the
			likelihood for the $n=2$ mode at $f_{\rm GW,0}$ (top), the $n=1$
			sideband at $f_{\rm GW,0}/2$ (middle), and the $n=3$ sideband at
			$3f_{\rm GW, 0}/2$ (bottom).  Apsidal motion splits the $n=1$ sideband into two pieces at $f_{\rm GW, 0}/2\pm\dot\omega/(2\pi)$ (orange and cyan), and shifts the
			$n=3$ sideband to $3f_{\rm GW, 0}/2-\dot\omega/(2\pi)$.  Black dashed
			lines mark the unshifted line positions ($\dot\omega=0$); red dashed
			lines mark the positions including the apsidal shift.  The injected waveform is the coherent sum of these pieces (red).  In each
			panel, the dWD confusion foreground adds to the noise and increases
			with orbital period, as visible in the $n=1$ panel; the analytic
			instrument noise floor is shown for reference (black dashed); and grey
			lines show confusion sources falling within the analysis window, which
			contribute unmodelled bias to the likelihood.}
		\label{fig:wavforms}
	\end{figure*}

\begin{figure*}[!htbp]
		\centering
		\includegraphics[width=1.0
		\textwidth]{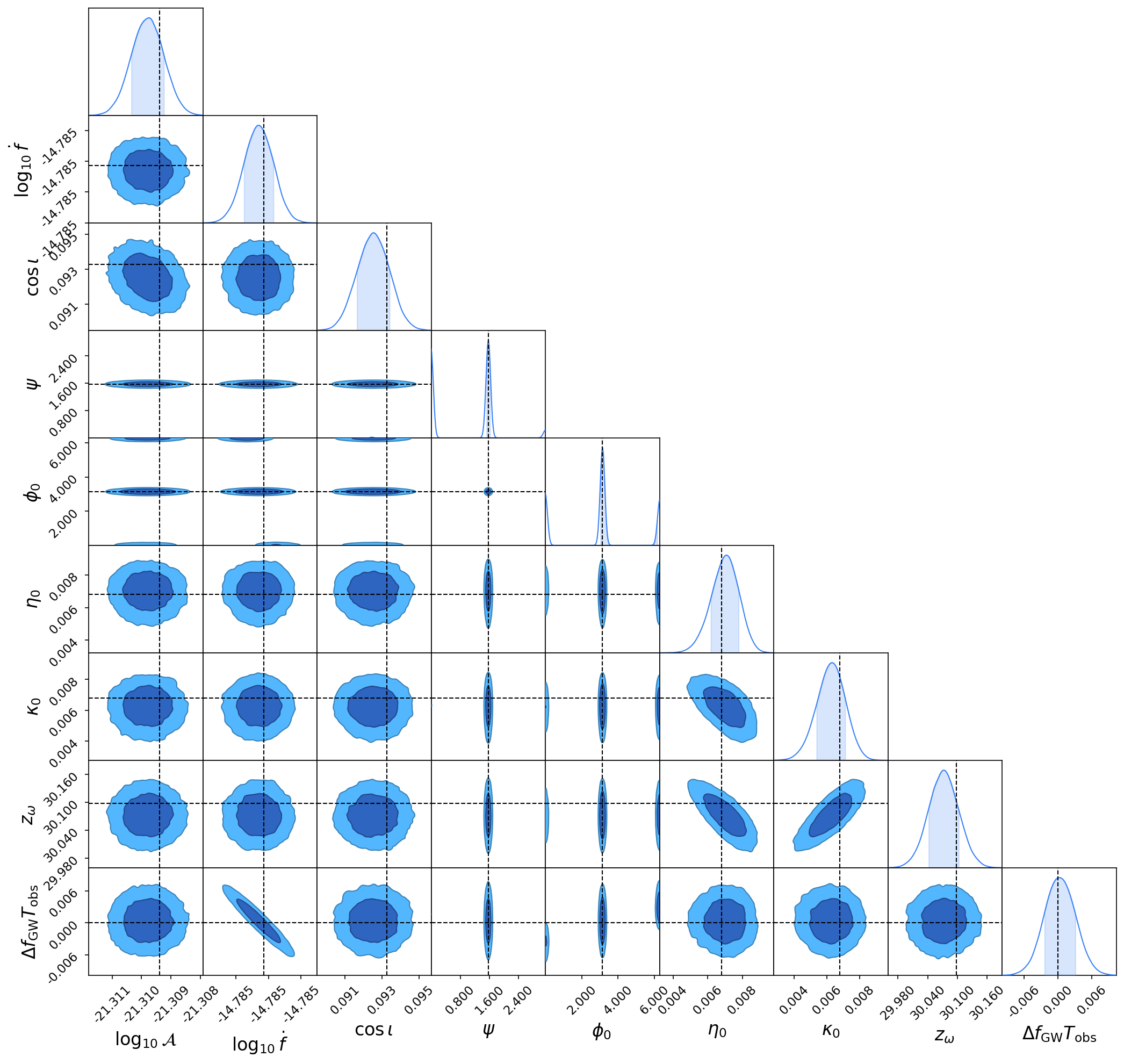}
		\caption{Four-year posterior distributions for the nine parameter Galactic binary waveform model with eccentricity and apsidal advance. The targeted MCMC fixed the sky position and
			sampled $(\log_{10}\mathcal A,\log_{10}\dot f_{\rm GW},\cos\iota,\psi,\phi_0,
			\eta_0,\kappa_0,z_\omega,f_{\rm GW, 0})$, where $z_\omega=(\dot\omega/2\pi)T_{\rm obs}$ is the apsidal frequency shift in
			Fourier bin units over the LISA observation. Dashed lines show the injected parameter values. The injected LL parameters
			$\eta_0=\kappa_0=0.0068$, corresponding to $(e,\omega_0)\simeq(0.00962,\pi/4)$, are recovered together with
			$z_\omega\simeq30.10$, demonstrating that the likelihood with shifted sidebands recovers both the small eccentricity and the apsidal advance
			for this source while allowing $f_{\rm GW, 0}$ to vary within a narrow targeted prior. The polarisation angle $\psi$ and initial phase $\phi_0$ show the
			expected mode degeneracy.}
		\label{fig:corner_ecc}
	\end{figure*}

\begin{figure*}[!htbp]
		\sidecaption
		\includegraphics[width=12cm]{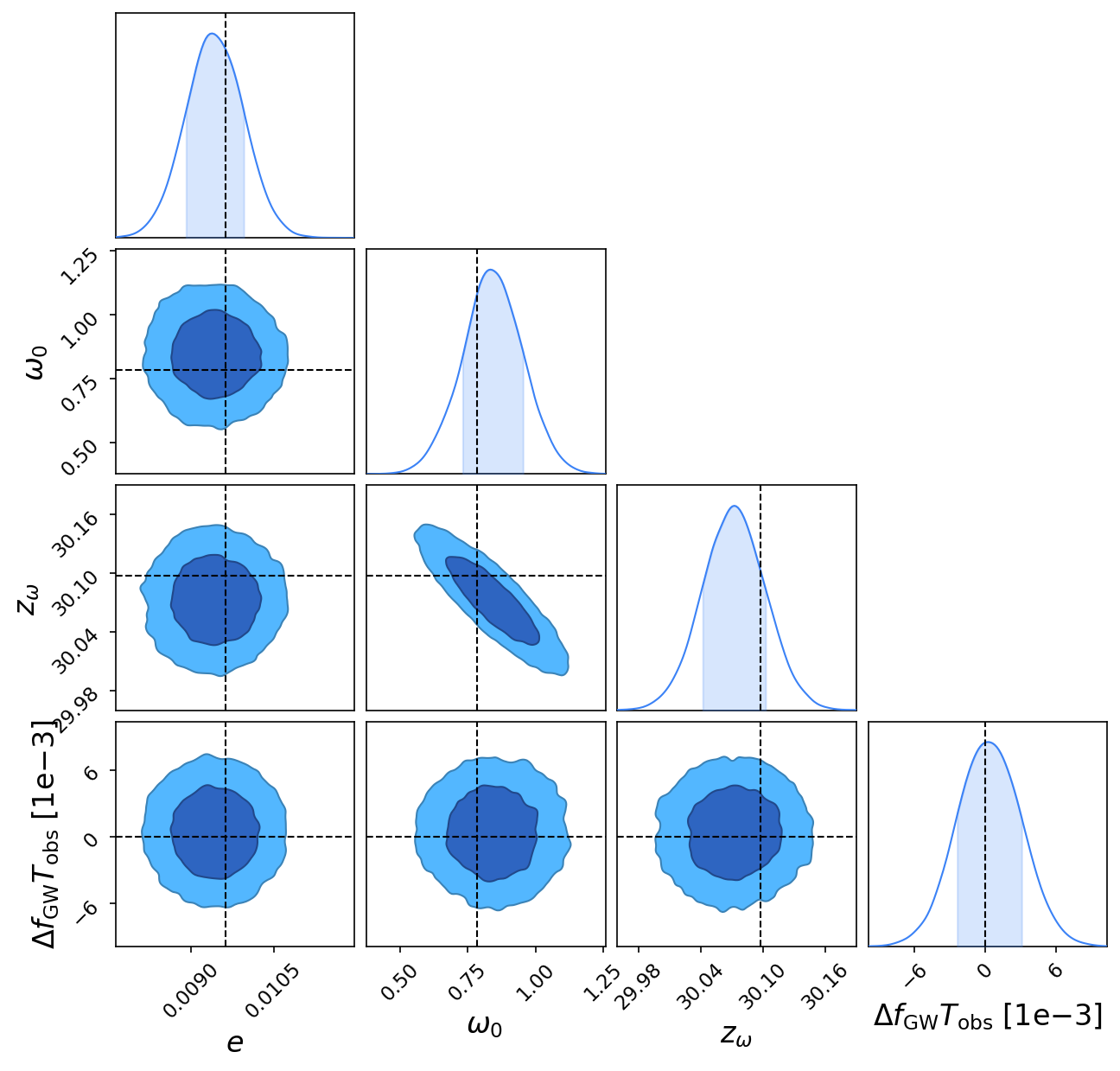}
		\caption{Marginal posterior distributions for the parameters inferred from the LISA analysis using the waveform model with eccentricity and apsidal advance. The corresponding waveforms are shown in Fig.~\ref{fig:wavforms}. The targeted MCMC held the sky position fixed and sampled the LL parameters
			$(\eta_0,\kappa_0)$ together with
			$z_\omega=(\dot\omega/2\pi)T_{\rm obs}$. We plot the derived eccentricity
			$e=(\kappa_0^2+\eta_0^2)^{1/2}$, the physical argument of periastron
			$\omega_0=\tan^{-1}(\eta_0/\kappa_0)$, and the apsidal shift parameter
			$z_\omega$. Filled regions show the $68\%$ and $95\%$ credible contours,
			and dashed lines mark the injected values. For the injected
			$\eta_0=\kappa_0=0.0068$, the physical longitude is
			$\omega_0=\pi/4$. The posterior localises both $(\eta,\kappa)$
			and the apsidal displacement of the sidebands, demonstrating that
			the apsidal likelihood recovers $(e,\omega_0,\dot\omega)$ for this
			source over a 4-year LISA baseline.}
		\label{fig:combined}
	\end{figure*}

	We used the small-$e$ $h_+$ and $h_\times$ waveforms, linear in the LL parameters $(\eta,\kappa)$, derived in Appendix~Sect.~\ref{app:appendixE}. The waveform model is described by the following parameters,
	\begin{equation}
		\boldsymbol{\lambda}_{\rm ecc}
		=
		\left(
		\log_{10}\mathcal A,~
		f_{\rm GW,0},~
		\log_{10}\dot f_{\rm GW},~
		\cos\iota,~
		\psi,~
		\phi_0,~
		\eta_0,~
		\kappa_0,~
		z_\omega
		\right).
		\label{eq:ecc_waveform_params}
	\end{equation}
	We used parameter $z_\omega$ to express the apsidal frequency $\dot\omega/(2\pi)$ in units of the Fourier frequency bin width $1/T_{\rm obs}$. In general, $\dot\omega$ is the total apsidal advance rate that rotates $(\eta_0,\kappa_0)=(e\sin\omega_0,e\cos\omega_0)$.   It contains the GR, tidal, and rotational contributions to the apsidal advance. For the proof of concept analysis below, we set $\dot\omega$ equal to the GR contribution only, $\dot\omega_{\rm GR}$.  The same analysis applies if $\dot\omega$ contains tidal, rotational, or other contributions to the apsidal advance. The simulated dominant waveform and sidebands are shown in Fig.~\ref{fig:wavforms}, and the
	recovered parameter posteriors are shown in Figures~\ref{fig:corner_ecc} and~\ref{fig:combined}. 
	
	For the simulated binary described at the start of this section, we adopted $T_{\rm obs}=4$~yr and $\dot\omega=\dot\omega_{\rm GR}$, and we referenced the waveform parameters to the start of the observation, $t_{\rm ref}=t_0=0$. At this epoch the injected parameter values are
	\begin{equation}
		\begin{aligned}
			\boldsymbol{\lambda}_{\rm ecc}^{\rm inj}
			=( 
			-{} & 21.309,\;
			8.406186457145\times 10^{-3},\;
			-14.7850265,\; \\
			&0.093,\;
			0.339,\;
			4.6100488,\; \\
			&0.0068,\;
			0.0068,\;
			30.097462
			),
			\label{eq:ecc_mcmc_inj}
		\end{aligned}
	\end{equation}
	in the ordering of Eq.~\eqref{eq:ecc_waveform_params}, with $\psi$ and $\phi_0$ in radians. The injection $\eta_0=\kappa_0=0.0068$ corresponds to $(e,\omega_0)=(9.62\times10^{-3},~\pi/4)$, and $z_\omega=30.097$ encodes
	$\dot\omega_{\rm GR}=1.498\times10^{-6}$~rad~s$^{-1}$ for the component
	masses.
	
	To forecast when spectra can be obtained at a chosen anchor $u_\star$, we propagated the posterior covariance of Eq.~\eqref{eq:phase_forecast_coords} into the predicted argument of latitude,
	\begin{equation}
		u(t)=\frac{1}{2}\Phi_{\rm GW}(t)+\epsilon+\delta_{e,\omega}(t),
		\label{eq:arg_latitude_main}
	\end{equation}
	where $\epsilon$ is the constant phase origin offset between the LISA GW phase and the chosen star's argument of latitude (see Appendix~Sect.~\ref{app:appendixA1}). In our chosen convention, we reference the orbital phase with respect to the ascending node $\Phi_{\rm asc}=\Phi_{\rm GW}/2$, so that $\epsilon=0$ for star~1 and $\epsilon=\pi$ for star~2. The eccentricity correction $\delta_{e,\omega}(t)$ is determined by the LL pair and the apsidal advance rate $\dot\omega$ (derived in Appendix~Sect.~\ref{app:appendixA1}),
	\begin{equation}
		\delta_{e,\omega}(t)
		=
		2\kappa(t)\sin\Phi_{\rm asc}(t)
		-
		2\eta(t)\cos\Phi_{\rm asc}(t).
		\label{eq:deltae_main}
	\end{equation}
	The resulting uncertainty $\sigma_u(t)$ is used in two ways. Compared with the phase tolerance $\propto \delta u_{\rm tol}$, it sets the horizon $T_{\rm valid}$ (Sect.~\ref{sec:ecc_horizon_example}). Divided by the rate at which the argument of latitude advances, it gives the timing uncertainty of a predicted epoch $t_\star$,
	\begin{equation}
		\delta t_u = \frac{\sigma_u(t)}{|\dot u(t)|},
		\label{eq:delta_t_u}
	\end{equation}
	which is the precision with which an observation can be scheduled at $u_\star$.
	
\begin{figure*}[t]
	\sidecaption
	\includegraphics[width=12cm]{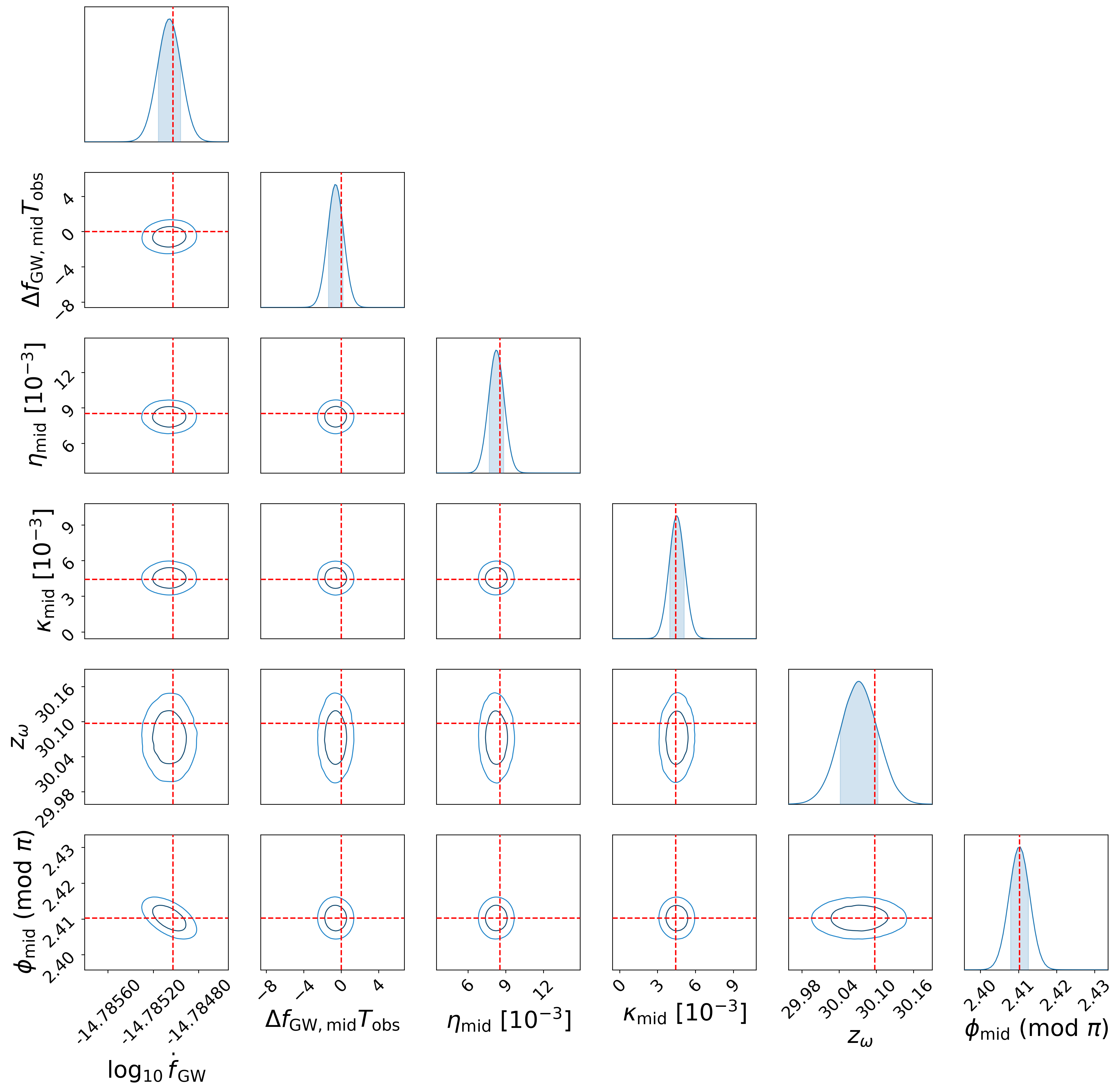}
		\caption{
			Corner plot of the MCMC posterior transformed to the midpoint reference parameters used for the phase forecast. The sampled chain used
			$t_{\rm 0}=0$, and the plotted quantities were obtained by transforming $\phi_0$, $f_{\rm GW,0}$, and the LL eccentricity pair $(\eta_0,\kappa_0)$ to $t_{\rm mid}=T_{\rm obs}/2$. The parameters shown are $\log_{10}\dot f_{\rm GW}$, $\Delta f_{\rm GW,mid}T_{\rm obs}$, $\eta_{\rm mid}$, $\kappa_{\rm mid}$, $z_\omega$, and $\phi_{\rm mid}$. Dashed lines mark the injected values. The midpoint reference acts as a pivot for the GW phase and for the apsidally rotating $(\eta,\kappa)$, reducing the off diagonal covariance terms between the phase forecast parameters that appear at the start epoch.}
		\label{fig:corner_ecc_midpoint}
	\end{figure*}

	The phase $\Phi_{\rm asc}$ advances at $\bar n_b$, while the LL pair rotates
	rigidly at $\dot\omega$, Eq.~\eqref{eq:kappa_lisa_prediction}.
	To determine the phase-coherence horizon, we first referenced the timing parameters to the midpoint of the LISA observation.\footnote{For the LISA analysis, one should choose the observation midpoint as the waveform reference epoch from the outset. In the present implementation, the sampled chain used $t_0=0$, so we transform $\phi_0$, $f_{\rm GW,0}$, and the eccentricity vector $(\kappa_0,\eta_0)$ to the midpoint $t_{\rm mid}$ in post-processing.} This is a choice of reference epoch, not a change in the physical signal. When the phase model is referenced to the beginning of the observation, inferred $f_{\rm GW,0}$ and $\dot f_{\rm GW}$ show a strong covariance because a small change in $\dot f_{\rm GW}$ is partially absorbed by a change in the initial frequency over the finite observing baseline. This parameter covariance is visible in Fig.~\ref{fig:corner_ecc}. Referencing the frequency at the midpoint,
	$$
	f_{\rm GW,mid}
	=
	f_{\rm GW,0}
	+
	\dot f_{\rm GW}\frac{T_{\rm obs}}{2},
	$$
	removes this artificial $f_{\rm GW,0}$-$\dot f_{\rm GW}$ correlation, reducing the covariance between the two parameters when propagating the phase to future observing epochs. Rotating $\kappa$ and $\eta$ to the same epoch likewise removes their correlations with $z_\omega$. The resulting posterior referenced to the observation midpoint time is shown in Fig.~\ref{fig:corner_ecc_midpoint}. The frequency offset is depicted in units of the Fourier resolution
	$1/T_{\rm obs}$, where the horizontal scale is expressed as a
	fraction of a frequency bin, so it does not depend on the assumed
	observation baseline. In these units, $2\pi\Delta f_{\rm GW}
	T_{\rm obs}$ is the phase error accumulated across the observation
	from the frequency uncertainty. The reference phase $\phi(t_{\rm mid})=\phi_{\rm mid}$ is plotted
	modulo $\pi$ because the waveform is invariant under
	$\psi\rightarrow\psi+\pi/2$ together with
	$\phi_{\rm ref}\rightarrow\phi_{\rm ref}+\pi$, a transformation that
	leaves $\kappa$ and $\eta$ unchanged. The two modes carry the same physical solution, and folding them onto
	the interval of width $\pi$ centred on the injected value gives the
	width used in the phase forecast.
	
	For the phase-coherence forecast, we adopted the covariance matrix of the parameters in Eq.~\eqref{eq:phase_forecast_coords}, evaluated at the midpoint reference epoch.
	Here $\phi_{\rm mid}$ and $f_{\rm GW,mid}$ are obtained by shifting the quadratic phase model to $T_{\rm obs}/2$, and $(\eta_{\rm mid},\kappa_{\rm mid})$ are obtained by rotating the LL parameter pair by the corresponding apsidal advance. We also converted the sampled variables $\log_{10}\dot f_{\rm GW}$ and $z_\omega$ to physical units before constructing the covariance matrix. Writing $\tau = t - t_{\rm mid}$ and defining $n_{\rm GW} = 2\pi f_{\rm GW} = 2\bar n_b$, with midpoint value $n_{\rm GW,mid}$, and $\dot n_{\rm GW} = 2\pi \dot f_{\rm GW}$, we propagated uncertainties using the GW phase model
	\begin{equation}
		\Phi_{\rm GW}(\tau) = \phi_{\rm mid} + n_{\rm GW,mid}~\tau + \tfrac12 \dot n_{\rm GW}\tau^2.
		\label{eq:GW_phase_t_mid}
	\end{equation}
	In the targeted MCMC, $f_{\rm GW}$ was sampled, so $(\phi_{\rm mid}, n_{\rm GW,mid}, \dot n_{\rm GW})$ all carry posterior variance. We label their covariance elements by the indices $\phi$, $n$, and $\dot n$, giving $C_{\phi\phi}$, $C_{\phi n}$, $C_{\phi\dot n}$, $C_{nn}$, $C_{n\dot n}$, and $C_{\dot n\dot n}$. The argument of latitude Eq.~\eqref{eq:arg_latitude_main} acquires a further uncertainty from the eccentricity correction Eq.~\eqref{eq:deltae_main}. 
	To expose the different time evolution of the GW phase and eccentricity contributions, we first considered them separately, neglecting covariances between the two parameter groups. In this decomposition,
	\begin{equation}
		\sigma_u^{2}(\tau)
		\simeq
		\tfrac{1}{4}\sigma_{\Phi_{\rm GW}}^{2}(\tau)
		+
		\sigma_{u,e}^{2}(\tau).
		\label{eq:sigma_u_from_PhiGW}
	\end{equation}
	The scheduling horizons themselves were computed by propagating the full posterior covariance of Eq.~\eqref{eq:phase_forecast_coords} through $u(t)$.
	The first term is then
	\begin{equation}
		\begin{aligned}
			\tfrac{1}{4}\sigma_{\Phi_{\rm GW}}^{2}(\tau)
			{} & = \\ & \tfrac{1}{4}\!\left[
			C_{\phi\phi}
			+ 2 C_{\phi n}\tau
			+ C_{nn}\tau^{2}
			+ C_{\phi\dot n}\tau^{2}
			+ C_{n\dot n}\tau^{3}
			+ \tfrac{1}{4}C_{\dot n\dot n}\tau^{4}
			\right],
			\label{eq:sigma_PhiGW_quartic}
		\end{aligned}
	\end{equation}
	and the second term arises from propagating the posterior uncertainties
	in $(\eta_{\rm mid},\kappa_{\rm mid},\dot\omega)$ through the first-order
	expression for the argument of latitude in
	Eq.~\eqref{eq:deltae_main}. The two grow differently. The carrier
	variance is quartic in $\tau$ through $C_{\dot n\dot n}$, so
	$\sigma_{\Phi_{\rm GW}}/2$ scales as $\tau^{2}$ at long baselines. In $\sigma_{u,e}$, the only secular time dependence comes from $\omega(\tau)=\omega_{\rm mid}+\dot\omega\tau$, so $\delta\dot\omega$ contributes linearly in $\tau$ and the envelope of $\sigma_{u,e}$ grows as $\tau$, with the term itself oscillating at $n_b$. For the source considered here, both contributions to $\sigma_u(\tau)$ are non-negligible across the relevant scheduling baselines and both must be retained; however, the carrier term dominates by the epochs at which $T_{\rm valid}$ is reached, so the modulation has little effect on the horizons.
	
	\subsection{Horizon as a constrained interval of a quartic}
	\label{sec:ecc_horizon_example}
	
	In the Appendix~Sect.~\ref{app:appendixC}, we derive upper limits for $\delta u$, such that the velocity error, $\delta {\rm v}$, caused by phase scheduling, Eq.~\eqref{ap:delta_v}, stays below the RV uncertainty of the chosen observable, $\sigma_\Delta$ (Eq.~\eqref{eq:sigma_delta_general}). We define $T_{\rm valid}$ as the latest time at which the phase uncertainty
	$\sigma_u(t)$ remains within the $\delta u$ tolerance,
	\begin{equation}
		T_{\rm valid} = \max\left\{~t \ge 0 \mid \sigma_u(t) \le \alpha~\delta u_{\rm tol}~\right\},
		\label{eq:T_valid_alpha}
	\end{equation}
	where $\alpha\leq1$ (Eq.~\eqref{eq:horizon_condition}) scales the tolerance. For $\alpha<1$, the velocity error from phase scheduling remains below $\sigma_\Delta$ up to $T_{\rm valid}$ even if the true phase error exceeds the forecast $\sigma_u(t)$ by up to a factor $1/\alpha$.
	
	If no $t \ge 0$ satisfies the condition in Eq.~\eqref{eq:T_valid_alpha}, we set $T_{\rm valid} = 0$. Because $\sigma_u(t)$ oscillates around the tolerance once $\dot\omega$ is included, we define $T_{\rm valid}$ as the closing time of the first admissible window after $T_{\rm obs}$.
	
	The multi-year phase forecast used the posterior covariance obtained from a dedicated simulated 4-year LISA observation of the source considered here, and is tied to the target source 4-year parameter uncertainties. The forecast is specified once the source posterior covariance, the adopted phase model, the RV semi-amplitude $K$ (derived from Eqs.~\eqref{eq:sb2_half_diff} or \eqref{eq:v1v2_ecc_rv}), and the spectroscopic precision $\sigma_{\rm v}$ are given.
	
	We adopted the conjunction and quadrature phase tolerances
	\begin{align}
		\delta u_{\mathrm{tol}}^{(\mathrm{conj})}
		&= \frac{\sigma_{\rm v}}{K},
		\label{eq:tol_conj_main}\\
		\delta u_{\mathrm{tol}}^{(\mathrm{quad})}
		&= \left(\frac{\sqrt{2}~\sigma_{\rm v}}{K}\right)^{1/2},
		\label{eq:tol_quad_conj_main}
	\end{align}
	where tolerances are derived for the single-epoch case and two-epoch velocity
	difference case in Appendix~Sect.~\ref{app:appendixC1}. Next, we present results for Eq.~\eqref{eq:T_valid_alpha} using $\alpha={0.5,1}$, where $\alpha=0.5$ imposes a conservative margin on the phase tolerance and $\alpha=1$ corresponds to the tolerance itself.
	
	We applied the phase-coherence forecast to construct a scheduling horizon. The source parameters adopted for the forecast are as follows:
	$M_1=0.347~M_\odot$,
	$M_2=0.285~M_\odot$,
	$P_{\rm b}=0.00275386$~d ($\simeq 3.97$~min),
	luminosity distance $D_L=0.577$~kpc, and inclination
	$\iota=1.4775$~rad ($\simeq84.7^\circ$).
	Apsidal advance carries the argument of latitude at
	$\bar n_{\rm b}=n_{\rm b}+\dot\omega$, so the injected carrier is
	$f_{\rm GW}=\bar n_{\rm b}/\pi=8.4061865\times10^{-3}$~Hz, exceeding
	$2/P_{\rm b}$ by $\dot\omega/\pi$. For the representative
	eccentricity used in the forecast, $e=0.0096$, the primary and secondary RV semi-amplitudes are $K_1\simeq585$~km~s$^{-1}$ and $K_2\simeq713$~km~s$^{-1}$. The
	targeted MCMC was performed at fixed sky location, using the posterior covariance from the simulated 4-year LISA
	observation as the input to the phase forecast. In our 4-year targeted MCMC analysis, the injected parameters, at the midpoint reference epoch, are given in Table~\ref{tab:table2},
	\begin{table}
		\centering
		\caption{Injected parameters adjusted to $t_{\rm mid}$.}
		\label{tab:table2}
		\begin{tabular}{lll}
			\toprule
			Parameter & Value & Unit \\
			\midrule
			$\phi_{\rm inj,mid}$    & $2.4102573$                    & rad          \\
			$f_{\rm GW,inj,mid}$    & $8.406289998784\times10^{-3}$  & Hz           \\
			$\dot f_{\rm GW,inj}$   & $1.6404897\times10^{-15}$      & Hz~s$^{-1}$  \\
			$\eta_{\rm inj,mid}$    & $8.5334116\times10^{-3}$       & \ldots       \\
			$\kappa_{\rm inj,mid}$  & $4.4340597\times10^{-3}$       & \ldots       \\
			$\dot\omega_{\rm inj}$  & $1.49809106\times10^{-6}$      & rad~s$^{-1}$ \\
			\bottomrule
		\end{tabular}
	\end{table}
	The LISA-inferred parameter values at the midpoint reference epoch are in Table~\ref{tab:table3}. 
	\begin{table}
		\caption{Parameter values inferred from LISA at the midpoint reference
			epoch $t_{\rm mid}=t_0+T_{\rm obs}/2$. Values are posterior medians, and
			uncertainties are standard deviations.}
		\label{tab:table3}
		\centering
		\begin{tabular}{lll}
			\toprule
			Parameter & Value & Unit \\
			\midrule
			$\phi_{\rm mid}$   & $2.410235 \pm 0.001978$                   & rad          \\
			$f_{\rm GW,mid}$   & $8.4062899940(55)\times10^{-3}$ & Hz     \\
			$\dot f_{\rm GW}$  & $(1.640367 \pm 0.000341)\times10^{-15}$    & Hz~s$^{-1}$  \\
			$\eta_{\rm mid}$   & $(8.249643 \pm 0.538244)\times10^{-3}$       & \ldots       \\
			$\kappa_{\rm mid}$ & $(4.524585 \pm 0.541031)\times10^{-3}$       & \ldots       \\
			$\dot\omega$       & $(1.496836 \pm 0.001505)\times10^{-6}$     & rad~s$^{-1}$ \\
			\bottomrule
		\end{tabular}
	\end{table}
	The GW frequency and orbital decay parameter are recovered with fractional
	precisions:
	\begin{equation}
		\frac{\sigma_{f_{\rm GW,mid}}}{f_{\rm GW,mid}}
		\simeq 6.6\times10^{-10},
		\qquad
		\frac{\sigma_{\dot f_{\rm GW}}}{|\dot f_{\rm GW}|}
		\simeq 2.1\times10^{-4}.
	\end{equation}
	
\begin{figure}[t]
	\resizebox{\hsize}{!}{\includegraphics{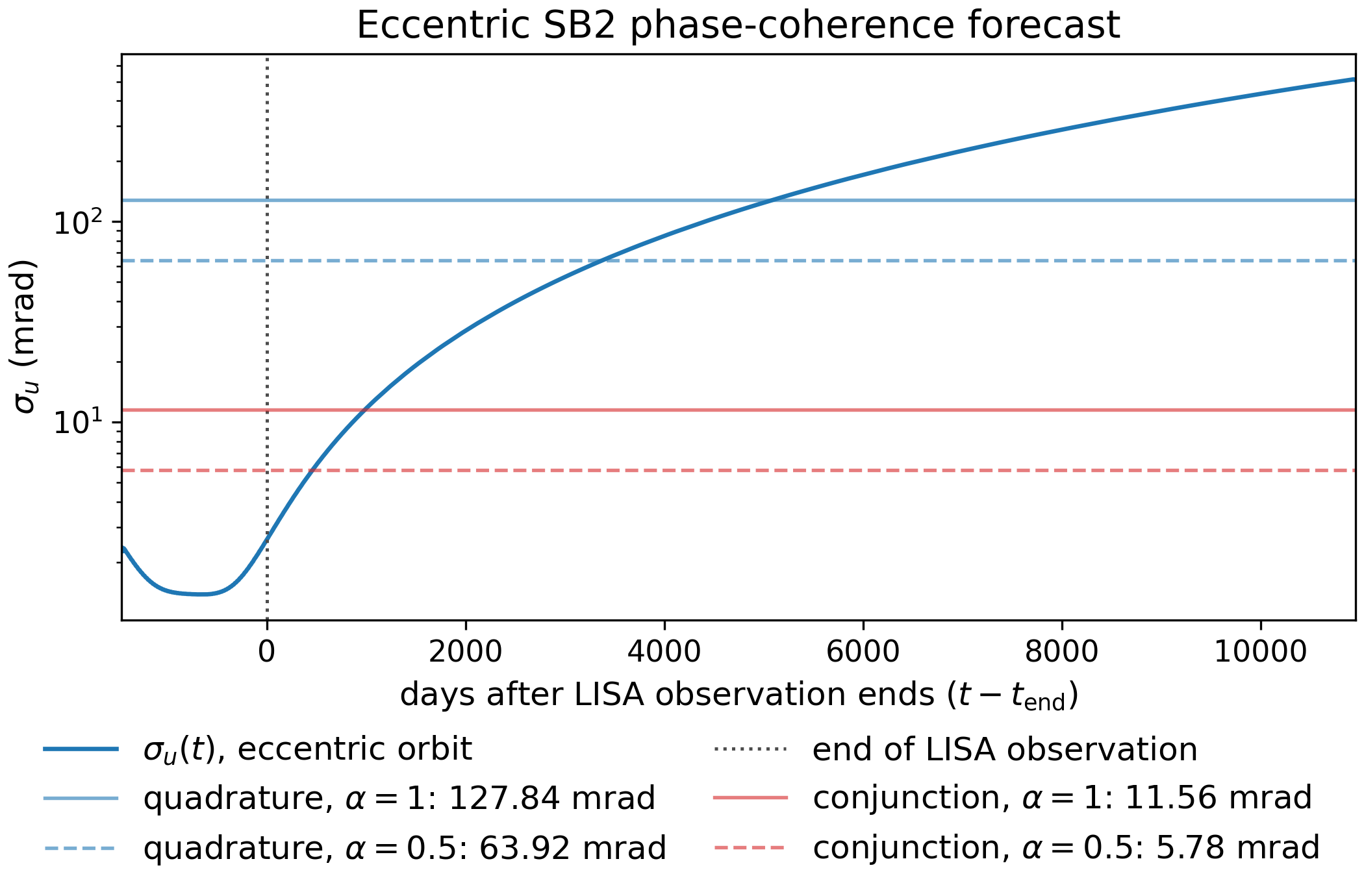}}
		\caption{Phase-coherence forecast from the MCMC posterior for the fiducial binary. The horizontal axis is measured relative to the end of the 4-year LISA observation, with the vertical dashed line marking the start of optical follow-up. The solid curve shows the conservative upper envelope of the quadrature $\sigma_u(t)$ forecast. Horizontal lines show the phase tolerances, $\alpha\delta u_{\rm tol}$. Conjunction and quadrature thresholds give the relevant criterion for the fixed-$u$ strategy.}
		\label{fig:horizon}
\end{figure}

	The resulting scheduling horizon is summarised in Table~\ref{tab:table4}, and the underlying phase uncertainty $\sigma_u(t)$ is shown in Fig.~\ref{fig:horizon}. The forecast gives admissible windows in both $\alpha\in\{0.5,1.0\}$ cases, starting at the end of the 4-year LISA observation.
	
	The GW phase timing uncertainty is
	\begin{equation}
		\delta t_{\Phi}(\tau)
		=
		\frac{\sigma_{\Phi}(\tau)}
		{2\pi f_{\rm GW}(\tau)},
		\qquad
		\tau=t-t_{\rm mid},
		\label{eq:timing_unc}
	\end{equation}
	where $t_{\rm mid}$ is the midpoint of the LISA observation.
	For the recovered midpoint-reference posterior, this gives
	\begin{equation}
		\delta t_{\Phi}(1~{\rm yr})\simeq0.04~{\rm s},
		\qquad
		\delta t_{\Phi}(4~{\rm yr})\simeq0.33~{\rm s}.
	\end{equation}
	The carrier-phase timing uncertainty remains below $0.4$~s at $\tau=4$~yr, two years after the end of the 4-year LISA observation. The timing uncertainties from the full forecast, Eq.~\eqref{eq:delta_t_u}, are given in Table~\ref{tab:table4}.
	
	\begin{table}[t]
	\centering
	\caption{Phase-coherence scheduling horizons for conjunction and quadrature
		anchors for the representative source with $P_b\simeq3.97$~min.}
	\label{tab:table4}
	\small
	\renewcommand{\arraystretch}{1.15}
	\setlength{\tabcolsep}{6pt}
		\begin{tabular}{@{}c c c c c c@{}}
			\hline
			LISA Obs. &
			anchor &
			$\alpha$ &
			$\alpha\delta u_{\rm tol}$ &
			$T_{\rm valid}$ &
			$\delta t_u(T_{\rm valid})$ \\
			(yr) & & & (mrad) & (days) & (s) \\
			\hline
			4 & conjunction & 0.5 & $5.78$   & $390$  & $0.21$ \\
			4 & conjunction & 1.0 & $11.56$  & $920$  & $0.43$ \\
			4 & quadrature  & 0.5 & $63.92$  & $3194$ & $2.38$ \\
			4 & quadrature  & 1.0 & $127.84$ & $4844$ & $4.76$ \\
			\hline
		\end{tabular}
		\tablefoot{For the SB2 velocity-difference observable, we used
		$K_{\rm eff}=K_1+K_2=1298~{\rm km~s^{-1}}$ and adopted
		$\sigma_{\rm v}=\sigma_{\rm spec}=15~{\rm km~s^{-1}}$ for each velocity measurement, corresponding to one spectrum per epoch.
		We propagated the covariance of the
		midpoint-reference parameters
		$(\phi_{\rm mid},f_{\rm GW,mid},\dot f_{\rm GW},
		\eta_{\rm mid},\kappa_{\rm mid},\dot\omega)$ through the
		first-order derivatives of the predicted scheduling phase $u(t)$,
		including apsidal motion.
		The admissible condition is
		$\sigma_u(t)\leq\alpha\delta u_{\rm tol}$, with
		$\alpha\in\{0.5,1.0\}$.
		The tolerances are
		$\delta u_{\rm tol}^{(\mathrm{conj})}=0.01156$~rad and
		$\delta u_{\rm tol}^{(\mathrm{quad})}=0.12784$~rad.
		We define $T_{\rm valid}$ as the closing time of the admissible interval
		beginning at the end of the LISA observation.
		The corresponding timing uncertainty is
		$\delta t_u=\sigma_u/|\dot u|$, evaluated at $T_{\rm valid}$.}
		\end{table}
	
For sources with an EM ephemeris established before the start of the LISA observation, the resulting optical baseline improves the phase forecast used for targeted multimessenger spectroscopy. At long extrapolation times, $T_{\rm valid}$ is limited by the uncertainty in the quadratic phase term, $\sigma_{\dot f_{\rm GW}}$. For sources at low frequency, an independent optical $\dot P_b$ prior from RV phase tracking constrains the same secular phase evolution, since $\dot f_{\rm GW}=-2\dot P_b/P_b^2$, extending the targeted-$u$ window. The early optical baselines emphasised by \citet{2020ApJ...892L..35B} contribute to the targeted-$u$ method by tightening the prior used to
propagate the phase forward, rather than by improving the semi-amplitude
measurements, since $K_1$, $K_2$, and $q$ are obtained within a single
session. 

A tightened prior extends the interval over which $\sigma_u(t)$ remains below the required tolerance, increasing the allowable separation $\Delta T=t_\beta-t_\alpha$ between two sessions at a common $u$. Such a pair measures $\Delta{\rm v}_{\rm CM}$ and $\Delta\kappa$ together, Eq.~\eqref{eq:dv_ecc}. The longer baseline increases the accumulated signal from $\Delta{\rm v}_{\rm CM}$, which grows secularly with $\Delta T$ (Sect.~\ref{sec:DeltaVCM}), and sharpens the check on $\dot\omega$, since $\partial\Delta\kappa/\partial\dot\omega=-\Delta T~\eta(t_\beta)$ (Sect.~\ref{sec:delta_kappa_consistency}).

\subsection{Circular orbits: Estimating a time horizon for HM~Cancri}
\label{sec:cir_horizon_example}

The semi-detached system HM~Cnc \citep{2002MNRAS.332L...7R,2002A&A...386L..13I,2010ApJ...711L.138R} is the shortest period dWD known and a guaranteed LISA source \citep{2010ApJ...711L.138R,2023MNRAS.524.5442S,2023MNRAS.518.5123M,2023PhRvD.108b3019S,2024ApJ...977..262C,2026PhRvD.113f3009S}. Two decades of optical timing were needed to trace HM~Cnc's orbital
phase to second order in frequency and establish its evolutionary state
\citep{2023MNRAS.518.5123M}. LISA will detect many more binaries like
it, and here we show that four years of LISA observation give an
ephemeris that stays phase coherent for years afterward
(Table~\ref{tab:table7}), enough to schedule spectroscopy at
a chosen orbital phase without an equivalent ground campaign.

LISA data for an HM~Cnc-like system were simulated with component masses
$M_1=1.0~M_\odot$ and $M_2=0.17~M_\odot$, orbital period
$P_b\simeq5.36$~min ($f_{\rm GW,0}=6.2204944297$~mHz), distance $d=2.0$~kpc, and
Galactic coordinates $(l,b)=(206.92^\circ,23.40^\circ)$. The frequency
derivative was set to the measured value
$\dot f_{\rm GW}=7.076\times10^{-16}~{\rm Hz~s^{-1}}$, twice the orbital
rate obtained from two decades of optical timing
\citep{2023MNRAS.518.5123M}, rather than to the chirp expected from gravitational radiation reaction alone.

The forecast also carries the second frequency derivative
$\ddot f_{\rm GW}$, which the 4-year LISA analysis does not
constrain. Its cubic phase contribution reaches the recovered phase
uncertainty only for
$|\ddot f_{\rm GW}|\gtrsim7\times10^{-26}~{\rm Hz~s^{-2}}$, about seven times the measured optical magnitude, so the marginal posterior returns the prior. We adopted the optical constraint
$\ddot f_{\rm GW}= -1.076\times10^{-26}~{\rm Hz~s^{-2}}$ with
$\sigma_{\ddot f_{\rm GW}}=4.2\times10^{-27}~{\rm Hz~s^{-2}}$
from \citet{2023MNRAS.518.5123M} and propagated it together with
$(\phi_{\rm mid},f_{\rm GW,mid},\dot f_{\rm GW})$.

In this system the donor fills its
Roche lobe and transfers mass by direct impact, which opposes the orbital decay
driven by GW emission. No inclination is measured for HM~Cnc. Spheres of radii $R_1$ and $R_2$
separated by $a$ eclipse when $\cos\iota<(R_1+R_2)/a$, and with radii
from \citet{2023MNRAS.518.5123M} this limit falls at $\iota\simeq70^\circ$.
For this non-eclipsing system, we adopted the median of an isotropic prior over
the allowed range, $\cos\iota=0.67$ ($\iota\simeq48^\circ$).
The injected GW parameters are shown in Table~\ref{tab:table5} and the LISA-inferred parameter values at the midpoint reference epoch are
given in Table~\ref{tab:table6}.

The scheduling tolerance is estimated through the ratio
$\sigma_{\rm v}/K_{\rm eff}$, where $K_{\rm eff}$ is the velocity
semi-amplitude of the spectral feature tracked.
\citet{2010ApJ...711L.138R} measured $K_{\rm eff}=390\pm40$~km~s$^{-1}$
in He~I 4471, the only semi-amplitude reported for this system. For the
masses and inclination adopted above, the component values are
$K_1\simeq157$~km~s$^{-1}$ and $K_2\simeq924$~km~s$^{-1}$, so the measured
feature arises in the accretion flow rather than in either photosphere.
\citet{2023MNRAS.518.5123M} reach the same conclusion and caution that
the helium line emission may not track the stars. We used $K_{\rm eff}$ only to convert a scheduling phase error
into a velocity error, not as a component semi-amplitude.

\begin{table}
	\centering
	\caption{Injected HM~Cnc parameters referenced to the midpoint of the
		LISA observation, $t_{\rm mid}$.}
	\label{tab:table5}
	\begin{tabular}{lll}
		\toprule
		Parameter & Value & Unit \\
		\midrule
		$\phi_{\rm inj,mid}$
		& $5.5107189$
		& rad \\
		$f_{\rm GW,inj,mid}$
		& $6.220494429661\times10^{-3}$
		& Hz \\
		$\dot f_{\rm GW,inj,mid}$
		& $7.0692087\times10^{-16}$
		& Hz~s$^{-1}$ \\
		$\ddot f_{\rm GW,inj}$
		& $-1.0760000\times10^{-26}$
		& Hz~s$^{-2}$ \\
		\bottomrule
	\end{tabular}
\end{table}

\begin{table}
	\centering
	\caption{HM~Cnc parameter values inferred from LISA at the midpoint
		reference epoch.}
	\label{tab:table6}
	\begin{tabular}{lll}
		\toprule
		Parameter & Value & Unit \\
		\midrule
		$\phi_{\rm mid}$
		& $5.500397 \pm 0.022993$
		& rad \\
		$f_{\rm GW,mid}$
		& $6.2204944286(77)\times10^{-3}$
		& Hz \\
		$\dot f_{\rm GW,mid}$
		& $(7.073958 \pm 0.004662)\times10^{-16}$
		& Hz~s$^{-1}$ \\
		$\ddot f_{\rm GW}$
		& $(-1.065650 \pm 0.367104)\times10^{-26}$
		& Hz~s$^{-2}$ \\
		\bottomrule
	\end{tabular}
\end{table}

The LISA GW parameter posteriors give conjunction scheduling horizons of
$T_{\rm valid}\simeq7$ and $10$~years, and quadrature horizons of
$12$ and $16$~years, after the completion of the 4-year LISA observation
for $\alpha=0.5$ and $1.0$ respectively
(Table~\ref{tab:table7}). The GW-derived ephemeris remains
usable for two to four times the length of the observation that produced
it, depending on the anchor and $\alpha$ combination used. In timing terms the orbital phase is known to $\sim0.6$~s in a
$321.5$~s period at the end of the observation, degrading to between 3.7 and 23.2~s at $T_{\rm valid}$, depending on the anchor and $\alpha$ (Table~\ref{tab:table7}). Spectroscopic follow up need not be
contemporaneous with the mission, nor prompt after it. A campaign
scheduled a decade later is still
anchored to the correct orbital phase.

At the end of the LISA observation, the phase uncertainty is $\sigma_u(t_{\rm end})\lesssim 12$~mrad. This already satisfies $\sigma_u(t)\leq\alpha~\delta u_{\rm tol}$ for every anchor and every $\alpha$ in Table~\ref{tab:table7}, so each window is open at the end of the LISA observation and $T_{\rm valid}$ measures how long the condition continues to hold. The conjunction horizons are shorter because the leading targeted-$u$ velocity error from phase scheduling is linear in phase mismatch near
conjunction, whereas near quadrature it begins at second order, so the same velocity budget
results in a tolerance larger by a factor $(\sqrt{2}K/\sigma_{\rm v})^{1/2}\simeq 3.1$, Eqs.~\eqref{eq:tol_conj_main} and~\eqref{eq:tol_quad_conj_main}.

The scheduling phase $u_\star$ depends on $\ddot f_{\rm GW}$ through a term
proportional to $\tau^{3}$. A 4-year LISA observation does not constrain
this parameter \citep{2023PhRvD.108b3019S}, and our posterior narrows the
optical prior by less than 1\%, so $\sigma_{\ddot f_{\rm GW}}$ is set
by the EM measurement. The resulting horizon scales as
$\sigma_{\ddot f_{\rm GW}}^{-1/3}$, and the horizons of Table~\ref{tab:table7} are $12$\% to $28$\% shorter than a
forecast that neglects $\ddot f_{\rm GW}$.
In HM~Cnc the frequency evolution is not purely due to GW emission. Mass transfer contributes to the
frequency evolution, $\ddot f_{\rm GW}$ becomes a free parameter, and
the optically determined value is negative. The cubic term is negligible over the 4-year observation, but it
grows as $\tau^{3}$ while the contributions from $\phi_{\rm mid}$,
$f_{\rm GW,mid}$, and $\dot f_{\rm GW,mid}$ grow no faster than $\tau^{2}$, so it dominates the phase
uncertainty on the baselines that matter for multimessenger scheduling.

\begin{table}[t]
	\centering
	\caption{Scheduling horizons over which phase coherence is retained for the HM~Cnc-like system of
		Sect.~\ref{sec:cir_horizon_example}, with $P_b\simeq5.36$~min, for the
		conjunction and quadrature anchors.}
	\label{tab:table7}
	\small
	\renewcommand{\arraystretch}{1.15}
	\setlength{\tabcolsep}{6pt}
	\begin{tabular}{@{}c c c c c c@{}}
		\hline
		LISA Obs. &
		anchor &
		$\alpha$ &
		$\alpha\delta u_{\rm tol}$ &
		$T_{\rm valid}$ &
		$\delta t_u(T_{\rm valid})$ \\
		(yr) & & & (mrad) & (days) & (s) \\
		\hline
		4 & conjunction & 0.5 & $73$  & $2554$ & $3.7$  \\
		4 & conjunction & 1.0 & $145$ & $3646$ & $7.4$  \\
		4 & quadrature  & 0.5 & $227$ & $4481$ & $11.6$ \\
		4 & quadrature  & 1.0 & $453$ & $6036$ & $23.2$ \\
		\hline
	\end{tabular}
	\tablefoot{With $K_{\rm eff}=390~{\rm km~s^{-1}}$ (see text), a precision per epoch
	$\sigma_{\rm v}=\sqrt{2}\sigma_K=56.6~{\rm km~s^{-1}}$, and
	$\sigma_\Delta=\sqrt{2}~\sigma_{\rm v}=80.0~{\rm km~s^{-1}}$, the tolerances are
	$\delta u_{\rm tol}^{(\mathrm{conj})}=\sigma_{\rm v}/K_{\rm eff}=0.145$~rad and $\delta u_{\rm tol}^{(\mathrm{quad})}=(\sigma_\Delta/K_{\rm eff})^{1/2}=0.453$~rad.
	The forecast propagated the midpoint reference covariance in
	$(\phi_{\rm mid},f_{\rm GW,mid},\dot f_{\rm GW},\ddot f_{\rm GW})$
	through the first-order derivatives of the predicted scheduling phase
	$u(t)$. The second frequency derivative is not constrained by the LISA data
	and carries the optical prior of
	Table~\ref{tab:table6}.
	The admissible condition is
	$\sigma_u(t)\leq\alpha~\delta u_{\rm tol}$, with
	$\alpha\in\{0.5,1.0\}$.
	The phase uncertainty at the end of the LISA observation is
	$\sigma_u=11.76$~mrad, below every threshold, corresponding to a scheduling
	precision of $0.60$~s.
	The value of $T_{\rm valid}$ is measured from the end of the LISA observation, and the
	timing uncertainty $\delta t_u=\sigma_u/\bar n_b$ is evaluated at
	$T_{\rm valid}$.}
	\end{table}

\subsection{Summary of forecasts}
\label{subsec:horizon_summary}

In this section, we used two representative binary systems, one circular and one eccentric, to quantify how long the ephemeris derived from LISA remains accurate enough to guide spectroscopic observations. The precision of the velocity at a single epoch $\sigma_{\rm v}$, the semi-amplitude, and the argument of latitude determine the phase tolerance $\delta u_{\rm tol}$, against which we compare the uncertainty in the argument of latitude, $\sigma_u$. A phase-connected solution at a later date requires $\sigma_u < \delta u_{\rm tol}$. The phase uncertainty $\sigma_u(t)$ grows with time and sets a `forecast horizon' $T_{\rm valid}$ in Eq.~\eqref{eq:T_valid_alpha}, defined as the time after the end of the LISA observation at which phase connection is assumed to be lost.

The effect of imperfect phase targeting on the measured velocity depends strongly on the chosen anchor $u_\star$, so we evaluate conjunction and quadrature as the limiting cases. At conjunction, the velocity derivative with respect to $u$ is maximal, so the phase targeting error contributes linearly and produces the most restrictive tolerance. At quadrature, the linear term vanishes and the leading error is quadratic, giving the least restrictive tolerance; generic anchors,~\eqref{ap:delu_gen}, lie between these limits. The conjunction and quadrature tolerances are given by Eqs.~\eqref{eq:tol_conj_main} and \eqref{eq:tol_quad_conj_main}. For the sources considered here, the quadrature tolerance is less restrictive by approximately a factor of 4 up to an order of magnitude. Consequently, the phase prediction remains valid for a few years at conjunction and for more than a decade at quadrature (Table~\ref{tab:table4}). We also find multi-year quadrature horizons for a system with the parameters of HM~Cnc (Table~\ref{tab:table7}). In timing terms, the GW ephemeris gives $\delta t_u\simeq0.3$~s for the eccentric source and $0.9$~s for the HM~Cnc-like system four years after the observation midpoint, against the $1-30$~s range found for optical ephemerides propagated over the same baseline (Sect.~\ref{subsec:magnitude_summary}). Neither exceeds the optical figures out to $T_{\rm valid}$ (Tables~\ref{tab:table4} and~\ref{tab:table7}).

The same phase tolerances at a single epoch apply to the individual velocities included in the ${\rm v}$--$\cos u$ and ${\rm v}$--$F$ regressions for circular and eccentric orbits, respectively. The two regressions differ in their phase uncertainty budgets, however. For eccentric orbits,
uncertainties in $(\eta_{\rm mid},\kappa_{\rm mid},\dot\omega)$ contribute to the
predicted uncertainty in $u(t)$ through Eq.~\eqref{eq:sigma_u_from_PhiGW}. This uncertainty from the inference of the eccentricity parameters is, of course, absent in the circular case.

\section{Mass--mass diagram}
\label{sec:mass_diagram_illustration}

The phase-horizon analysis establishes how long after the LISA
observation the ephemeris remains accurate enough to schedule
spectroscopic observations at a targeted orbital phase. We now show how the
velocities obtained from these observations constrain the component masses of an
eccentric SB2 binary.

\begin{figure}[t]
	\resizebox{\hsize}{!}{\includegraphics{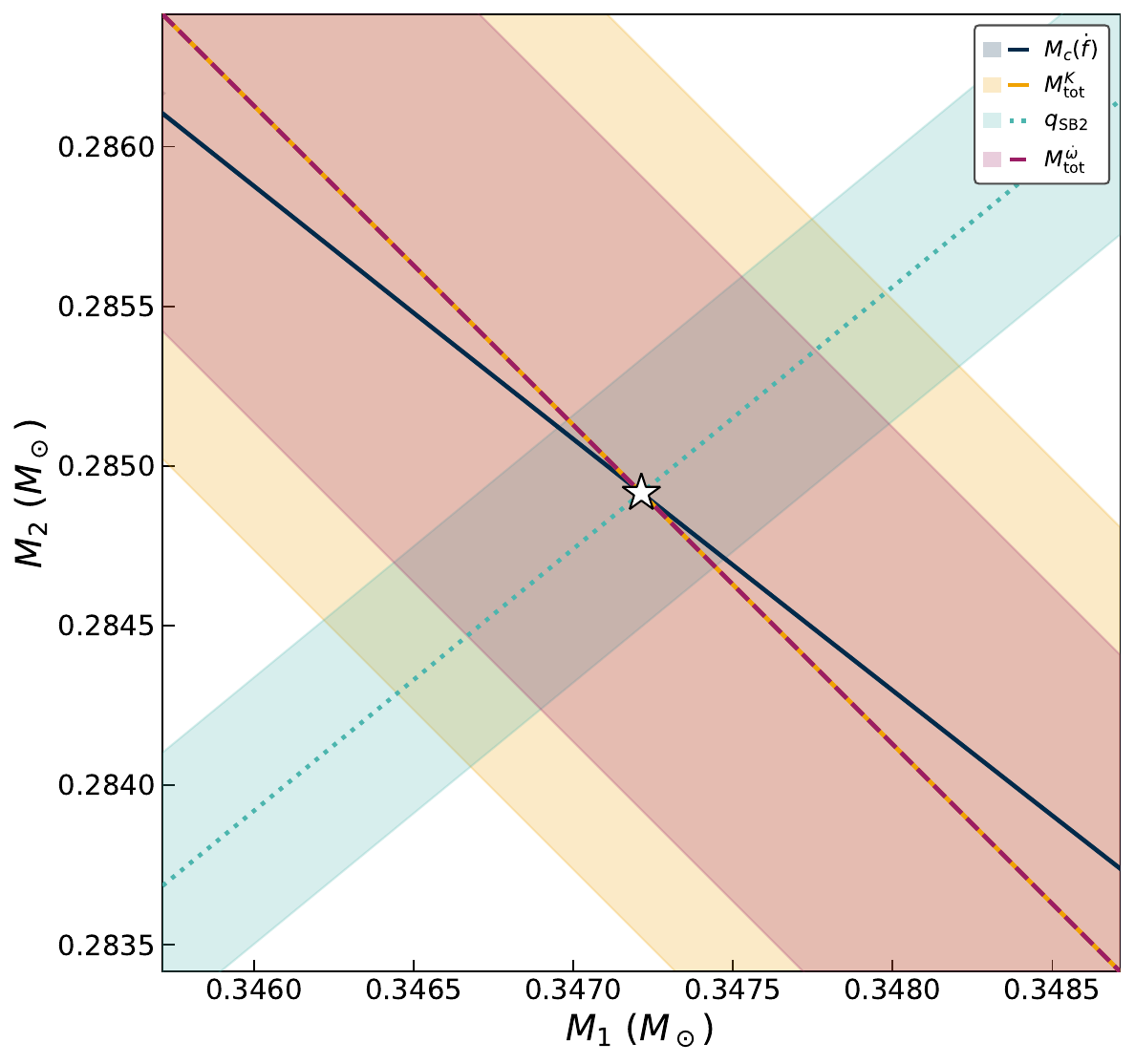}}
	\caption[Mock mass diagram for the simulated fiducial source]{Mock mass
		diagram for the simulated fiducial source in the $(M_1,M_2)$ plane, with
		$M_1=0.347~M_\odot$, $M_2=0.285~M_\odot$, $P_b=3.97~{\rm min}$, and
		$\iota=84.7^\circ$. The $M_{\rm tot}^{K}$ and $q_{\rm SB2}$ loci are
		distinct combinations of the same targeted-$u$ measurements, obtained within one observing session.
		In this mock presentation, the semi-amplitudes are centred on their injected
		values. Assuming an uncertainty of $\sigma_{\rm spec}=15~{\rm km~s^{-1}}$ per spectrum,
		480 spectra distributed in the orbital phase give
		$\sigma_{K_j}=\sigma_{\rm spec}\sqrt{2/N}\simeq0.97~{\rm km~s^{-1}}$
		for each semi-amplitude. The LISA posterior supplies
		$f_{\rm GW}$, $\cos\iota$, and the eccentricity components for each
		sample; the injection has $\eta_0=\kappa_0=0.0068$ at mission start, giving
		$e=0.0096$ and the independent variable $F=\cos u+\kappa(t)$. The intersection of
		the $M_{\rm tot}^{K}$ and $q_{\rm SB2}$ regions gives the targeted-$u$
		component masses. The $M_c$ locus assumes that $\dot f_{\rm GW}$ is
		driven by gravitational radiation alone, while
		$M_{\rm tot}^{\dot\omega}$ assumes $\dot\omega=\dot\omega_{\rm GR}$.
		The shaded regions show the corresponding $1\sigma$ uncertainty bands,
		and the centre star marks the injected component masses. All loci are
		centred on the injected solution; their widths reflect the adopted
		measurement uncertainties.}
	\label{fig:mass_mass_diagram}
\end{figure}

Figure~\ref{fig:mass_mass_diagram} combines two fixed-$u$ constraints derived from the relations in Sect.~\ref{sec:caseSB2} with two
constraints obtained from the LISA measurement of the secular terms $\dot f_{\rm GW}$ and $\dot\omega$. All four loci describe the same binary, but they use
different information. The fixed-$u$ measurements provide two loci; a mass-ratio constraint from $K_1/K_2$ and a total-mass constraint from $K_1+K_2$. Both are constructed from the orbital velocity amplitudes measured within one observing session, with the total-mass locus additionally using the inclination, period, and eccentricity recovered by LISA. The remaining pair of loci use the measured secular evolution, $\dot f_{\rm GW}$ and $\dot\omega$, whose conversion into mass constraints requires assumptions about the physical processes contributing to each. The
figure is illustrative rather than a measurement. Every locus is centred on the injected masses of the fiducial source, so that only the widths differ. For the two spectroscopic loci, we assumed eight exposures per orbit, $T_{\rm exp}=P_b/8\simeq30$~s, over a block of sixty orbits, giving $N=480$ spectra in just under $4$~h. With $\sigma_{\rm spec}=15~{\rm km~s^{-1}}$ per spectrum and the orbital phase of each spectrum supplied by the LISA ephemeris, the ${\rm v}$--$F$
regression returns $\sigma_{K_j}=\sigma_{\rm spec}\sqrt{2/N}\simeq0.97~{\rm km~s^{-1}}$, Appendix~Sect.~\ref{app:appendixC}. At this exposure the smearing factor
is $\Lambda_K={\rm sinc}(\pi/8)=0.974$, Eq.~\eqref{eq:Lambda_K}, and is
divided out of the recovered semi-amplitudes. This precision is
propagated with the LISA posterior samples of $\cos\iota$, $P_b$, and $e$
through Eqs.~\eqref{eq:semimajor_axis_j}--\eqref{eq:M1M2_cir_and_ecc}. The two GW
loci use the posterior samples of $\dot f_{\rm GW}$ and $\dot\omega$.

The four loci are detailed in the following list:
\newline
(1)~The SB2 mass ratio,
$q_{\rm SB2}=K_1/K_2$, from the semi-amplitudes measured by the
${\rm v}$--$F$ fits, Eq.~\eqref{eq:v1v2_ecc_rv}. It defines the straight
line $M_2=q_{\rm SB2}M_1$ through the origin. The ratio is independent of
the inclination and orbital period, and any effect that scales both
semi-amplitudes equally cancels. Its width comes from $\sigma_{K_j}$
alone.
\newline
(2)~The total mass $M_{\rm tot}^{K}$, obtained from
the sum of the same semi-amplitudes, $K_1+K_2$. The projected semi-major
axis follows from Eq.~\eqref{eq:semimajor_axis_j}, and the LISA inclination,
period, and eccentricity then give $M_{\rm tot}^{K}$ through Kepler's
third law, Eq.~\eqref{eq:Kepler_mtot}. This defines the straight line
$M_2=M_{\rm tot}^{K}-M_1$ and makes no assumptions about the physical origin
of $\dot f_{\rm GW}$ or $\dot\omega$. It is the widest of the four
because $M_{\rm tot}\propto(K_1+K_2)^3$ triples the fractional error on
the summed semi-amplitudes.
\newline
(3)~The chirp-mass curve,
\begin{equation}
	M_c=\frac{(M_1M_2)^{3/5}}{(M_1+M_2)^{1/5}},
\end{equation}
obtained from the recovered $f_{\rm GW}$ and $\dot f_{\rm GW}$ by
interpreting the frequency evolution as driven by gravitational radiation alone.
Its width follows from the recovered $\dot f_{\rm GW}$, with
$M_c\propto\dot f_{\rm GW}^{3/5}$. The locus requires $\dot f_{\rm GW}>0$ and
assumes that the measured frequency
evolution is dominated by gravitational radiation from a detached binary.
If tides, rotation, mass transfer, or other astrophysical effects
contribute appreciably to $\dot f_{\rm GW}$, the $M_c$ locus no longer
gives a reliable constraint on the component masses.
\newline
(4)~The apsidal mass $M_{\rm tot}^{\dot\omega}$, obtained from
the recovered $f_{\rm GW}$ and $\dot\omega_{\rm LISA}$ by attributing the
whole apsidal advance to $\dot\omega_{\rm GR}$, since this is how the waveform was modelled for simplicity. 
It defines another line of constant total mass and is recovered to approximately $0.16\%$, since
$M_{\rm tot}\propto\dot\omega^{3/2}$ and the apsidal shift is measured to
$\sigma_{z_\omega}=0.03$ Fourier bins. For UCBs the measured apsidal advance includes 
the point-mass relativistic contribution, together with tidal and rotational 
contributions \citep{2008PhRvL.100d1102W}. The tidal and
rotational terms of Eq.~\eqref{eq:omega_dot_tot} are summed over the two stars
\begin{equation}
	\dot\omega_{\rm LISA}
	=
	\dot\omega_{\rm GR}
	+
	\sum_i \dot\omega_{{\rm tide},i}
	+
	\sum_i \dot\omega_{{\rm rot},i}
	+\cdots .
	\label{eq:omega_dot_full}
\end{equation}
Tides and rotation can displace this locus from the
$M_{\rm tot}^{K}$ locus. For example, \citet{2012ApJ...745..137V} give the
Newtonian tidal and rotational contributions in terms of the WD structure
quantities $k_iR_i^5$. So, more generally, Eq.~\eqref{eq:omega_dot_full} provides
the relation required to evaluate a model-dependent contour in the
$(M_1,M_2)$ plane.

The two spectroscopic loci are distinct combinations of the same two semi-amplitudes and cross at a large angle, which determines the component masses. The component masses determined in Sect.~\ref{sec:caseSB2} do not
require the binary dynamics to be dominated by GW emission. For the
fiducial binary shown here, the fixed-$u$ velocity differences are only a
few km~s$^{-1}$ because $e\lesssim0.01$. The two GW loci are narrower but
depend on physical interpretations of $\dot f_{\rm GW}$ and
$\dot\omega$. Comparing $M_{\rm tot}^{\dot\omega}$ with $M_{\rm tot}^{K}$ isolates the non-relativistic contribution to the
apsidal advance. Here we injected only $\dot\omega_{\rm GR}$, so the two
coincide by construction.

\section{Conclusions}
\label{sec:conclusions}

For dWD UCBs, the orbital periods are short enough that a single observing session spanning a few hours captures many complete orbits. In favourable bright systems, sufficiently dense spectroscopy over a single observing sequence is sufficient to reconstruct the full RV curve without any GW information. The role of LISA is to supply a barycentric, phase-coherent orbital reference that ties sparse or multi-epoch spectra to known orbital phases. For bright LISA sources, it is possible to carry this reference across widely separated epochs, which turns a repeated phase-resolved spectroscopic campaign into a smaller set of targeted measurements.

The central role of the LISA-derived ephemeris is to supply the orbital coordinate that is normally missing from sparse spectroscopy. In an SB2, the WIL41 velocity--velocity relation determines the mass ratio without reference to orbital phase \citep{1941ApJ....93...29W}. Restoring that phase information changes what can be measured. For circular orbits, Eq.~\eqref{eq:v1v2_circular_rv} makes the radial velocity of each star a linear function of the known quantity $\cos u$, while for eccentric orbits, Eqs.~\eqref{eq:v1v2_ecc_rv} and \eqref{eq:F_definition} replace $\cos u$ by the LISA-predicted quantity $F=\cos u+\kappa$. The corresponding fits yield the individual semi-amplitudes $K_1$ and $K_2$, and hence both the mass ratio and the orbital velocity scale.
Together with the LISA inclination, orbital period, and eccentricity, these measurements give the total and component masses through Eqs.~\eqref{eq:semimajor_axis_j}--\eqref{eq:M1M2_cir_and_ecc}. The resulting mass determination does not require $\dot f_{\rm GW}$ to be driven by gravitational radiation, and it remains applicable to interacting systems, including semi-detached UCBs with $\dot f_{\rm GW}<0$.

The same ephemeris provides an independent long-baseline use of the spectroscopy. Measurements repeated at a fixed $u$ eliminate the leading phase-dependent orbital term. For an eccentric SB2, subtracting the velocity changes of the two stars gives the spectroscopic eccentricity evolution $\Delta\kappa_{\rm spec}$ through Eq.~\eqref{eq:delta_kappa_spec}, which can be compared with the LISA prediction $\Delta\kappa_{\rm LISA}$. The differenced velocity--velocity relation, Eq.~\eqref{eq:SB2_fixedu_diff}, separately provides a consistency check on the SB2 mass ratio and recovers the barycentric velocity change from its intercept. These inter-session observables test the LISA apsidal solution and search for barycentric acceleration rather than setting the primary mass scale.

The main multimessenger gain is that the component masses can be determined from the orbital velocity amplitudes. The SB2 semi-amplitudes, together with the orbital period, inclination, and eccentricity, define a solution in the $(M_1,M_2)$ plane without using $\dot f_{\rm GW}$ or $\dot\omega$ as mass constraints Sect.~\ref{sec:mass_diagram_illustration}. The measured $\dot f_{\rm GW}$ and $\dot\omega$ can then be used to test the physical processes driving the orbital evolution. For $\dot f_{\rm GW}$, the independently determined masses predict the contribution from gravitational radiation, allowing additional effects from tides, mass transfer, or acceleration of the barycentre to be constrained. The last of these is degenerate with $\dot f_{\rm GW}$ in the GW data when the acceleration is approximately constant, unless higher-order frequency evolution such as $\ddot f_{\rm GW}$ can be measured and modelled. Fixed-$u$ spectroscopy constrains it from $\Delta{\rm v}_{\rm CM}$, needing only that same near-constancy over the inter-epoch baseline (Sect.~\ref{sec:DeltaVCM}). For apsidal advance, the masses determine $\dot\omega_{\rm GR}$, while LISA measures the total $\dot\omega$. Their difference isolates the sum of the tidal and rotational contributions, which in dWDs may be dominated by tides and thus constrain the tidal response of the stars. In this way, the GW phase information does more than improve optical timing: It allows the masses to be determined without using $\dot f_{\rm GW}$ or $\dot\omega$ as mass constraints, leaving these measurements available as separate probes of the underlying astrophysics.

A practical advantage comes from anchoring the measurements at quadrature. At quadrature, the velocity error from a finite phase targeting offset appears only at the second order, rather than linearly. The allowed scheduling error is larger by a factor of the order of $(\sqrt{2}K/\sigma_{\rm v})^{1/2}$ compared with conjunction orbital phases. This quadrature anchoring is what makes the method practical over long baselines. For the bright systems considered here, it extends the LISA scheduling horizon to years. The same framework supplies the systemic velocity from which $\Delta{\rm v}_{\rm CM}$ is formed. For SB2 systems the component sum isolates the systemic velocity at each epoch. For SB1 systems, averaging the two RV zero crossings at each epoch achieves the same to quadratic accuracy in a common phase error (Sect.~\ref{sec:DeltaVCM}).

The phase-coherence horizon $T_{\rm valid}$ quantifies how far in time from the LISA reference epoch the predicted orbital phase remains accurate to within the spectroscopic tolerance. For the fiducial binary system with finite eccentricity, conjunction phase anchors yield $T_{\rm valid}\sim1$-$2.5$~years, and quadrature anchors yield $T_{\rm valid}\sim9$-$13$~years. For an HM Cnc-like binary system, conjunction phase anchors yield $T_{\rm valid}\sim7$-$10$~years, and quadrature anchors yield $T_{\rm valid}\sim12$-$16$~years. We conclude that multi-epoch campaigns are feasible for the bright, optically detected members of the dWD population. The range of $T_{\rm valid}$ is due to more and less conservative tolerance threshold factors of $\alpha={0.5,1}$, respectively.

The required precision of the GW phase forecast is modest by the standards of precision pulsar timing in radio. In the timing solution for PSR B1913+16 spanning 35 years, the reference epoch uncertainty is $3\times10^{-8}$ d ($\approx 2.6$ ms), while the binary period uncertainty of $3\times10^{-12}$ d corresponds to $\approx 0.26~\mu$s. Accumulated over the full data span, the period term contributes only $\sim 10$ ms. \citep{2016ApJ...829...55W} Likewise, for the double pulsar PSR~J0737$-$3039A, a 16-year timing solution gives $\sigma_{T_0}\approx 1.1\mu$s and $\sigma_{P_b}\approx 8.6\times10^{-9}$ s, implying $\sim 0.5$ ms accumulated uncertainty from the period term over the full span \citep{2021PhRvX..11d1050K}. In contrast, the per-epoch scheduling precision delivered here is $\delta t \approx 5$~s at the maximum $T_{\rm valid}$ for the source in Table~\ref{tab:table4}, set by the ratio between RV uncertainty and the UCB semi-amplitude through Eq.~\eqref{eq:tol_quad_conj_main}. The optical fixed-$u$ condition is one of seconds-level scheduling, not millisecond-level timing.

The $\sim9$ and $13$ year horizons in Table~\ref{tab:table4}, for $\alpha=0.5$ and 1, give the baseline over which the LISA ephemeris remains accurate enough for fixed-$u$ spectroscopy at quadrature. These horizons are set primarily by the uncertainty in the orbital period derivative. For $\alpha=1$, applying the same phase error criterion to binary pulsars gives formal horizons of $\sim5\times10^3$ yr for PSR~B1913+16 and PSR~J0737$-$3039A. The LISA ephemeris is shorter lived by a factor of a few hundred, rather than the four orders of magnitude suggested by the orbital period uncertainties alone. In practice, unmodelled orbital evolution shortens these extrapolation times for both the LISA and pulsar ephemerides.

\begin{acknowledgements}
	The author thanks the participants of the MIAPbP programme `Enabling Future Gravitational Wave Astrophysics in the Milli-Hertz Regime'
	(Munich, 2025) and colleagues at DZA and MPIfR for enlightening discussions. We thank the developers of \texttt{Eryn} for making the code publicly available. We also thank Natalia Korsakova and Valeriya Korol for providing the Galactic binary
	population used to simulate the GW UCB source parameters in this study. We especially thank Thomas Kupfer for reviewing and helping to improve this paper. 
	The author gratefully acknowledges the financial support provided by the German Federal Ministry of Research, Technology and Space (BMFTR) in the framework of the Knowledge creates perspectives for the region!, for the project StStG – DZA – Aufbauphase: Deutsches Zentrum für Astrophysik, Großforschungszentrum in der sächsischen Lausitz: Aufbauphase 2026, grant number 03WSP1746.
\end{acknowledgements}

\bibliographystyle{aa}
\bibliography{refs}

\begin{appendix}
\nolinenumbers

\section{Scheduling algorithm: Computing $t_\star$ from $u_\star$}
\label{app:appendixA}

For nearly circular orbits, the dominant quadrupole GW emission occurs at
$f_{\rm GW}=\bar n_{\rm b}/\pi$.
The carrier phase model is
\begin{equation}
	\begin{aligned}
		\Phi_{\rm GW}(t)
		={}& \phi_{\rm ref}
		+ 2\pi\!\left[
		f_{\rm GW,ref}(t-t_{\rm ref})
		+ \tfrac{1}{2}\dot f_{\rm GW}(t-t_{\rm ref})^2
		\right.
		\\
		&\left.
		+ \tfrac{1}{6}\ddot f_{\rm GW}(t-t_{\rm ref})^3
		\right].
		\label{app:phase_model_app}
	\end{aligned}
\end{equation}
The cubic term is small over the observation baseline but grows as $T^3$ under extrapolation. For the fiducial binary (Sect.~\ref{sec:horizon_forecast_ecc}), we set $\ddot f_{\rm GW}=0$ in the injected waveform, with $\dot f_{\rm GW}$ fixed by radiation reaction alone. The corresponding value would be $\ddot f_{\rm GW}=11/3\times\dot f_{\rm GW}^2/f_{\rm GW, ref}=1.1\times10^{-27}~{\rm Hz~s^{-2}}$, contributing $\Delta\Phi_{\ddot f}=(\pi/3)\ddot f_{\rm GW}T^3=2.5\times10^{-3}$ rad over four years and a shift of $4.5\times10^{-2}$~s in the epoch at which a targeted-$u$ is reached. The shift is far below the integration times of seconds to minutes required for optical spectroscopy of UCBs, and the phase contribution is below the uncertainty in the GW phase (Fig.~\ref{fig:horizon}). For HM Cnc the injected $\dot f_{\rm GW}$ and $\ddot f_{\rm GW}$ were taken from the measured cubic ephemeris of \cite{2023MNRAS.518.5123M}, which includes the accretion torque and departs from the radiation reaction value. Times are expressed in TCB, following the LISA convention \citep{2026arXiv260322377B}, and $t_{\rm ref}$ is an arbitrary reference epoch, which can be at the start of the LISA observation $t_0$ or the LISA observation midpoint $t_{\rm mid}$. In the following, we establish the GW $\rightarrow$ optical phase mapping $u(t) = \tfrac{1}{2}\Phi_{\rm GW}(t) + \delta_{e,\omega}(t)$ and then the phase-to-time mapping $u_\star\rightarrow t_\star$.

\subsection{Scheduling observations at fixed $u_1$}
\label{app:appendixA1}
The fixed-$u$ strategy requires predicting star~1's argument of latitude $u_1 = \omega_1 + \nu_1$ at future epochs with sufficient accuracy to place both
spectra at the same orbital phase. 
For LISA sources, the coherent GW phase model predicts $u_1(t)$ with timing precision of order $\lesssim 0.1\%P_b$
(Eq.~\ref{eq:timing_precision_Cutler}) over multi-year baselines. LISA measures the binary's GW phase $\Phi_{\rm GW}(t)$ continuously at twice the rate of $\Phi_{\rm asc}$. This phase encodes the orbital motion of the system and maps to star~1's argument of latitude through
\begin{equation}
	u_1(t) = \tfrac{1}{2}\Phi_{\rm GW}(t) + \epsilon_1 + \delta_{e,\omega}(t) + \mathcal{O}\Big(e^2\Big),
	\label{eq:u_mapping_simple}
\end{equation}
where the eccentricity correction $\delta_{e,\omega}(t) = \mathcal{O}(e)$ is defined below and the constant $\epsilon$ aligns the phase origin of $\Phi_{\rm GW}$ (set by the LISA inferred GW parameters) with that of $u_1$, \ie~$\epsilon_1 = 0$ in our convention, and $\epsilon_2 = \pi$ for the secondary argument of latitude defined by $u_2$.  
The orbital frame is conventionally aligned with the line of nodes (\citealp{2003PhRvD..67b4015C}; \citealp[Eq.~2.18 of][]{2009agwd.book.....J}), so the polar angle in that
frame is the argument of latitude $u$. The quadrupole nature of the emission
gives the GW phase as twice the orbital phase \citep[Eq.~6.138 of][]{2009agwd.book.....J}. For $e\neq0$ the $\mathcal{O}(e)$ part of this is carried by the sidebands, leaving the carrier at twice the mean orbital phase measured from the ascending node, $\Phi_{\rm asc}$,
\begin{equation}
	\Phi_{\rm GW}(t)=2\Phi_{\rm asc}(t).
	\label{app:PhiGW}
\end{equation}
Since $\Phi_{\rm asc}=\mathcal M+\omega$, with $\mathcal M$ the mean anomaly,
its rate is $\bar n_b=n_b+\dot\omega=\pi f_{\rm GW}$.
For constant $\bar n_b$, this gives
$\Phi_{\rm asc}(t)=\bar n_b(t-T_{\rm asc})$, with
$T_{\rm asc}=t_{\rm ref}-\phi_{\rm ref}/(2\pi f_{\rm GW,ref})$
obtained from the LISA-inferred parameters, where
$\phi_{\rm ref}=\Phi_{\rm GW}(t_{\rm ref})$.

The LISA posterior for $\phi_{\rm ref}$ has two modes separated by $\pi$ radians. Because the GW quadrupole repeats at twice the orbital frequency, dividing the GW phase by two propagates this into a $\pi/2$ ambiguity in $u_1(t)$. For a circular binary, the two modes correspond to orientations rotated by $\pi/2$ on the sky, which \citet{2025PhRvD.111h3051S} show are distinguished by radial velocity or eclipse timing. Scheduling at a specified $u_\star$ yields two candidate observing times per orbit, separated by approximately $P_b/4$. A single targeted spectrum resolves this branch ambiguity, since the two candidates correspond to spectroscopic phases differing by $\pi/2$ and predict different radial velocities. When quadrature is the targeted phase, the wrong branch lands near conjunction, with the radial velocity close to zero rather than at its extremum, so the two differ by $K$. 

The correction $\delta_{e,\omega}$ follows from the Keplerian expansion $\nu_1 = \mathcal M + 2e\sin \mathcal M + \mathcal{O}\Big(e^2\Big)$, where $\mathcal M = \Phi_{\rm asc} - \omega_1$
is the mean anomaly. Substituting into $u_1 = \nu_1 + \omega_1$ gives $u_1 = \Phi_{\rm asc} + \delta_{e,\omega} + \mathcal{O}\Big(e^2\Big)$ with $\delta_{e,\omega} = 2e\sin\mathcal M$, the leading term of the equation of the centre $\nu_1-\mathcal M$, which varies at the anomalistic rate $n_b$ while $\Phi_{\rm asc}$ advances at $\bar n_b$. Expanding with $(\eta,\kappa)$,
\begin{equation}
	\delta_{e,\omega}(t) \;=\; 2\kappa(t)\sin\Phi_{\rm asc}(t) - 2\eta(t)\cos\Phi_{\rm asc}(t),
	\label{eq:deltae}
\end{equation}
consistent with Eq.~\eqref{eq:deltae_main}. When $(\eta, \kappa)$ are poorly constrained by LISA,
one sets $\delta_{e,\omega} = 0$, and the effect is to enlarge the scheduling window by $\mathcal{O}(e)$.

For reference, the first-order small-$e$ l.o.s. velocity of component $i$ in the LL parameterisation is
\begin{equation}
	\begin{aligned}
		\frac{{\rm v}_{r,i}}{c} {}&= (-1)^{i-1} x_i n_b\left[\cos\Phi_{\rm asc}
		+ \kappa\cos 2\Phi_{\rm asc} + \eta\sin 2\Phi_{\rm asc}
		+ \mathcal{O}\Big(e^2\Big)\right],
		\\ K_i &= c n_b x_i,
		\label{eq:vr-asc-app}
	\end{aligned}
\end{equation}
where $\Phi_{\rm asc} = \Phi_{\rm GW}/2$ as defined by Eq.~\eqref{app:PhiGW}, and
$(\eta,\kappa)=(e\sin\omega_1,e\cos\omega_1)$ are referred to star~1 for both
components, so that the sign $(-1)^{i-1}$ accounts for $\omega_2=\omega_1+\pi$.
The prefactor $K_i=cn_bx_i$ is Keplerian and carries the anomalistic $n_b$. The
eccentricity correction is $\kappa\cos 2\Phi_{\rm asc}+\eta\sin 2\Phi_{\rm
	asc}=e\cos(2\Phi_{\rm asc}-\omega)$, a second harmonic at fixed $\omega$.
Equation \eqref{eq:vr-asc-app} follows from differentiating the Roemer delay \citep{10.1046/j.1365-8711.2001.04606.x}
\begin{equation}
	\Delta R(\Phi_{\rm asc}) = x\left[\sin\Phi_{\rm asc}
	+ \tfrac{\kappa}{2}\sin 2\Phi_{\rm asc}
	- \tfrac{\eta}{2}\cos 2\Phi_{\rm asc}\right]
	\label{eq:roemer-ell1}
\end{equation}
with respect to time at fixed $(\eta,\kappa)$, confirming that the Keplerian phase correction $\delta_{e,\omega}$ and the LL parameterisation are consistent representations of the same orbital motion.\footnote{Equation~\eqref{eq:roemer-ell1} omits a first-order term $-\tfrac32x\eta$, which \citet{10.1046/j.1365-8711.2001.04606.x} drop as constant in time. At fixed $(\eta,\kappa)$ its derivative vanishes and it does not contribute to Eq.~\eqref{eq:vr-asc-app}.} 

From Eq.~\eqref{eq:vr-asc-app}, the circular limit
${\rm v}_r/c = x n_b\cos\Phi_{\rm asc}$ has its maximum at $\Phi_{\rm asc}=0$,
at $t = T_{\rm asc}$, and its zero crossing, with star~1 moving from receding to
approaching, at $\Phi_{\rm asc}=\pi/2$, at $t = T_{\rm asc} + \pi/(2\bar n_b)$.
These are respectively the ascending node and conjunction epochs in the
circular limit. With eccentricity included, the zero crossing no longer coincides with conjunction. The radial velocity depends on the argument of latitude only through $\cos u$ plus the constant $\kappa$, so the velocity extremum lies at $u=0$ for any eccentricity.
To first order in $e$, it occurs at
$t = T_{\rm asc} + 2\eta/\bar n_b$, which is also the node crossing. The
additive $\kappa$ displaces the zero crossing to
$t = T_{\rm asc} + \pi/(2\bar n_b) - \kappa/\bar n_b$, while conjunction, the
geometric event $u = \pi/2$, is at
$t_{\rm conj} = T_{\rm asc} + \pi/(2\bar n_b) - 2\kappa/\bar n_b$; the two
differ by $\kappa/\bar n_b$.

For a target phase $u_\star$ (e.g. conjunction at $u_\star = \pm\pi/2$ or
quadrature at $u_\star = 0,\pi$), the observing epoch satisfies $u_1(t) =
u_\star$. The inversion for the corresponding barycentric time $t_\star$ is
given in Appendix~Sect.~\ref{app:appendixA2}.

\subsection{Inverting the ephemeris for $t_\star$}
\label{app:appendixA2}

\begin{figure}
	\centering
	\includegraphics[width=\columnwidth]{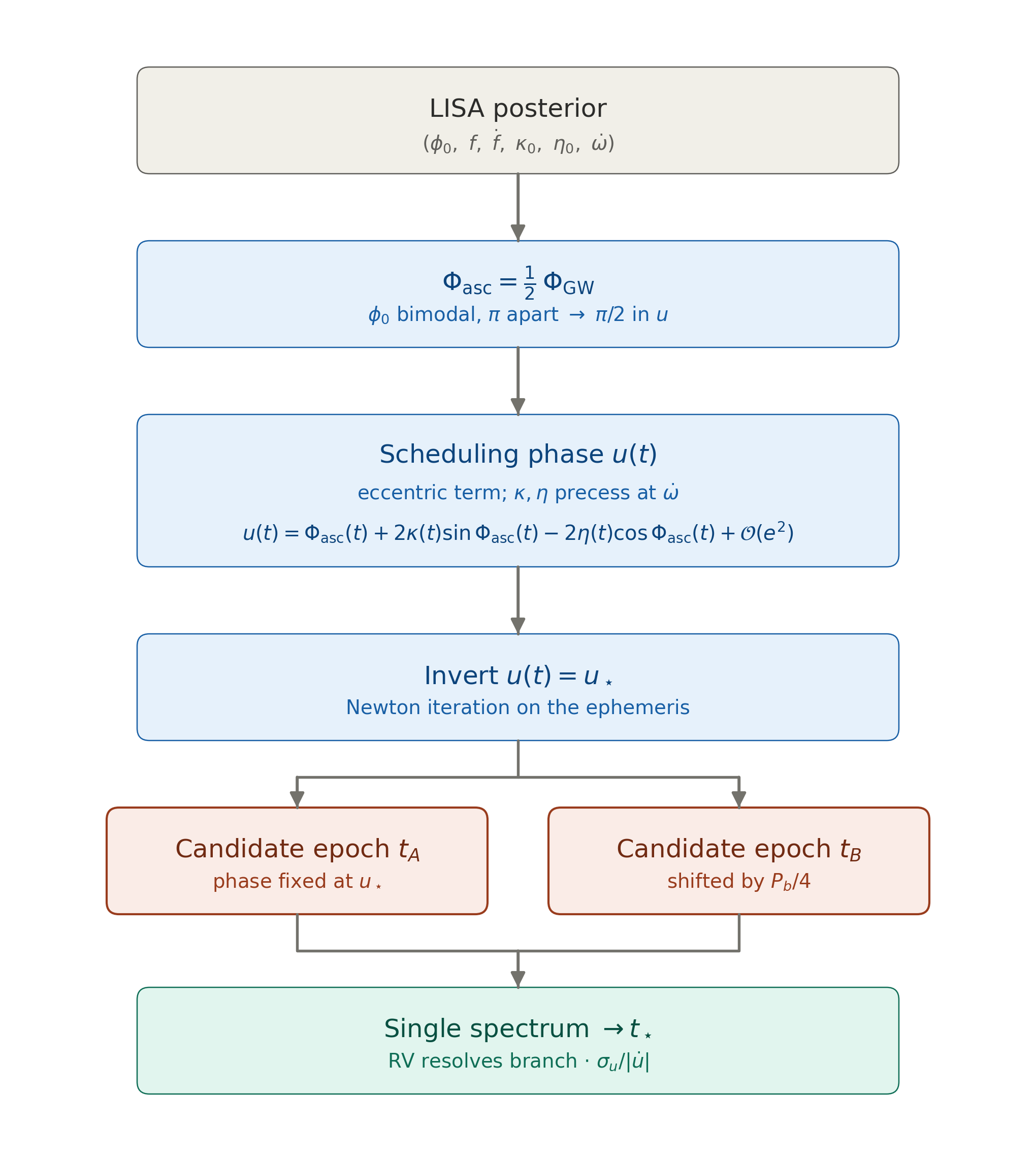}
	\caption{Mapping the LISA GW phase to a scheduled
		spectroscopic epoch. The GW phase is halved to the mean orbital
		phase. The $\phi_{\rm ref}$ multimodality of the LISA posterior (two modes
		$\pi$ apart) becomes a $\pi/2$ ambiguity in $u$, producing two
		candidate epochs separated by $P_b/4$ that one spectrum resolves
		through the radial velocity. The inversion $u(t)=u_\star$ is performed
		by Newton iteration on the ephemeris.}
	\label{fig:lisa_tstar_pipeline}
\end{figure}

Given a target argument of latitude $u_\star$ and the GW phase model $\Phi_{\rm GW}$, the observing epoch is the barycentric time $t_\star$ that satisfies $u(t_\star)=u_\star \pmod{2\pi}$. Fig.~\ref{fig:lisa_tstar_pipeline} summarises the mapping. The carrier phase is related to the mean orbital phase $\Phi_{\rm asc}=\Phi_{\rm GW}/2$, the
eccentricity term is applied with the LL pair $(\eta,\kappa)=(e\sin\omega,e\cos\omega)$ rotated to time $t$ at the apsidal rate $\dot\omega$,
\begin{equation}
	u(t)=\Phi_{\rm asc}(t)
	+2\kappa(t)\sin\Phi_{\rm asc}(t)
	-2\eta(t)\cos\Phi_{\rm asc}(t)
	+\mathcal{O}\Big(e^2\Big),
	\label{eq:u_precessing_app}
\end{equation}
and $u(t)=u_\star$ is solved by Newton iteration seeded from the planned window. In the RV convention used here, quadrature corresponds to the velocity extrema at
$u_\star=0$ or $\pi$, while conjunction occurs at the geometric phases
$u_\star=\pm\pi/2$, with the RV zero crossings displaced from these phases by
$\mathcal{O}(e)$. The two quadratures are physically distinct, separated by half an orbit, with opposite RV extrema. At Newtonian quadrupole order, the gravitational waveform depends on $2u$ and is invariant under $u\rightarrow u+\pi$. The sign of the orbital radial velocity of an identified component resolves this ambiguity. This is separate from the $\phi_{\rm ref}$ multimodality of Appendix~Sect.~\ref{app:appendixA1}, which displaces a single chosen target by $\pi/2$ in $u$, a quarter of an orbit, and is broken by one spectrum.

The scheduled epoch carries the timing uncertainty
\begin{equation}
	\sigma_{t_\star}\simeq\frac{\sigma_u(t_\star)}{\dot u(t_\star)},
	\qquad
	\dot u \simeq \dot\Phi_{\rm asc}=\bar n_{\rm b}=\pi f_{\rm GW}
	\label{eq:sigma_tstar}
\end{equation}
where the $\mathcal{O}(e)$ term $\dot\delta_{e,\omega}=2n_{\rm b}(\kappa\cos\Phi_{\rm asc}+\eta\sin\Phi_{\rm asc})$
and the chirp $\dot f_{\rm GW}(t-t_{\rm ref})$ are dropped from $\dot u$, contributing
at the $2e$ and $\tau/\tau_{\rm chirp}$ levels respectively. Both were retained in the
numerical forecast. The uncertainty
$\sigma_u(t)$ is propagated from the posterior covariance in
$(\phi_{\rm mid}, f_{\rm GW, mid},\dot f_{\rm GW},\kappa_{\rm mid},\eta_{\rm mid},\dot\omega)$ (Eq.~\eqref{eq:sigma_u_from_PhiGW} and Eq.~\eqref{eq:timing_unc}). The usable scheduling horizon is the span of future times over which $\sigma_u(t)$ stays below the tolerance set by the RV measurement (see Appendix~Sect.~\ref{app:appendixC1}). With apsidal advance modelled as a rotation of $(\eta,\kappa)$ at fixed eccentricity, the static conjunction timing scale $\Delta t_e\simeq P_b~e\cos\omega/\pi=P_b\kappa/\pi$ \citep{2010exop.book...55W,2010ApJ...717L.108K} holds locally with $\kappa\rightarrow\kappa(t)$ as $\omega$ evolves over the scheduling baseline.

\section{Corrections to measured velocity amplitudes from finite phase averaging}
\label{app:appendixB}

A spectrum records a velocity averaged over the orbital phase it spans, not the instantaneous velocity at its centre. A single short exposure spans a negligible phase range, but faint targets are co-added into phase bins to reach usable S/N, where a bin now covers an appreciable fraction of the orbit. We write $\Delta t_{\rm bin}$ for the width of this window.

For a circular orbit, the radial velocity varies as $\cos u$ at the orbital frequency $n_b$. Averaging $\cos u$  over a temporal window of width $\Delta t_{\rm bin}$ centred on $u_\star$ attenuates $K$ by
\begin{equation}
	\Lambda_K = \mathrm{sinc}(\bar n_b \Delta t_{\rm bin}/2),
	\label{eq:Lambda_K}
\end{equation}
the same form as the smear factor of Sect.~\ref{subsec:expected_magnitudes}, now evaluated over the bin width. If uncorrected, $\Lambda_K$ lowers the velocity amplitudes inferred from circular RV curves and biases the recovered masses. Because it depends only on $P_b$ and $\Delta t_{\rm bin}$, it is known in advance, and the amplitudes are recovered by dividing by it. For matched windows the same factor multiplies both components, so it cancels in the mass ratio $q$.

This correction is applied in practice. For the seven-minute binary ZTF~J1539+5027, \citet{2019Natur.571..528B} measured radial velocities from spectra co-added over one third of the orbit, each built from individual exposures spanning about one eighth of the orbit, the wider co-addition being necessary to reach a sufficient S/N. The authors increased both measured semi-amplitudes by $20\%$, consistent with $1/\Lambda_K = 1.21$ at a bin width of $P_b/3$, and applied it equally to both components, so the smearing correction leaves the mass ratio unchanged.

For an eccentric orbit, finite phase averaging acts differently on the
fundamental and eccentricity terms. Over one observing session, $\kappa$
and $\eta$ are assumed to be constant, and a uniformly weighted window of
duration $\Delta t_{\rm bin}$ centred at
$\Phi_{\rm asc}=\Phi_c$ gives, denoting the phase-averaged radial velocity by
$\bar{\rm v}_{r,j}$,
\begin{equation}
	\bar{\rm v}_{r,j} = \gamma_j + (-1)^{j-1}K_j\big[\Lambda_K\cos\Phi_c
	+\Lambda_\kappa\big(\kappa\cos2\Phi_c+\eta\sin2\Phi_c\big)\big],
	\label{eq:vbar_master}
\end{equation}
to first order in eccentricity. Here the fundamental term is attenuated by Eq.~\eqref{eq:Lambda_K},
while the eccentricity terms, which vary at twice the orbital phase rate,
are attenuated by
\begin{equation}
	\Lambda_\kappa
	=
	\mathrm{sinc}\!\left(
	\bar n_b\Delta t_{\rm bin}
	\right).
	\label{eq:Lambda_kappa}
\end{equation}
The eccentricity terms are more strongly attenuated. For
$\Delta t_{\rm bin}=P_b/3$, $\Lambda_K\simeq0.83$ and
$\Lambda_\kappa\simeq0.41$. Both factors assume a uniformly weighted
window.

The window is centred on the epoch $t_\star$ at which
$u=u_\star$, so $\Phi_c$ is displaced from the anchor by the
$\mathcal{O}(e)$ difference between $u$ and $\Phi_{\rm asc}$. Inverting
Eq.~\eqref{eq:u_precessing_app},
\begin{equation}
	\Phi_c = u_\star - 2\kappa\sin u_\star + 2\eta\cos u_\star
	+\mathcal{O}\Big(e^2\Big),
	\label{eq:Phi_c}
\end{equation}
which gives $\Phi_c=2\eta$ at $u_\star=0$ and
$\pi-2\eta$ at $u_\star=\pi$. The displacement is first order in $e$, but at
both quadratures $\cos\Phi_{\rm asc}$ is stationary, $\cos2\Phi_c=1$, and the
$\eta\sin2\Phi_c$ term is itself first order in $e$, so all three corrections
enter Eq.~\eqref{eq:vbar_master} at $\mathcal{O}(e^2)$. At a quadrature anchor,
the averaged velocity is
\begin{equation}
	\bar{\rm v}_{r,j} = \gamma_j + (-1)^{j-1}K_j
	\big(\pm\Lambda_K + \Lambda_\kappa\kappa\big)+\mathcal{O}\Big(e^2\Big),
	\label{eq:vbar_quad}
\end{equation}
with the upper sign at $u_\star=0$. The leading term
changes sign between the two quadratures while the eccentricity term does not,
and this asymmetry sets what each measurement returns. Within a session, the
eccentricity term is common to the two anchors and is absorbed into the fitted
intercept, leaving slope $K_j\Lambda_K$, attenuated exactly as for a circular
orbit.

Between sessions, take two epochs $t_\alpha$ and
$t_\beta$ at the same quadrature anchor and with the same bin width, so that
$\Lambda_K$ and $\Lambda_\kappa$ are matched. In the difference of
Eq.~\eqref{eq:vbar_quad}, the constant $\pm K_j\Lambda_K$ is identical at both
epochs and cancels, as does any component offset that is constant in time,
including the gravitational redshift. What survives is the shared barycentric
drift and the change in $\kappa$ at fixed $K_j$,
\begin{equation}
	\Delta\bar{\rm v}_{r,j}
	= \Delta{\rm v}_{\rm CM}
	+ (-1)^{j-1}K_j\Lambda_\kappa\Delta\kappa
	+ \mathcal{O}\Big(e^2\Big),
	\label{eq:dvbar_general}
\end{equation}
which written out for the two components is
\begin{equation}
	\Delta\bar{\rm v}_1
	=
	\Delta{\rm v}_{\rm CM}
	+
	K_1\Lambda_\kappa\Delta\kappa ,
	\label{eq:dvbar1}
\end{equation}
and
\begin{equation}
	\Delta\bar{\rm v}_2
	=
	\Delta{\rm v}_{\rm CM}
	-
	K_2\Lambda_\kappa\Delta\kappa ,
	\label{eq:dvbar2}
\end{equation}
where
\begin{equation}
	\Delta\kappa
	=
	[e\cos\omega]_{t_\beta}-[e\cos\omega]_{t_\alpha}.
\end{equation}
Subtracting the differences between the two components, Eq.~\eqref{eq:dvbar2} from Eq.~\eqref{eq:dvbar1}, removes the common barycentric
drift, giving the finite-bin form of
Eq.~\eqref{eq:delta_kappa_spec},
\begin{equation}
	\Delta\kappa_{\rm spec}
	=
	\frac{\Delta\bar{\rm v}_1-\Delta\bar{\rm v}_2}
	{(K_1+K_2)\Lambda_\kappa}.
	\label{eq:delta_kappa_spec_binned}
\end{equation}
Here $K_1$ and $K_2$ are determined from the intra-session regressions on $F$. Each regression returns a slope of magnitude $K_j\Lambda_K$, so $K_j$ follows on division by $\Lambda_K$.

For $\Delta t_{\rm bin}\ll P_b$, $\Lambda_\kappa\simeq1$ and
Eq.~\eqref{eq:delta_kappa_spec_binned} reduces to the instantaneous
fixed-$u_\star$ result, Eq.~\eqref{eq:delta_kappa_spec}. Binning over a
phase range that is a sizable fraction of the orbit drives
$\Lambda_\kappa$ below unity, and this attenuation must be included when
recovering $\Delta\kappa_{\rm spec}$. For matched windows,
$\Lambda_\kappa$ multiplies the apsidal terms in the velocity differences
of both stars and cancels from the slope that determines $q$ (as discussed in Sect.~\ref{sec:theory_ecc}), but it does
not cancel from the spectroscopic measurement of $\Delta\kappa$.

\section{Error model}
\label{app:appendixC}

Every velocity in this paper is measured at a target argument of latitude, with the observing epoch fixed by the LISA-derived ephemeris. For both circular and eccentric orbits, the semi-amplitudes follow from measurements at opposite quadratures, $u=0$ and $u=\pi$, within a single session. The two-epoch difference at a common $u$, taken at $t_\alpha$ and $t_\beta$, instead isolates the barycentric drift and the apsidal advance. The error propagation below applies in each case, with $K$ and $\sigma_\Delta$ denoting the velocity scale and the uncertainty of the chosen observable.

Each single spectrum carries an uncertainty $\sigma_{\rm spec}$ arising
from photon noise, wavelength calibration, template mismatch, and
systematic effects. For the ultracompact systems considered here, the
orbital period is short enough that a single exposure covers an
appreciable fraction of an orbit, so individual spectra are necessarily
short and their velocities correspondingly imprecise. We adopt
$\sigma_{\rm spec}\sim15~{\rm km~s^{-1}}$ as a fiducial per spectrum
value. Combining $N_{\rm RV}$ spectra scheduled at a common argument of
latitude gives the epoch precision
\begin{equation}
	\sigma_{\rm v} = \sigma_{\rm spec}/\sqrt{N_{\rm RV}},
	\label{eq:sigma_epoch}
\end{equation}
when photon noise dominates. It is $\sigma_{\rm v}$, not
$\sigma_{\rm spec}$, that sets the tolerances below. For comparison,
high resolution VLT/UVES observations from the SPY survey achieved
WD velocities at the few kilometre per second level per epoch,
with sub-$\mathrm{km~s^{-1}}$ precision for brighter targets
\citep{2020A&A...638A.131N}, on systems whose longer periods permit far
longer individual exposures.

Because every velocity contributes to the observable with coefficient $\pm1$,
mutually independent uncertainties give, for $N$ measurements,
\begin{equation}
	\sigma_\Delta^2
	=
	\sum_{i=1}^{N}\sigma_{{\rm v},i}^2 .
	\label{eq:sigma_delta_general}
\end{equation}
For comparable precision in all measurements,
$\sigma_{{\rm v},i}=\sigma_{\rm v}$, and negligible correlations,
Eq.~\eqref{eq:sigma_delta_general} reduces to
\begin{equation}
	\sigma_\Delta
	=
	\sqrt{N}~\sigma_{\rm v},
	\label{eq:sigma_delta_uncorr}
\end{equation}
where $N$ is the number of independent velocity measurements combined
with unit coefficients in the observable, rather than the number of
spectra included in a regression.

Two cases are relevant here. For $N=1$, the uncertainty is simply
$\sigma_{\rm v}$. This applies to each velocity used in the
intra-session ${\rm v}$--$\cos u$ regression for circular orbits and
in the corresponding regression for eccentric orbits, in which the $K_j\kappa$ contribution is included. For $N=2$,
\begin{equation}
	\sigma_\Delta=\sqrt{2}~\sigma_{\rm v},
	\label{app:unc_of_Delta_v}
\end{equation}
which applies to the single-component difference
$\Delta{\rm v}_j={\rm v}_j(t_\beta)-{\rm v}_j(t_\alpha)$, used to recover $\Delta{\rm v}_{\rm CM}$ in Sect.~\ref{sec:DeltaVCM}. These epochs are separated by months to
years and have independent instrument configurations, wavelength
calibrations, and observing conditions, so correlations between their
spectroscopic errors are neglected.

Velocities placed at the quadratures constrain the slope more strongly than the same number spread over the orbit. For $N$ velocity measurements for each component divided between $u=0$ and $u=\pi$, the regression recovers each semi-amplitude with $\sigma_{K_j}=\sigma_{\rm v}/\sqrt N$, a factor of $\sqrt2$ better than the same $N$ distributed uniformly in $u$.

For $N=4$,
\begin{equation}
	\sigma_\Delta=2\sigma_{\rm v},
\end{equation}
which applies to the two-component difference
$\Delta{\rm v}_1-\Delta{\rm v}_2$ appearing in
Eq.~\eqref{eq:delta_kappa_spec}. Equivalently, if each
single-component difference has uncertainty
$\sigma(\Delta{\rm v}_j)=\sqrt{2}\sigma_{\rm v}$, the uncertainty
of their difference is $2\sigma_{\rm v}$.

\subsection{Tolerance definitions}
\label{app:appendixC1}

Having established the error model, we define the phase tolerance
$\delta u_{\mathrm{tol}}$ by requiring that the velocity error from phase scheduling remain below the measured RV uncertainty $\sigma_\Delta$.
Let the phase error be Gaussian,
$\delta u\sim\mathcal{N}(0,\sigma_u^2)$. Expanding the velocity about
the intended phase gives
\begin{equation}
	\delta {\rm v}
	=
	K\left[\cos(u_\star+\delta u)-\cos u_\star\right]
	\simeq
	-K\sin u_\star~\delta u
	-\frac{K}{2}\cos u_\star~\delta u^2 .
	\label{ap:delta_v}
\end{equation}
Away from quadrature, the linear term dominates and has rms
\begin{equation}
	\sigma(\delta {\rm v})
	=
	K|\sin u_\star|~\sigma_u .
\end{equation}
Requiring this contribution to remain below the velocity uncertainty
$\sigma_{\rm v}$ gives the generic phase tolerance
\begin{equation}
	\delta u_{\rm tol}^{\rm (gen)}
	=
	\frac{\sigma_{\rm v}}
	{K|\sin u_\star|}.
	\label{ap:delu_gen}
\end{equation}

At conjunction, $|\sin u_\star|=1$, and the generic expression reduces
to
\begin{equation}
	\delta u_{\rm tol}^{\rm (conj)}
	=
	\frac{\sigma_{\rm v}}{K}.
\end{equation}
This is the tightest tolerance in the regime governed by the linear
term.

At quadrature, $\sin u_\star=0$ and $|\cos u_\star|=1$, so the leading
velocity error is quadratic. Since
$\langle\delta u^2\rangle=\sigma_u^2$, requiring
$K\sigma_u^2/2\lesssim\sigma_{\rm v}$ gives
\begin{equation}
	\delta u_{\rm tol}^{\rm (quad)}
	=
	\left(\frac{2\sigma_{\rm v}}{K}\right)^{1/2}.
\end{equation}
For $\sigma_{\rm v}/K\ll1$, the quadrature tolerance is looser than
the conjunction tolerance by
$\sqrt{2K/\sigma_{\rm v}}$. It is similarly looser than the generic
tolerance by
$|\sin u_\star|\sqrt{2K/\sigma_{\rm v}}$ wherever the linear
approximation applies.

\subsection{Tolerance for the two-epoch velocity difference}
\label{app:appendixC11}

Let the target argument of latitude be $u_\star$ and the two
LISA-scheduled observations occur at
\begin{equation}
	u_\alpha=u_\star+\delta u_\alpha,
	\qquad
	u_\beta=u_\star+\delta u_\beta,
\end{equation}
with $|\delta u_i|\ll1$. The physical contributions
$\Delta{\rm v}_{\rm CM}$ and $K\Delta\kappa$ are signal terms in the
two-epoch velocity difference and are independent of the shared anchor
$u_\star$. We use $\Delta{\rm v}_{\rm res}$ exclusively for the
contamination arising from the phase-scheduling errors
$\delta u_\alpha$ and $\delta u_\beta$.

If $\delta u_\alpha=\delta u_\beta$, the common phase offset cancels
from the fixed-$u$ velocity difference. This does not require the phase
error at either epoch to vanish, but equality is not generally maintained
between widely separated epochs. Uncertainties in $f_{\rm GW, mid}$ and
$\dot f_{\rm GW}$, or equivalently in
$\bar n_b=\pi f_{\rm GW, ref}$ and $\dot{\bar n}_b=\pi\dot f_{\rm GW}$, affect the
predicted phases at the two epochs by different amounts.
Expanding the Kepler term about $u_\star$ gives
\begin{equation}
	\begin{aligned}
		\Delta{\rm v}_{\rm res}
		&=
		K\left[
		\cos(u_\star+\delta u_\beta)
		-\cos(u_\star+\delta u_\alpha)
		\right]
		\\
		&\simeq
		-K\sin u_\star~\delta u_{\rm rel}
		-\frac{K}{2}\cos u_\star
		\left(\delta u_\beta^2-\delta u_\alpha^2\right),
		\label{app:leakage_series_app}
	\end{aligned}
\end{equation}
where
$\delta u_{\rm rel}=\delta u_\beta-\delta u_\alpha$.

The velocity error from phase scheduling is quantified in terms of the single-epoch phase
uncertainty $\sigma_u(t)$ propagated from the LISA posterior, which is
also the quantity used in the horizon condition below. For the
analytic estimates, we approximate $\delta u_\alpha$ and
$\delta u_\beta$ as independent zero-mean Gaussian variables with
common variance $\sigma_u^2$. Under this approximation,
$\delta u_{\rm rel}$ has standard deviation $\sqrt{2}\sigma_u$, while
$\delta u_\beta^2-\delta u_\alpha^2$ has zero mean and standard
deviation $2\sigma_u^2$.

In practice, the two phase errors arise from the same realisation of
the posterior in
$(\phi_{\rm ref},f_{\rm GW, mid},\dot f_{\rm GW})$ and are
correlated. In particular, an error in $\phi_{\rm ref}$ shifts both
epochs identically and cancels from the difference. Neglecting this
positive covariance gives conservative velocity error estimates. When the orbit is eccentric, uncertainties in $(\eta_{\rm ref},\kappa_{\rm ref},\dot\omega)$ also enter $\sigma_u$
through their contribution to the predicted argument of latitude.

Away from quadrature, the linear term in
Eq.~\eqref{app:leakage_series_app} dominates. Its rms is
\begin{equation}
	\sigma(\Delta{\rm v}_{\rm res})
	=
	\sqrt{2}~K|\sin u_\star|~\sigma_u .
\end{equation}
Requiring this velocity error to remain below the uncertainty
$\sigma_\Delta$ of the two-epoch velocity difference gives
\begin{equation}
	\delta u_{\rm tol}^{\rm (gen)}
	=
	\frac{\sigma_\Delta}
	{\sqrt{2}~K|\sin u_\star|}.
	\label{eq:tol_gen}
\end{equation}
The apparent divergence at quadrature marks the zeroing of the
linear term in Eq.~\eqref{app:leakage_series_app} rather than an unbounded tolerance. At
conjunction, $u_\star=\pm\pi/2$, the generic result reduces to
\begin{equation}
	\delta u_{\rm tol}^{\rm (conj)}
	=
	\frac{\sigma_\Delta}{\sqrt{2}~K}.
	\label{eq:tol_conj}
\end{equation}

At quadrature, $u_\star=0$ or $\pi$, the linear velocity error vanishes.
Because
$\delta u_\beta^2-\delta u_\alpha^2$ has standard deviation
$2\sigma_u^2$, the quadratic term has rms
\begin{equation}
	\sigma(\Delta{\rm v}_{\rm res})
	=
	K\sigma_u^2 .
\end{equation}
Requiring this contribution to remain below $\sigma_\Delta$ gives
\begin{equation}
	\delta u_{\rm tol}^{\rm (quad)}
	=
	\left(\frac{\sigma_\Delta}{K}\right)^{1/2}.
	\label{eq:tol_quad}
\end{equation}
The neglected secular term
$\Delta K\cos u_\star$ is maximal at quadrature but remains far below
the spectroscopic floor (Eq.~\eqref{eq:DeltaK_bound}) and does not
affect this tolerance.

\subsection{Summary}

For $N=1$, $\sigma_\Delta = \sigma_{\rm v}$, and the generic,
conjunction, and quadrature tolerances are
\begin{equation}
	\frac{\sigma_{\rm v}}{K|\sin u_\star|},
	\qquad
	\frac{\sigma_{\rm v}}{K},
	\qquad
	\left(\frac{2\sigma_{\rm v}}{K}\right)^{1/2},
\end{equation}
respectively. For the $N=2$ velocity difference,
$\sigma_\Delta=\sqrt{2}\sigma_{\rm v}$. The generic and conjunction
tolerances are unchanged, while the quadrature tolerance becomes
\begin{equation}
	\left(\frac{\sqrt{2}\sigma_{\rm v}}{K}\right)^{1/2},
\end{equation}
smaller than the $N=1$ value by a factor of $2^{-1/4}$.

We use the tolerance appropriate to the observation ($N=1$ or $N=2$) and the target phase $u_\star$ to enforce the horizon condition
\begin{equation}
	\sigma_u(T_{\mathrm{valid}})
	\leq
	\alpha~\delta u_{\mathrm{tol}},
	\qquad
	\alpha\in\{0.5,~1\}.
	\label{eq:horizon_condition}
\end{equation}
Both sides of Eq.~\eqref{eq:horizon_condition} are phase uncertainties at a single observing epoch. The left side is the forecast uncertainty of the predicted argument of latitude. On the right side, $\delta u_{\mathrm{tol}}$ is the largest such uncertainty for which the velocity error caused by phase scheduling stays below $\sigma_\Delta$, and $\alpha$ scales it. For $N=2$, the effect of combining two epochs is already included in $\delta u_{\mathrm{tol}}$ through $\sigma_\Delta$, so no further factor is applied to $\sigma_u$. To leading order, $\sigma_u(t)/\bar n_{\rm b}$ is the corresponding uncertainty in the predicted calendar time of $u_\star$. The adopted values of $\alpha$ are discussed further in Sect.~\ref{sec:ecc_horizon_example}.

\section{From the multipole expansion to the $n=1,2,3$ small-$e$ frequency components}
\label{app:appendixD}

At Newtonian order the leading source moment is the mass quadrupole. Writing $a$ for the relative semi-major axis, the binary obeys Kepler's third law,
\begin{equation}
	n_b^2 a^3 = G(M_1+M_2) = GM_{\rm tot}.
\end{equation}
Here $G$ is Newton's gravitational constant, $M_{\rm tot} = M_1 + M_2$ is the total mass, and $r$ is the instantaneous separation between the two components. The Keplerian equations of motion are
\begin{equation}
	r\ddot r = (r\dot u)^2 - \frac{GM_{\rm tot}}{r},
	\qquad
	\frac{d}{dt}\!\left(r^2\dot u\right) = 0.
	\label{eq:kepler-eom}
\end{equation}
Using these, the wave-zone multipole expansion reduces to the polarisation functions
\begin{align}
	h_+(t)
	&=
	\frac{A_+}{2(an_b)^2}\Big(\mathcal Q\cos 2u-2\dot r(r\dot u)\sin 2u\Big)  \nonumber\\
	&\quad + \frac{\mathcal A\sin^2\iota}{2(an_b)^2}\Big(\mathcal Q+2(r\dot u)^2\Big),
	\label{eq:hplus_exact}\\[4pt]
	h_\times(t)
	&=
	\frac{A_\times}{2(an_b)^2}\Big(\mathcal Q\sin 2u+2\dot r(r\dot u)\cos 2u\Big).
	\label{eq:hcross_exact}
\end{align}
We define the GW amplitude $\mathcal A$ in terms of the chirp mass $M_c$, luminosity distance $D_L$, and the GW frequency $f$
\begin{equation}
	\mathcal A
	= \frac{2GM_1M_2(an_b)^2}{c^4M_{\rm tot}D_L}
	\simeq \frac{2G^{5/3}M_c^{5/3}(\pi f_{\rm GW, ref})^{2/3}}
	{c^4D_L}.
	\label{f_amp}
\end{equation}
and
\begin{equation}
	A_+=\mathcal A(1+\cos^2\iota),
	\qquad
	A_\times=2\mathcal A\cos\iota.
\end{equation}
We also introduce
\begin{equation}
	\mathcal{Q}=\dot r^{2} - (r\dot u)^2 - \frac{GM_{\rm tot}}{r}.
	\label{eq:mathQ}
\end{equation}

For a circular orbit, $\dot r=0$, $r=a$, and $\dot u=n_b$, so
$$
\mathcal Q=-(r\dot u)^2-\frac{GM_{\rm tot}}{r}=-2a^2n_b^2,
\qquad
\mathcal Q+2(r\dot u)^2=0.
$$
Hence the trace term in $h_+$ vanishes identically, and one recovers
the usual circular orbit quadrupole pattern
$h_+\propto \cos 2u$, $h_\times\propto \sin 2u$,
up to the overall polarisation sign convention.
A generic orientation of the polarisation basis in the sky plane is then
obtained by the usual spin-2 rotation
$$
h_+\rightarrow h_+\cos2\psi-h_\times\sin2\psi,\qquad
h_\times\rightarrow h_+\sin2\psi+h_\times\cos2\psi.
$$

Here we evaluate the orbital coefficients at Newtonian order, treating $a$ and $e$ as constant on the timescale of the observation and neglecting secular evolution from radiation reaction \citep{1964PhDT........51P,2009agwd.book.....J}. The Keplerian relations for $r$, $\dot r$, and $\dot u$ then apply,
with the secular change in the argument of periastron $\dot\omega$ also neglected in the amplitude coefficients. The mean anomaly $\mathcal{M}=n_b(t-T_0)$, with $T_0$ the time of periastron passage, advances uniformly in time, unlike the true anomaly $\nu$. To linear order in $e$, the needed quantities are
\begin{equation}
	\begin{aligned}
		(r\dot u)^2 + \dot r^{2} &= (an_b)^2\big(1+2e\cos\mathcal M\big), \\
		\frac{GM_{\rm tot}}{r} &= (an_b)^2\big(1+e\cos\mathcal M\big), \\
		\dot r(r\dot u) &= (an_b)^2~ e\sin\mathcal M,
	\end{aligned}
	\label{eq:inv-expand}
\end{equation}
all up to $\mathcal{O}\Big(e^2\Big)$. 
We expand the orbital quantities in $\mathcal{M}$, in which every periodic
orbital quantity can be expanded as a sum of harmonics, and write the result in the
mean orbital phase $\Phi_{\rm asc}=\mathcal{M}+\omega$, abbreviated $\Phi$
throughout this appendix. The true and mean orbital phases differ at first order
in $e$, giving
\begin{equation}
	u=\Phi+2\kappa\sin\Phi-2\eta\cos\Phi+\mathcal{O}(e^2),
	\label{app:u_deltae}
\end{equation}
the equation of the centre, linking the mean phase to the argument of latitude.
The mean orbital phase advances at the rate $\bar n_b=\pi f_{\rm GW}$, with $f_{\rm GW}$ measured by LISA.
The orbital elements change little over one orbit, $\dot\omega P_b=3.6\times10^{-4}$~rad, so we hold them fixed while deriving the $n=1,2,3$ frequency components, and restore their evolution in Appendix~Sect.~\ref{app:apsidal_mcmc}.\footnote{Equivalently, $\Phi=\bar n_b(t-T_0)+\omega(T_0)=\bar n_b(t-T_{\rm asc})$, where $T_{\rm asc}=T_0-\omega(T_0)/\bar n_b$ is defined by $\Phi(T_{\rm asc})=0$. This holds when $n_b$ is treated as constant between $T_0$ and $T_{\rm asc}$, leaving $\bar n_b=n_b+\dot\omega$. If $n_b$ is referenced at $T_0$ while $\bar n_b$ is referenced at $T_{\rm asc}$, retaining the chirp gives the epoch-conversion term
	$\bar n_b=n_b+\dot\omega-\dot n_b(T_0-T_{\rm asc})$
	\citep{10.1046/j.1365-8711.2001.04606.x}, which we neglect, giving a fractional error of $5\times10^{-11}$ in $\bar n_b$.}
The quantity defined in Eq.~\eqref{eq:mathQ}  expands as
\begin{equation}
	\mathcal{Q}
	= -(an_b)^2\big(2+3e\cos \mathcal M\big)+\mathcal{O}\Big(e^2\Big)= \mathcal{Q}_0+\mathcal{Q}_1+\mathcal{O}\Big(e^2\Big),
	\label{eq:Q-expand}
\end{equation}
and the second term of Eq.~\eqref{eq:hplus_exact}  expands as
\begin{equation}
	\mathcal{Q}+2(r\dot u)^2
	= \dot r^{2}+(r\dot u)^2-\frac{GM_{\rm tot}}{r}
	= (an_b)^2e\cos \mathcal M+\mathcal{O}\Big(e^2\Big).
	\label{eq:trace-expand}
\end{equation}

Using the Kepler equation at small-eccentricity,
$$
\nu = \mathcal M + 2e\sin \mathcal M + \mathcal{O}\Big(e^2\Big),
$$
so that
$$
u = \omega + \mathcal M + 2e\sin \mathcal M + \mathcal{O}\Big(e^2\Big),
$$
we have
\begin{align}
	\cos 2u
	&{}= \cos(2\omega+2\mathcal M) \nonumber\\
	&{}+ 2e\!\left[\cos(2\omega+3\mathcal M)-\cos(2\omega+\mathcal M)\right] + \mathcal{O}\Big(e^2\Big), \label{eq:cos2u-expand}\\
	\sin 2u
	&{}= \sin(2\omega+2\mathcal M) \nonumber\\
	&{}+ 2e\!\left[\sin(2\omega+3\mathcal M)-\sin(2\omega+\mathcal M)\right] + \mathcal{O}\Big(e^2\Big). \label{eq:sin2u-expand}
\end{align}
We insert Eqs.~\eqref{eq:cos2u-expand}-\eqref{eq:sin2u-expand}, \eqref{eq:Q-expand}, and \eqref{eq:trace-expand} into Eqs.~\eqref{eq:hplus_exact} and \eqref{eq:hcross_exact}, retaining only $\mathcal O(e)$ terms. For $h_+$ the dominant mode arises from the $\mathcal O(e^0)$ component of $\mathcal Q\cos2u$, namely $\mathcal Q_0\cos2\Phi$. The $\mathcal O(e)$ sidebands in $h_+$ receive contributions from four sources: (i) the $\mathcal O(e)$ term from $\mathcal Q_0\cos2u$, (ii) the $\mathcal O(e)$ term from $\mathcal Q_1\cos2u$, (iii) the $\mathcal O(e)$ term from $-2\dot rr\dot u\sin 2u$, and (iv) the $\sin^2\iota$ term, which contributes $(\mathcal A\sin^2\iota/2)~e\cos\mathcal M$.

The cosine term in Eq.~\eqref{eq:hplus_exact} gives
\begin{equation}
	\begin{aligned}
		\mathcal{Q}\cos 2u
		{}\simeq{}& -2(an_b)^2\Big(\cos 2\Phi+2e\cos(3\Phi-\omega) \\
		&\qquad\qquad -2e\cos(\Phi+\omega)\Big) \\
		&-3e(an_b)^2\cos\mathcal{M}\cos 2\Phi \\[4pt]
		{}={}& \mathcal{Q}_0\big(\cos 2\Phi+2e\cos(3\Phi-\omega) \\
		&\qquad\quad -2e\cos(\Phi+\omega)\big) \\
		&+\mathcal{Q}_1\cos 2\Phi+\mathcal{O}\!\left(e^2\right),
		\label{eq:Qcos2u}
	\end{aligned}
\end{equation}
and the sine term in Eq.~\eqref{eq:hplus_exact} gives
\begin{equation}
	\begin{aligned}
		-2\dot r r\dot u\sin 2u
		{}\simeq{}& -2(an_b)^2 e\sin\mathcal{M}\sin 2\Phi+\mathcal{O}\!\left(e^2\right) \\
		{}={}& e~\mathcal{Q}_0\sin\mathcal{M}\sin 2\Phi+\mathcal{O}\!\left(e^2\right).
		\label{eq:rdotsin2u}
	\end{aligned}
\end{equation}
Reducing the products with $\mathcal M=\Phi-\omega$ and the identities
\begin{align*}
	\cos(\Phi{-}\omega)\cos 2\Phi &= \tfrac12\big[\cos(3\Phi{-}\omega)+\cos(\Phi{+}\omega)\big], \\
	\sin(\Phi{-}\omega)\sin 2\Phi &= \tfrac12\big[\cos(\Phi{+}\omega)-\cos(3\Phi{-}\omega)\big],
\end{align*}
the four $\mathcal O(e)$ contributions to $h_+/A_+$ combine as follows:
\begin{center}
	\begingroup
	\setlength{\tabcolsep}{4pt}
	\begin{tabular}{lccc}
		\hline
		Source & $\cos(3\Phi{-}\omega)$ & $\cos(\Phi{+}\omega)$ & $\cos(\Phi{-}\omega)$\\
		\hline
		$\mathcal Q_0\cos2u$ term of $\mathcal O(e)$ & $-2e$ & $+2e$ & $0$ \\
		$\mathcal Q_1\cos2u$ term of $\mathcal O(e)$ & $-\tfrac34 e$ & $-\tfrac34 e$ & $0$ \\
		$-2\dot rr\dot u\;\sin 2u$ term of $\mathcal O(e)$ & $+\tfrac12 e$ & $-\tfrac12 e$ & $0$ \\
		$\sin^2\iota$ term in $h_+$            & $0$ & $0$ & $\tfrac{e\sin^2\iota}{2(1+\cos^2\iota)}$ \\
		\hline
		Sum & $-\tfrac94 e$ & $\tfrac34 e$ & $\tfrac{e\sin^2\iota}{2(1+\cos^2\iota)}$ \\
		\hline
	\end{tabular}
	\endgroup
\end{center}
Finally, we have
\begin{align}
	\frac{h_+(t)}{A_+} &{}= -\cos 2\Phi \nonumber\\
	&\quad{}- e\Big[\tfrac94\cos(3\Phi-\omega) - \tfrac34\cos(\Phi+\omega)\nonumber\\
	&\quad{}
	- \tfrac{\sin^2\iota}{2(1+\cos^2\iota)}\cos(\Phi-\omega)\Big] + \mathcal{O}\Big(e^2\Big), \label{eq:hplus-final}\\
	\frac{h_\times(t)}{A_\times} &{}= -\sin 2\Phi \nonumber\\
	&\quad- e\Big[\tfrac94\sin(3\Phi-\omega) - \tfrac34\sin(\Phi+\omega)\Big]
	+ \mathcal{O}\Big(e^2\Big). \label{eq:hcross-final}
\end{align}
The analogous calculation for $h_\times$ proceeds in similar fashion. The orbital bracket is obtained from the $h_+$ bracket by the replacement $\cos 2u\to \sin 2u$, $\sin 2u\to -\cos 2u$. This gives the same $9/4$ and $-3/4$ sideband coefficients in the $n=3$ and $n=1$ terms, but without the additional $\sin^2\iota$ contribution present only in $h_+$.

Comparing Eqs.~\eqref{eq:hplus-final}-\eqref{eq:hcross-final} with the generic form
\begin{align}
	h_+(t) =& -A_+\Big[(1+a_2 e)\cos 2\Phi \nonumber\\
	+ &\quad e\big(a_3\cos(3\Phi{-}\omega)+a_1\cos(\Phi{+}\omega)
	- b_1\cos(\Phi{-}\omega)\big)\Big] \nonumber\\
	+ &\quad
	\mathcal{O}\Big(e^2\Big), \label{eq:hplus-a123}\\
	h_\times(t) =&-A_\times\Big[(1+\tilde a_2 e)\sin 2\Phi \nonumber\\
	+ &\quad e\big(a_3\sin(3\Phi{-}\omega)+a_1\sin(\Phi{+}\omega)\big)\Big] + \mathcal{O}\Big(e^2\Big). \label{eq:hx-a123}
\end{align}
we obtain, at Newtonian order,
\begin{align}
	a_3=+\frac{9}{4},\qquad a_1=-\frac{3}{4},\qquad
	\nonumber \\ b_1=\frac{\sin^2\iota}{2(1+\cos^2\iota)}, \qquad a_2=\tilde a_2=0.
	\label{eq:a123-track1}
\end{align}
The small-$e$ Newtonian waveform for $h_+$ contains four phase components $\{\Phi{-}\omega,\Phi{+}\omega,2\Phi,3\Phi{-}\omega\}$, while $h_\times$
contains three $\{\Phi{+}\omega,2\Phi,3\Phi{-}\omega\}$. The dominant mode is unchanged at $\mathcal O(e)$, the sidebands satisfy $|a_3/a_1|=3$, and the additional $\cos(\Phi{-}\omega)$ component in $h_+$ carries an inclination-dependent amplitude $0\le b_1\le 1/2$, vanishing for face-on systems and reaching its maximum for edge-on systems.

Our Newtonian waveform expressions Eqs.~\eqref{eq:hplus_exact}-\eqref{eq:hcross_exact} and their small-$e$ reduction Eqs.~\eqref{eq:hplus-final}-\eqref{eq:hcross-final}
agree with the literature \citep{10.1093/mnras/274.1.115,2001MNRAS.325..358P,2004PhRvD..70f4028D,2016PhRvD..93l4061M, 2009agwd.book.....J}. Any apparent differences are attributed to conventions, including the definition of the quadrupole moment, the overall polarisation normalisation, the notation, or the choice of orbital phase variable. In particular, the comparison with \cite{2009agwd.book.....J} requires rewriting their orbital phase angle $\phi=u=\nu+\omega$ in a mean phase basis $\Phi=\mathcal M+\omega$ used here; the choice of $\Phi$ generates additional $\mathcal O(e)$ sidebands, so the coefficients look different even though the waveform is mathematically the same.

\section{Small-$e$ rewrite in $(\eta, \kappa)$}
\label{app:appendixE}

It is convenient to replace $(e,\omega)$ with the LL pair
\citep{10.1046/j.1365-8711.2001.04606.x}
\begin{equation}
	\eta=e\sin\omega,\qquad \kappa=e\cos\omega.
	\label{eq:eta_kappa}
\end{equation}
The orbital phase is then $\Phi_{\rm asc}(t) = \Phi_{\rm GW}(t)/2$, which advances uniformly in time (apart from the small chirp $\dot{\bar n}_b$) and  provides a reference for small-eccentricity expansions. The same phase $u=\omega+\nu$ governs conjunctions, radial velocities, and eclipse timing in optical observations, so the phase correction derived here applies
generally when converting a measured GW orbital phase into an orbital phase appropriate for EM observables. This is not the first appearance of the LL eccentric parameters in the GW literature, where earlier applications describe an orbit rather than an emitted signal. Searches for continuous GWs from spinning neutron stars in binary systems use the parameter pair that describes the Roemer delay and its modulation of the signal phase \citep{2015PhRvD..91j2003L}. \cite{2022PhRvD.105f4021B} use $(e\sin\omega,e\cos\omega)$ as osculating orbital elements of a nearly circular binary that responds to a stochastic GW background, with applications to the timing of binary pulsars and to laser ranging. Here the pair parameterises the GW signal emitted by the binary itself.

We start from the small-$e$ waveform derived in the previous section, in a polarisation frame where the circular limit has the form
\begin{align}
	h_+(t) &= -A_+\Big[\cos 2\Phi \nonumber\\
	&\quad + e\Big(\tfrac{9}{4}\cos(3\Phi-\omega) - \tfrac{3}{4}\cos(\Phi+\omega)
	- b_1\cos(\Phi-\omega)\Big)\Big] + \nonumber\\
	&\quad\mathcal{O}\Big(e^2\Big), \label{eq:hplus_smalle}\\
	h_\times(t) &= -A_\times\Big[\sin 2\Phi \nonumber\\
	&\quad + e\Big(\tfrac{9}{4}\sin(3\Phi-\omega) - \tfrac{3}{4}\sin(\Phi+\omega)\Big)\Big] + \mathcal{O}\Big(e^2\Big). \label{eq:hcross_smalle}
\end{align}
where $b_1=\sin^2\iota/[2(1+\cos^2\iota)]$ and $\Phi$ is the mean orbital phase from Eq.~\eqref{eq:phib}.
Gravitational radiation reaction drives the binary's inspiral, so $\bar n_b$ evolves with time. Writing $\bar n_{b,\rm ref}=\pi f_{\rm GW, ref}$ for its value at $t_{\rm ref}$, time integration of $\bar n_b(t)=\bar n_{b,\rm ref}+\dot{\bar n}_b(t-t_{\rm ref})+\cdots$ gives the polynomial
\begin{equation}
	\Phi=\Phi_{\rm ref}+\bar n_{b,\rm ref}(t-t_{\rm ref})
	+\tfrac{1}{2}\dot{\bar n}_b(t-t_{\rm ref})^2+\cdots,
	\label{eq:phi_chirp}
\end{equation}
with $\Phi_{\rm ref}=\Phi(t_{\rm ref})$ fixed below by Eq.~\eqref{eq:phib}. For the constant $\dot\omega$ model adopted here, $\dot{\bar n}_b=\dot n_b$.
At $\mathcal O(e)$, the $2\Phi$ carrier has no correction that is linear in $e$. Terms linear in $e$ appear in additional harmonic components instead.

Now we write Eqs.~\ref{eq:hplus_smalle} and \ref{eq:hcross_smalle} in terms of the LL pair $(\eta, \kappa)$, Eq.~\ref{eq:eta_kappa},
\begin{align*}
	e\cos(\Phi+\omega)  &= \kappa\cos\Phi - \eta\sin\Phi, \\
	e\sin(\Phi+\omega)  &= \kappa\sin\Phi + \eta\cos\Phi, \\
	e\cos(3\Phi-\omega) &= \kappa\cos3\Phi + \eta\sin3\Phi, \\
	e\sin(3\Phi-\omega) &= \kappa\sin3\Phi - \eta\cos3\Phi, \\
	e\cos(\Phi-\omega)  &= \kappa\cos\Phi + \eta\sin\Phi.
\end{align*}
Using $e\cos\omega=\kappa$ and $e\sin\omega=\eta$, the  factors of $e$ are absorbed into the LL parameters, and the $\mathcal O(e)$ strains become linear in $(\eta,\kappa)$:
\begin{align}
	h_+(t) &= -A_+\Big[\cos 2\Phi \nonumber\\
	&\quad + \kappa\Big(\tfrac{9}{4}\cos3\Phi - \big(\tfrac{3}{4}+b_1\big)\cos\Phi\Big) \nonumber\\
	&\quad + \eta\Big(\tfrac{9}{4}\sin3\Phi + \big(\tfrac{3}{4}-b_1\big)\sin\Phi\Big)\Big] + \mathcal{O}\Big(e^2\Big),\\
	h_\times(t) &= -A_\times\Big[\sin 2\Phi \nonumber\\
	&\quad + \kappa\Big(\tfrac{9}{4}\sin3\Phi - \tfrac{3}{4}\sin\Phi\Big) \nonumber\\
	&\quad - \eta\Big(\tfrac{9}{4}\cos3\Phi + \tfrac{3}{4}\cos\Phi\Big)\Big] + \mathcal{O}\Big(e^2\Big).
\end{align}
Now we recall that $A_+$ and $A_\times$ are expressed in terms of the single amplitude $\mathcal A$ and inclination $\iota$
$$
A_+ = \mathcal A (1+\cos^2\iota),\qquad A_\times = 2\mathcal A \cos\iota.
$$
Inserting these gives the final form
\begin{align}
	h_+(t) &= -\mathcal A (1+\cos^2\iota)\Big[\cos 2\Phi \nonumber\\
	&\quad + \kappa\Big(\tfrac{9}{4}\cos3\Phi - \big(\tfrac{3}{4}+b_1\big)\cos\Phi\Big) \nonumber\\
	&\quad + \eta\Big(\tfrac{9}{4}\sin3\Phi + \big(\tfrac{3}{4}-b_1\big)\sin\Phi\Big)\Big] + \mathcal{O}\Big(e^2\Big),
	\label{app:final_form_h_plus}\\
	h_\times(t) &= -2\mathcal A \cos\iota\Big[\sin 2\Phi \nonumber\\
	&\quad + \kappa\Big(\tfrac{9}{4}\sin3\Phi - \tfrac{3}{4}\sin\Phi\Big) \nonumber\\
	&\quad - \eta\Big(\tfrac{9}{4}\cos3\Phi + \tfrac{3}{4}\cos\Phi\Big)\Big] + \mathcal{O}\Big(e^2\Big).
	\label{app:final_form_h_cross}
\end{align}
where $b_1=\sin^2\iota/[2(1+\cos^2\iota)]$.
Here $(\eta,\kappa)$ are evaluated at $t_{\rm ref}$, and the argument of periastron is held fixed. Apsidal advance rotates the pair rigidly, which displaces each sideband in frequency without altering the coefficients above. Appendix~Sect.~\ref{app:apsidal_mcmc} gives the shifted frequencies and the sampled parameterisation.

Following the plane wave treatment of \citet{2004PhRvD..69h2003R} and the GW phase parameterisation of \citet{2007PhRvD..76h3006C}, we write the waveform in terms of the wave variable
\begin{equation}
	\chi \;=\; t-\hat{\boldsymbol k}\!\cdot\!\boldsymbol r(t)/c,
	\label{eq:chi_def}
\end{equation}
where $t$ is TCB, $\hat{\boldsymbol k}$ is the GW propagation direction, and $\boldsymbol r(t)$ is the barycentric position of the detector. Constant $\chi$ labels a surface of constant phase for the incoming plane wave. 
The LISA GW phase is anchored at $t_{\rm ref}$ (the analysis reference point
for the phase variable), and the phase of GW frequency component $n$ is
\begin{align}
	\Psi_n(\chi)&= 2\pi f_n~(\chi-t_{\rm ref})
	+ \pi \dot f_n~(\chi-t_{\rm ref})^2
	+ \tfrac{n}{2}\phi_{\rm ref},
	\nonumber\\
	&\quad f_n=\tfrac{n}{2} f_{\rm GW,ref},\
	\dot f_n=\tfrac{n}{2}\dot f_{\rm GW},
	\label{eq:psi_n}
\end{align}
with $f_{\rm GW,ref}$ evaluated at $t_{\rm ref}$ and
$\phi_{\rm ref}=\Psi_2(t_{\rm ref})$. The dominant mode is $n=2$; the
eccentricity sidebands are $n=1$ and $n=3$, displaced from $f_n$ by the apsidal
rate as given in Appendix~\ref{app:apsidal_mcmc}. The mean orbital phase
$\Phi_{\rm asc}$, abbreviated $\Phi$ in this appendix, is
\begin{equation}
	\Phi(\chi)=\frac{1}{2}\Phi_{\rm GW}=\tfrac{1}{2}\Psi_2(\chi)
	= \bar n_{b,\rm ref}(\chi-t_{\rm ref})+\tfrac{1}{2}\dot{\bar n}_b(\chi-t_{\rm ref})^2+\tfrac{1}{2}\phi_{\rm ref},
	\label{eq:phib}
\end{equation}
and the argument of latitude follows from
$u(\chi)=\Phi(\chi)+\delta_{e,\omega}(\chi)
=\Phi_{\rm GW}(\chi)/2+\delta_{e,\omega}(\chi)$,
recovering Eq.~\eqref{eq:u_mapping_simple}.
With the convention of Appendix~Sect.~\ref{app:appendixA1}, $T_{\rm asc}$ is the circular limit ascending node epoch, defined by the phase condition $\Phi(T_{\rm asc})=0$,
\begin{equation}
	T_{\rm asc}=t_{\rm ref}-\phi_{\rm ref}/(2 \bar n_{b,\rm ref})+\mathcal{O}(\dot{\bar n}_b/\bar n_{b,\rm ref}^3),
	\label{eq:Tasc_derived}
\end{equation} modulo the $u\!\to\!u+\pi$ ambiguity discussed in Appendix~Sect.~\ref{app:appendixA2}. We check consistency with \citet{2004PhRvD..69h2003R} and Appendix~A of \cite{2007PhRvD..76h3006C}
(CL07). \citet{2007PhRvD..76h3006C} use only the dominant ($n=2$) mode $\Psi_{\rm CL07}(\xi)=2\pi f_{\rm ref}\xi+\pi\dot f_{\rm ref}\xi^2+\varphi_{\rm ref}$. This corresponds to our $\Psi_2(\chi)$ (Eq.~\eqref{eq:psi_n} with $n=2$) under $(f_0,\dot f_0,\varphi_0)\leftrightarrow(f_{\rm GW,ref},\dot f_{\rm GW},\phi_{\rm ref})$ and $\xi\leftrightarrow(\chi - t_{\rm ref})$. The extension to eccentric orbits adds the $n=1,3$ sidebands separated from the dominant mode by approximately half its frequency, encoded linearly in $(\eta,\kappa)=(e\sin\omega,e\cos\omega)$, which share the underlying orbital phase via the $n\phi_{\rm ref}/2$ offset of Eq.~\eqref{eq:psi_n}. In the circular limit $(e\to 0)$, only the $n=2$ mode survives at $f_2=f_{\rm GW, ref}$, recovering the leading-order quadrupole waveform of CL07.

\subsection{Small-$e$ sidebands}
We define the small-$e$ coefficients as
\begin{equation}
	\begin{aligned}
		C_2^{+} &= 1, \\
		C_1^{+} &= -\tfrac{3}{4}(\kappa_{\rm ref}+i\eta_{\rm ref}) \\
		&\quad - b_1(\kappa_{\rm ref}-i\eta_{\rm ref}), \\
		C_3^{+} &= \tfrac{9}{4}(\kappa_{\rm ref}-i\eta_{\rm ref}), \\
		C_2^{\times} &= 1, \\
		C_1^{\times} &= -\tfrac{3}{4}(\kappa_{\rm ref}+i\eta_{\rm ref}), \\
		C_3^{\times} &= \tfrac{9}{4}(\kappa_{\rm ref}-i\eta_{\rm ref}),
	\end{aligned}
	\label{eq:coefficients}
\end{equation}
where $b_1=\sin^2\iota/[2(1+\cos^2\iota)]$,
which vanishes for face-on systems.
With $(\eta_{\rm ref}, \kappa_{\rm ref})=(e\sin\omega_{\rm ref},e\cos\omega_{\rm ref})$, the CL07 quadrature pair reads,
\begin{align}
	h_+(\chi) &= -\Re\!\left[A_+\sum_{n=1}^{3} C_n^{+}~e^{i\Psi_n(\chi)}\right], \label{eq:hplus-phasor}\\
	h_\times(\chi) &= \Re\!\left[e^{i\pi/2}A_\times\sum_{n=1}^{3} C_n^{\times}~e^{i\Psi_n(\chi)}\right], \label{eq:hcross-phasor}
\end{align}
$$
A_+=\mathcal{A}\left(1+\cos^2\iota\right),\qquad A_\times=2\mathcal{A}\cos\iota.
$$

\subsection{Detector response (component-wise slow-fast split)}
If we let $d^{+,\times}_{lm}(t)$ be the arm-projected antenna factors for the link from spacecraft $l$ to spacecraft $m$ and $\mathcal T(f,t,\hat{\boldsymbol k})$ the one-arm transfer (CL07, Appendix~A). We write
$$
y_{lm}(t)=\Re\!\Bigg[\sum_{n=1}^{3} y^{\rm slow}_{lm,n}(t)e^{i2\pi f_n t}\Bigg],
$$
with slow envelopes
\begin{equation}
	y^{\rm slow}_{lm,n}(t)
	= \mathcal T_{lm}(f_n,t,\hat{\boldsymbol k})~\mathcal C_{lm,n}(t)~e^{i\Psi_n(t)},
	\label{eq:perharm_slow_env}
\end{equation}
where
\begin{align}
	\mathcal C_{lm,n}(t) &= \tfrac14\Big[
	d^{+}_{lm}(t)\big(-A_+ C_n^{+}\cos2\psi + iA_\times C_n^{\times}\sin2\psi\big) \nonumber\\
	&\quad + d^{\times}_{lm}(t)\big(A_+ C_n^{+}\sin2\psi + iA_\times C_n^{\times}\cos2\psi\big)
	\Big], \\
	\Psi_n(t) &= \pi\dot f_n(\chi-t_{\rm ref})^2
	+ \tfrac{n}{2}\phi_{\rm ref} \nonumber\\
	&\quad - 2\pi f_n
	\left(t_{\rm ref}+\frac{\hat{\boldsymbol k}\cdot\boldsymbol r(t)}{c}\right),
	\end{align}
Rapid oscillations are in the carriers $e^{i2\pi f_n t}$. The remaining factors in $y^{\rm slow}_{lm,n}(t)$ evolve slowly with time. The antenna functions and the transfer function follow the spacecraft motion, and the residual phase $e^{i\Psi_n(t)}$ carries the chirp and the barycentric Doppler delay.
The $n=1,2,3$ components share an orbital phase, so their reference phases are $n\phi_{\rm ref}/2$, and $\dot f_n=n\dot f_{\rm GW}/2$. When the argument of periastron advances, the coefficients of Eq.~\eqref{eq:coefficients} are no longer constant. The LL pair rotates at $\dot\omega$, slowly compared with the carriers. The rotation appears in $\mathcal C_{lm,n}(t)$ and shifts the sideband frequencies without affecting the $n=2$ mode.

\subsection{Circular limit and recovery of CL07}
For $e\to0$ the sidebands vanish $\left(C_{1,3}^{+,\times}\to0\right)$, leaving only $n=2$:
\begin{equation}
	y^{\rm slow}_{lm,2}(t)
	= \mathcal T_{lm}(f_{\rm GW, ref},t,\hat{\boldsymbol k})~\mathcal C_{lm,2}(t)~e^{i\Psi_2(t)}.
	\label{eq:perharm_slow_circ}
\end{equation}
\begin{align}
	\mathcal C_{lm,2}(t) &= \tfrac14\Big[
	d^{+}_{lm}(t)\big(-A_+\cos2\psi + iA_\times\sin2\psi\big) \nonumber\\
	&\quad + d^{\times}_{lm}(t)\big(A_+\sin2\psi + iA_\times\cos2\psi\big)
	\Big], \label{eq:Clm2}\\
	\Psi_2(t) &= \pi\dot f_{\rm GW}(\chi-t_{\rm ref})^2
	+ \phi_{\rm ref} \nonumber\\
	&\quad - 2\pi f_{\rm GW,ref}
	\left(t_{\rm ref}+\frac{\hat{\boldsymbol k}\cdot\boldsymbol r(t)}{c}\right).
	\label{eq:Pslmk2}
\end{align}
This has the same fast-slow detector-response structure as CL07. The polarisation axis is aligned with the ascending node and $\Phi$ is measured
from the node \citep{2009agwd.book.....J};
the circular limit is $h_+=-\mathcal A(1+\cos^2\iota)\cos2\Phi$ and
$h_\times=-2\mathcal A\cos\iota~\sin2\Phi$. With this choice the polarisation
angle $\psi$ is the position angle of the ascending node on the sky (modulo
$\pi$), and $T_{\rm asc}$ follows from the carrier phase with no offset.
Expanded to $\mathcal{O}(e)$ and rewritten in $(\Phi,\eta,\kappa)$ via the
equation of the centre $\xi\to\Phi-\omega+2e\sin(\Phi-\omega)$, the polarisations of
\citet{2026arXiv260322377B} match
Eqs.~\eqref{eq:hplus_smalle}--\eqref{eq:hcross_smalle} term by term, differing
only in the sign of $b_1$. This reflects the polarisation vector handedness.
Their basis and the ascending node convention used here differ by a $\pi/2$
rotation, flipping the sign of the $\cos(\Phi-\omega)$ term that $b_1$ multiplies.

\subsection{Including apsidal advance in the targeted search}
\label{app:apsidal_mcmc}

For UCBs, the argument of periastron need not remain fixed over a multi-year LISA observation.  We promote it to
$$
\omega(t)=\omega_{\rm ref}+\dot\omega(t-t_{\rm ref}),
$$
with constant $\dot\omega$.  Since $\kappa+i\eta=e\exp(i\omega)$, this amounts to a rigid rotation of the LL parameters,
$$
(\kappa_{\rm ref}\pm i\eta_{\rm ref}) \;\longrightarrow\; (\kappa_{\rm ref}\pm i\eta_{\rm ref})~
e^{\pm i\dot\omega(t-t_{\rm ref})}.
$$
In the Fourier domain, the apsidal rotation shifts each sideband by the rate at which its coefficient phase advances. With $f_{\rm GW, ref}$ denoting the dominant $n=2$ frequency mode, the sideband frequencies become
$$
\tfrac{1}{2}f_{\rm GW, ref} \pm \frac{\dot\omega}{2\pi},
\qquad
\tfrac{3}{2}f_{\rm GW, ref} - \frac{\dot\omega}{2\pi}.
$$
The two $n=1$ frequencies correspond to the two pieces of $C_1^+$ in Eq.~\eqref{eq:coefficients}, which rotate in opposite senses under $\dot\omega$. The lower of the two lies at $f_{\rm GW, ref}/2-\dot\omega/2\pi=n_b/2\pi$, the anomalistic orbital frequency\footnote{This line arises from the $b_1$ piece of $C_1^+$, which is present in $h_+$ only and vanishes for face-on systems.}, so that it and the carrier measure $n_b$ and $\bar n_b$ separately; $f_{\rm GW, ref}/2$ is not the orbital frequency.  The $n=2$ mode amplitude is independent of $(\eta,\kappa)$ at this order and is unaffected.

In the targeted MCMC analysis, we fixed the sky position $(\theta,\phi)$ at its value from the preceding broad LISA search, and sampled
$$
\boldsymbol\theta = \big(
\log_{10}\mathcal A,~
f_{\rm GW, ref},~
\log_{10}\dot f_{\rm GW},~
\cos\iota,~
\psi,~
\phi_{\rm ref},~
\kappa_{\rm ref},~
\eta_{\rm ref},~
z_\omega
\big),
$$
where $$z_\omega= \frac{\dot\omega}{2\pi}T_{\rm obs}.$$ We used a narrow targeted prior on $f_{\rm GW, ref}$, centred on the search value and spanning $\pm 10/T_{\rm obs}$, corresponding to
$\pm 10$ Fourier bins. We also used a uniform prior on $z_\omega\in[0,50]$ frequency bins, wider than the injected apsidal advance ($z_\omega \approx 30.1$ frequency bins). The range $[0,50]$ reflects the positivity of $\dot\omega$ in GR. For each proposed $z_\omega$, the sideband pieces were
generated at the shifted frequencies and passed through the same time-dependent LISA response as the dominant mode.  The sideband templates are not single Fourier bins. Their envelope is broadened by the LISA response and chirp, and their central frequency is displaced by the apsidal advance.  The posterior on $z_\omega$ then reflects whether the data are consistent with a static apsidal line ($\dot\omega=0$) or require nonzero apsidal advance.

\end{appendix}
\end{document}